\PassOptionsToPackage{unicode}{hyperref}
\PassOptionsToPackage{hyphens}{url}
\documentclass[
]{article}
\usepackage{lmodern}
\usepackage{amsmath}
\usepackage{ifxetex,ifluatex}
\ifnum 0\ifxetex 1\fi\ifluatex 1\fi=0 
  \usepackage[T1]{fontenc}
  \usepackage[utf8]{inputenc}
  \usepackage{textcomp} 
  \usepackage{amssymb}
\else 
  \usepackage{unicode-math}
  \defaultfontfeatures{Scale=MatchLowercase}
  \defaultfontfeatures[\rmfamily]{Ligatures=TeX,Scale=1}
\fi
\IfFileExists{upquote.sty}{\usepackage{upquote}}{}
\IfFileExists{microtype.sty}{
  \usepackage[]{microtype}
  \UseMicrotypeSet[protrusion]{basicmath} 
}{}
\makeatletter
\@ifundefined{KOMAClassName}{
  \IfFileExists{parskip.sty}{%
    \usepackage{parskip}
  }{
    \setlength{\parindent}{0pt}
    \setlength{\parskip}{6pt plus 2pt minus 1pt}}
}{
  \KOMAoptions{parskip=half}}
\makeatother
\usepackage{xcolor}
\IfFileExists{xurl.sty}{\usepackage{xurl}}{} 
\IfFileExists{bookmark.sty}{\usepackage{bookmark}}{\usepackage{hyperref}}
\hypersetup{
  pdftitle={A Road Segment Prioritization Approach for Cycling Infrastructure},
  pdfauthor={Hussein Mahfouz1,; Elsa Arcaute1; Robin Lovelace2},
  hidelinks,
  pdfcreator={LaTeX via pandoc}}
\usepackage[margin=2cm]{geometry}
\usepackage{longtable,booktabs}
\usepackage{calc} 
\usepackage{etoolbox}
\makeatletter
\patchcmd\longtable{\par}{\if@noskipsec\mbox{}\fi\par}{}{}
\makeatother
\IfFileExists{footnotehyper.sty}{\usepackage{footnotehyper}}{\usepackage{footnote}}
\makesavenoteenv{longtable}
\usepackage{graphicx}
\makeatletter
\def\maxwidth{\ifdim\Gin@nat@width>\linewidth\linewidth\else\Gin@nat@width\fi}
\def\maxheight{\ifdim\Gin@nat@height>\textheight\textheight\else\Gin@nat@height\fi}
\makeatother
\setkeys{Gin}{width=\maxwidth,height=\maxheight,keepaspectratio}
\makeatletter
\def\fps@figure{htbp}
\makeatother
\providecommand{\tightlist}{%
  \setlength{\itemsep}{0pt}\setlength{\parskip}{0pt}}
\usepackage{subcaption}
\usepackage{longtable}
\usepackage{float}
\usepackage{multirow}
\usepackage{placeins}
\usepackage{multicol}
\usepackage{booktabs}
\usepackage{array}
\usepackage{wrapfig}
\usepackage{colortbl}
\usepackage{pdflscape}
\usepackage{tabu}
\usepackage{threeparttable}
\usepackage{threeparttablex}
\usepackage[normalem]{ulem}
\usepackage{makecell}
\ifluatex
  \usepackage{selnolig}  
\fi
\newlength{\cslhangindent}
\newlength{\csllabelwidth}
\newenvironment{CSLReferences}[2] 
 {
  \setlength{\parindent}{0pt}
  \ifodd #1 \everypar{\setlength{\hangindent}{\cslhangindent}}\ignorespaces\fi
  \ifnum #2 > 0
  \setlength{\parskip}{#2\baselineskip}
  \fi
 }%
 {}
\usepackage{calc}

\title{A Road Segment Prioritization Approach for Cycling Infrastructure}
\author{Hussein Mahfouz\textsuperscript{1,*} \and Elsa Arcaute\textsuperscript{1} \and Robin Lovelace\textsuperscript{2}}
\date{}

\begin{document}
\maketitle
\begin{abstract}
Understanding the motivators and deterrents to cycling is essential for creating infrastructure that gets more people to adopt cycling as a mode of transport. This paper demonstrates a new approach to support the prioritization of cycling infrastructure and cycling network design, accounting for cyclist preferences and the growing emphasis on `filtered permeability' and `Low Traffic Neighborhood' interventions internationally. The approach combines distance decay, route calculation, and network analysis methods to examine where future cycling demand is most likely to arise, how such demand could be accommodated within existing street networks, and how to ensure a fair distribution of investment. Although each of these methods has been applied to cycling infrastructure prioritization in previous research, this is the first time that they have been combined, creating an integrated road segment prioritization approach. The approach, which can be applied to other cities, as shown in the Appendix, is demonstrated in a case study of Manchester, resulting in cycling networks that balance directness against the need for safe and stress-free routes under different investment scenarios. A key benefit of the approach from a policy perspective is its ability to support egalitarian and cost-effective strategic cycle network planning.

\par

\end{abstract}

\textsuperscript{1} The Centre for Advanced Spatial Analysis, University College London, London, UK\\
\textsuperscript{2} Institute of Transport Studies, University of Leeds, Leeds, UK

\textsuperscript{*} Correspondence: \href{mailto:h.mahfouz.19@ucl.ac.uk}{Hussein Mahfouz \textless{}\href{mailto:h.mahfouz.19@ucl.ac.uk}{\nolinkurl{h.mahfouz.19@ucl.ac.uk}}\textgreater{}}

\hypertarget{introduction}{%
\section{Introduction}\label{introduction}}

The 2015 Paris agreement (UN, 2015) acknowledged that
fundamental changes to societies and economies are necessary to mitigate
climate change. Like other sectors, transport is under substantial
pressure to decarbonise, resulting in a number of technical innovations
including electric vehicles. But new vehicle technologies can only go so
far and do not tackle parallel problems such as congestion, road traffic
casualties and physical inactivity (Brand et al., 2020).

In this context, interest and investment in active modes are growing.
The benefits extend beyond congestion and the environment, as active
travel can also help alleviate what is referred to as the pandemic of
global inactivity; physical inactivity is on the rise and has become the
4th highest cause of death globally (Kohl 3rd et al., 2012). Various
studies have documented the association between active transport and
lower risk of disease, including cancer and cardiovascular disease
(Celis-Morales et al., 2017; Jarrett et al., 2012; Patterson et al., 2020). In the wake of the Covid-19 pandemic, and
the resulting reduced capacity of public transport, the UK government
has pledged to invest billions of pounds to improve walking and cycling
infrastructure across the country. While this unprecedented sum is an
opportunity to reshape cities in a way that improves the well-being of
citizens, it does come with a warning:

\begin{quote}
``Inadequate cycling infrastructure discourages cycling and wastes
public money. Much cycling infrastructure in this country is
inadequate. It reflects a belief, conscious or otherwise, that hardly
anyone cycles, that cycling is unimportant and that cycles must take
no meaningful space from more important road users, such as motor
vehicles and pedestrians'' (DfT, 2020a).
\end{quote}

The funding on its own is therefore no guarantee of a change in
commuting across the country; it must be used to design adequate cycling
infrastructure that is based on motivators and deterrents to cycling.

\hypertarget{motivators-and-deterrents-to-cycling}{%
\subsection{Motivators and Deterrents to Cycling}\label{motivators-and-deterrents-to-cycling}}

Much research has been done to understand what gets people to cycle.
Segregated cycling infrastructure\footnote{\emph{Segregated cycling infrastructure}
  refers to road space that is allocated to cyclists only, with physical
  separation to protect cyclists from other modes of transport.} has been
shown to increase cycling uptake (Aldred et al., 2019; Goodman et al., 2014; Marqués et al., 2015), with the separation from motorized
vehicles being key (Winters et al., 2011). Revealed preference of
cyclists shows that they are willing to deviate from the most efficient
routes in order to commute on safer roads (Crane et al., 2017).
However, such deviations are only considered if they do not considerably
increase route circuitry; behaviour studies have found that the
probability of choosing a route decreases in proportion to its length
relative to the shortest route (Broach et al., 2011; Winters et al., 2010).
Another defining feature for cycling infrastructure is how well
connected it is; cyclists prefer cohesive infrastructure, particularly
when cycling on arterial roads with high levels of motorized traffic
(Stinson and Bhat, 2003). The lack of well-connected cycling
infrastructure is one of the main obstacles to increasing cycling uptake
(Caulfield et al., 2012). While direct and cohesive cycling networks
have been shown to positively impact cycling rates, density\footnote{making an
  area's bicycle network denser means adding more cycling routes in the
  area and thereby giving cyclists more route options.} of the cycling
network is also vital (Schoner and Levinson, 2014).

\hypertarget{network-level-approaches}{%
\subsection{Network-Level Approaches}\label{network-level-approaches}}

The studies outlined above lay out the fundamentals for designing
cycling networks that generate significant cycling uptake, but they do
not propose network-level interventions. In this section we outline
methods used in past studies, namely optimization and network
analysis methods, such as connected components and community
detection, and examine how they are leveraged to suggest cycling network
designs. We compare the effectiveness of these network-level studies in
incorporating the fundamentals outlined above.
Our proposed approach is inspired by these methods, but it attempts to add to them
by ensuring that all of the outlined fundamentals are accounted for. It
also goes further by attempting to factor in ethical considerations
related to the distribution of investment.

\emph{Optimization} methods have been used to propose improvements to
cycling networks. Mesbah et al. (2012) propose a bi-level formulation to
optimize allocation of cycling lanes to the network without exceeding a
set budget. They account for the effect of cycling lanes on car traffic,
and attempt to maximize utilization of said lanes with minimal impact on
car travel times. Safeguarding against increased car traffic may be counter-productive if the goal is to create a mode-shift, as research has shown that reducing road space for cars leads to less cars on the road, a phenomenon referred to as ``traffic evaporation'' (Nello-Deakin, 2020).
On the other hand, Mauttone et al. (2017) developed an optimization framework aiming at minimizing the total
user cost of cycling on the network without considering car usage. The aggregate flow\footnote{\emph{flow} is
  used throughout this research to refer to the cycling demand when it is
  routed onto the road network. The flow on any road segment is the
  cumulative demand on it, resulting from cyclists commuting between
  various Origin-Destination (OD) pairs.} on road segments (links) is obtained by using
shortest paths to route existing cycling demand onto the road network. The solution is a proposed set of links where cycling infrastructure
should be added in order to minimize the overall travel cost of cyclists
across the network. The cost of traversing a link is given as a function
of its length and whether or not it has cycling infrastructure. The
problem has also been solved by attempting to find the minimum cost of
improving roadway links to meet a desired level of service (LOS)
(Duthie and Unnikrishnan, 2014). In this formulation, all origin-destination
(OD) pairs need to be connected by roads that meet the desired LOS. A
limitation of these approaches is that they do not explicitly solve for
continuity. The latter is addressed using either a constraint specifying
that each link with a bike lane should be connected to at least one
destination (Mesbah et al., 2012), a constraint on maximum deviation
from shortest paths (Duthie and Unnikrishnan, 2014), or a discontinuity
penalty to prioritize connected road segments (Mauttone et al., 2017).

In this paper, continuity is analysed by looking at the connectivity of
the network through the graph-theoretic concept of \emph{connected
components}. Natera et al. (2019) for example,
study the existing cycling network in terms of its disconnected
components. They propose two different algorithms to connect disconnected components; one connects the two largest components, and the other connects the largest component to its nearest neighbor. By measuring the growth of the largest connected component as a function of the kilometers of network added, they determine that these approaches are more effective at improving cycling network connectivity than random allocation of cycling infrastructure.
The concept of connected components is
also at the core of the methodology proposed by Olmos et al. (2020). After
routing the cycling demand onto the network links, they use percolation
theory to filter out the links based on the aggregate flow passing through them, varying the flow threshold for
filtering to identify the minimum flow at which the whole city is
connected by a giant component. While these approaches deal with
continuity better, they look at the network as a whole when attempting
to improve it, and in doing so fail to account for equitable
distribution of infrastructure.

\hypertarget{ethical-underpinnings-and-proposed-approach}{%
\subsection{Ethical Underpinnings and Proposed Approach}\label{ethical-underpinnings-and-proposed-approach}}

All of these network-level methodologies are underpinned by ethical
principles, even though these principles are not explicitly acknowledged
by the authors. This is important since different ethical principles
constitute different problem formulations and targets. Broadly speaking,
transport appraisal can be based on either utilitarian or egalitarian
principles. The former seeks to maximize the overall benefit, while the
latter is concerned with a fair distribution of benefits
(Jafino et al., 2020).
The utilitarian approach, historically popular in transport planning, has been criticised for focusing on the bigger picture and failing to account for the distribution of investments on the different communities of the study area (Lucas et al., 2016; Nahmias-Biran et al., 2017).
Pereira et al. (2017) emphasize the need for a more
egalitarian approach to transport planning. They highlight accessibility
as a cornerstone of distributive justice, and contend that policies
should aim to distribute investments in a way that minimizes spatial
variations in accessibility.

This research attempts to propose an egalitarian framework for cycling
network design. This is done by identifying the different sub-networks
that exist within the larger network, and ensuring that each gets a fair
share of investment. Trip patterns in a city are not uniformly
distributed geographically, and \emph{community finding} methods have been
used to partition study areas into localized areas that experience a
disproportionate number of trips within them. Akbarzadeh et al. (2018)
use a modularity maximization approach (Blondel et al., 2008) on taxi trip
data to identify 7 different communities in the city of Isfahan, Iran.
An optimization problem is then formulated to connect nodes within each
community with cycling infrastructure, with the emphasis being on
connectivity within the communities, not between them. Bao et al. (2017)
adopt a similar methodology, first identifying communities and then
using a greedy network expansion algorithm to simultaneously add links
to each community. The link with the highest benefit-cost ratio in each
community is selected, and the network is grown by adding neighboring
links to the solution until a budget limit is met. The benefit is the
flow on the link, and each link is assigned a cost based on current road
conditions.

Our work builds on these \emph{community finding} approaches by proposing a
similar greedy network expansion algorithm for cycle network expansion
within communities. We incorporate community finding methods for study
area partitioning with weighted routing to avoid links that are
stressful to cycle on. In doing so, we propose an approach that
accounts for motivators and deterrents to cycling. We propose three
sub-methods that address some of the limitations of previous
studies. These limitations include (a) an inherent bias of basing network
design solely on existing cycling demand, (b) the proposal of routes that may
not correspond to studies on cyclist preference and government policies,
and (c) an insufficient consideration of the ethical principles underlying
the analysis. The current work addresses these limitations through the following organisation. Section \ref{calculating-potential-cycling-demand}
focuses on calculating potential cycling demand. Section \ref{routing}
focuses on routing the demand onto the road network while accounting for
cyclist preferences and government priorities. Section
\ref{road-segment-prioritization} outlines a method for partitioning the
study area based on a community finding algorithm and routed cycling
demand. It then introduces the network expansion algorithms, and compares an approach grounded in
`egalitarianism' to one grounded in `utilitarianism.'

\hypertarget{calculating-potential-cycling-demand}{%
\section{Calculating Potential Cycling Demand}\label{calculating-potential-cycling-demand}}

Many of the cycling network studies mentioned above use demand data for cycling as a starting point.
Some use existing cycling demand, some calculate potential cycling demand, and others ignore demand completely.
Duthie and Unnikrishnan (2014) note that relying only on existing cycling activity to prioritize cycling
infrastructure can reinforce existing cycling patterns and ignore potential cycling demand that could be satisfied by a connected network.
To avoid this issue they choose to ignore existing demand completely, and focus on creating a network that connects the
entire study area.
Olmos et al. (2020) opt to calculate potential demand instead; they obtain the distance distribution of cyclists using a smartphone-based bicycle GPS data, and then use a rejection-sampling algorithm on the OD data of the study area to match
the potential demand distribution to the distribution obtained from GPS data.

\begin{figure}

{\centering \includegraphics[width=1\linewidth]{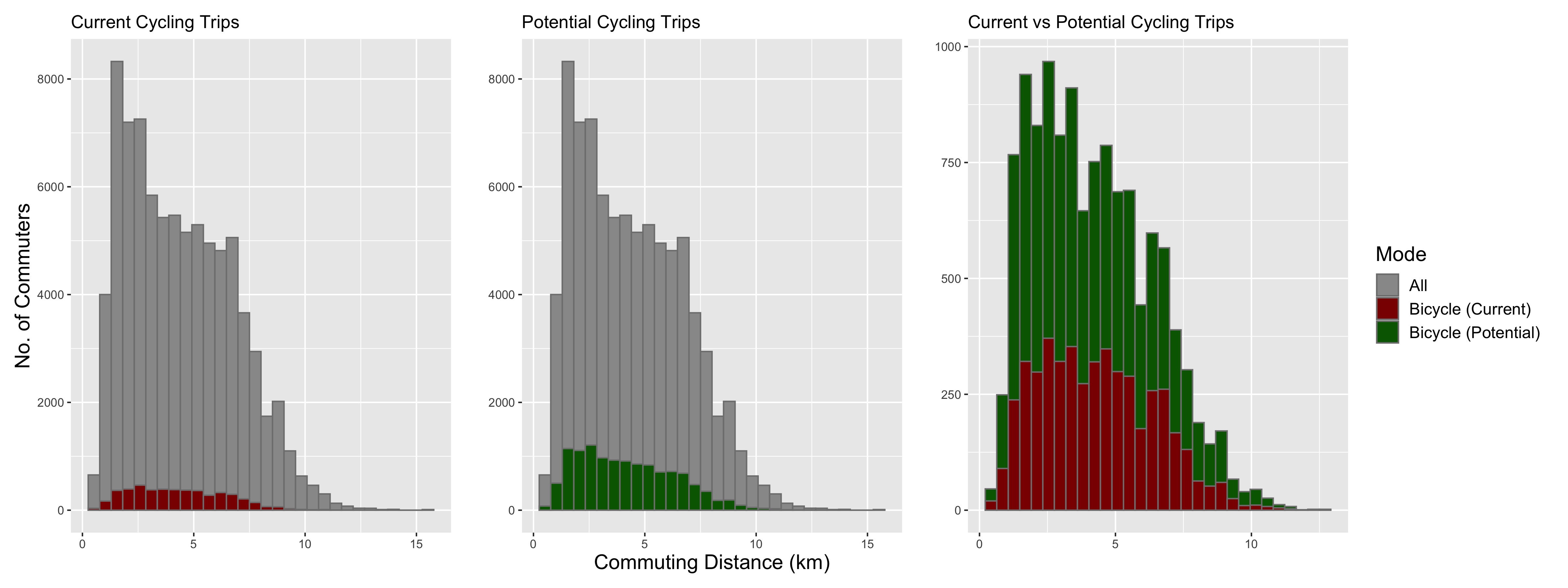} 

}

\caption{Distribution of potential cycling demand}\label{fig:potdemhistograms}
\end{figure}

\begin{figure}[H]

{\centering \includegraphics[width=0.65\linewidth]{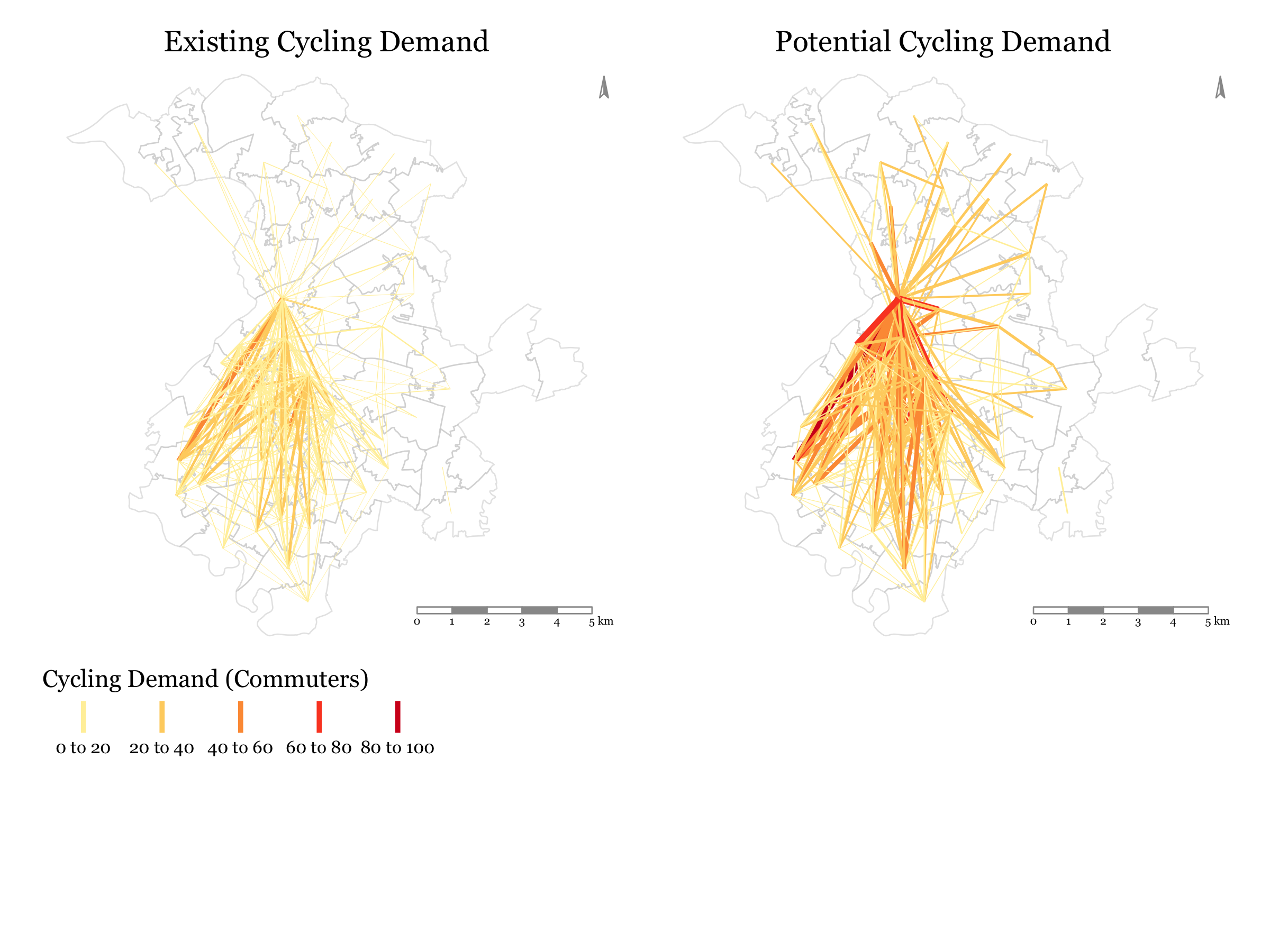} 

}

\caption{Current and potential cycling demand}\label{fig:desirefacetcycling}
\end{figure}

\emph{OD} data can be obtained from a range of sources,
including GPS data, household travels surveys and Census data on work
locations. In areas where observed OD data is unavailable, modelling
techniques such as spatial interaction models can be used to estimate
travel volumes between zones (Black, 1995; Martínez and Viegas, 2013; Wilson, 1971). In this paper we use open access data from the UK
census (ONS, 2011), which contains aggregate statistics on
number of commuters between administrative zones --- Middle layer Super
Output Areas (MSOA) --- by mode of travel. MSOAs have an average
population of just over 8000 (ONS, 2018).\footnote{ See
  \url{https://wicid.ukdataservice.ac.uk/} for open access to the OD data.}
Figures \ref{fig:potdemhistograms} and \ref{fig:desirefacetcycling}
illustrate the proportion of trips cycled by distance and the
geographic extent of the input OD dataset used in this paper.

For our purposes, we use a logistic regression model to calculate
potential cycling demand. The model was adapted from the Propensity to
Cycle Tool (PCT), which estimates the proportion of trips
(\(\boldsymbol{C_{p}}\)) for each OD pair that would cycle under
different scenarios of change as a
function of distance and hilliness (Lovelace et al., 2017). We used the Government Target scenario,
indicating a nationwide target of doubling cycling by 2025. The
logistic regression model used to calculate \(\boldsymbol{C_{p}}\) has
the following parameters:

\begin{align}\label{eq:pcteqn}
     logit(C_{p}) = & \alpha + \beta_1 d + \beta_2 \sqrt{d} + \beta_3 d^2 + \beta_4 s + \beta_5 ds + \beta_6 \sqrt{d}s 
\end{align}

\noindent where \(d\) and \(s\) are the distance and slope associated with each
OD pair, and \(\alpha\) and all \(\beta\)s are parameters calculated by a regression
model on the training data. The square and square-root distance terms
``capture the non-linear impact of distance on the likelihood of
cycling,'' and interaction terms to capture the combined effect of slope
and distance (Lovelace et al., 2017). Alternative cycling uptake
models could be `plugged in' to our approach for different contexts or scenarios of change.

The potential demand calculations show that the current and potential
number of cyclists both follow a bell-shaped distribution, with the
number of trips peaking around the 3-4km commuting distance and then
going back down for longer distances (see Figures
\ref{fig:potdemhistograms} and \ref{fig:desirefacetcycling}).

The uptake model used in this paper assumes uptake of traditional bicycles,
which are affected by topography and distance due to physical effort.
As discussed in Section \ref{discussion-and-conclusions}, using adapted uptake
models could enable the approach to help plan for solutions such as e-bikes
that enable trips covering longer distances and hillier roads.

\hypertarget{routing}{%
\section{Routing}\label{routing}}

The next step in our approach is to route the potential cycling demand
(\(\boldsymbol{C_{p}}\)) between all OD pairs onto the road network. We
choose not to use the PCT approach for routing as it is reliant on
external and therefore inflexible routing services.

\begin{longtable}{lll}
\caption{OSM road types}
\label{table:osmroadtypes}\\
\textbf{OSM Road Type } & \textbf{Description}                                                                                                                                            & \textbf{UK Equivalent}                                                                                                   \endfirsthead 
\hline
Motorway                & \begin{tabular}[c]{@{}l@{}}Road open to high-speed \\vehicles only\end{tabular}                                                                                 & Motorway                                                                                                                 \\* 
\hline
Trunk                   & \multirow{5}{*}{\begin{tabular}[c]{@{}l@{}}Roads that don't \\meet motorway criteria, \\in descending order \\of importance and \\through-traffic\end{tabular}} & A-Roads with Primary Status                                                                                              \\* 
\cline{3-3}
Primary                 &                                                                                                                                                                 & A-Roads with Non-Primary Status                                                                                          \\* 
\cline{3-3}
Secondary               &                                                                                                                                                                 & B-roads                                                                                                                  \\* 
\cline{3-3}
Tertiary                &                                                                                                                                                                 & \begin{tabular}[c]{@{}l@{}}Classified unnumbered roads \\OR unclassified busy through roads\end{tabular}                 \\* 
\cline{3-3}
Unclassified            &                                                                                                                                                                 & \multirow{2}{*}{\begin{tabular}[c]{@{}l@{}}Unclassified (Intended for \\local traffic - 60\% of UK roads)\end{tabular}}  \\* 
\cline{1-2}
Residential             & Function is purely residential                                                                                                                                  &                                                                                                                          \\ 
\hline
Service                 & \begin{tabular}[c]{@{}l@{}}Road that provides \\access to a facility\end{tabular}                                                                               &                                                                                                                          \\ 
\hline
Track                   & \begin{tabular}[c]{@{}l@{}}Unpaved - suitable for \\two track vehicles (mostly rural)\end{tabular}                                                              &                                                                                                                          \\ 
\hline
Cycleway                & \begin{tabular}[c]{@{}l@{}}Designated cycleway - \\open to cyclists only\end{tabular}                                                                           &                                                                                                                          \\ 
\hline
Path                    & \begin{tabular}[c]{@{}l@{}}Unpaved - open to \\non-motorized traffic only\end{tabular}                                                                          &                                                                                                                          \\
\hline
\end{longtable}

To conduct routing, the following is considered:

\begin{enumerate}
\def\labelenumi{\arabic{enumi}.}
\tightlist
\item
  \textbf{Cyclist Preference}: Work done by Dill and McNeil (2013) on examining
  cyclist typologies determined that around 60\% of Portland residents
  fit under the \emph{interested but concerned} category. These were people
  that enjoyed cycling but avoided it due safety concerns. The key to
  encouraging this group was to create a low-stress cycling network,
  not only though segregated infrastructure but also by planning
  routes that passed through residential streets.
\item
  \textbf{Low-Traffic Neighbourhoods (LTNs)}: The UK Department for Transport is
  allocating funding to local authorities to invest in Active
  Transport, partially through the creation of LTNs
  (DfT, 2020a). This includes closing off residential
  streets to motorized traffic.
\item
  \textbf{Existing Cycling Infrastructure}: Utilizing existing cycling
  infrastructure makes economic sense, as small investments may lead
  to large connectivity gains as the disconnected cycling
  infrastructure gets joined together.
\end{enumerate}

The above points are accounted for by using a weighted road network for
routing. This has been previously done by multiplying all road segments
without cycling infrastructure by a fixed impedance factor
(Mauttone et al., 2017), or by assigning different weights to the road
segments proportional to the investment cost of bringing them to an
acceptable level of stress for cycling (Duthie and Unnikrishnan, 2014).
Perceived stress for cyclists has been found to be a function of vehicular traffic
volume and speed (Sorton and Walsh, 1994), both of which vary predictably
with road type. Gehrke et al. (2020) build on this information to use
routing impedance factors for road segments that are based on road type
and the existence of cycling infrastructure.

For the purposes of this paper, we created a similar weighting profile
that is adjusted to favor less stressful roads (based on the definition
by Sorton and Walsh (1994) and information from Table
\ref{table:osmroadtypes}), and roads with existing cycling
infrastructure. We believe this to be more appropriate than the simplistic
approach adopted by Mauttone et al. (2017), as it makes use of a hierarchy
of road preference based on perceived stress levels, going beyond simply
favoring roads with existing cycling infrastructure.

\begin{longtable}{lrrr}
\caption{Weighting profiles}
\label{table:weightprofiles}\\
\multirow{2}{*}{\textbf{OSM Road Type}} & \multicolumn{3}{c}{\textbf{Weighting Profile}}                                                 \\*
                                        & \multicolumn{1}{l}{\textit{Unweighted}} & \multicolumn{1}{l}{\textit{Weighted}} & \multicolumn{1}{l}{\textit{Weighted\_2}}  \endfirsthead
Cycleway                                & 1          & 1                                     & 1                                         \\
Path                                    & 1          & 0.9                                   & 0.9                                       \\
Residential                             & 1          & 0.9                                   & 0.9                                       \\
Service                                 & 1          & 0.9                                   & 0.9                                       \\
Tertiary                                & 1          & 0.9                                   & 0.9                                       \\
Track                                   & 1          & 0.9                                   & 0.9                                       \\
Unclassified                            & 1          & 0.9                                   & 0.9                                       \\
Secondary                               & 1          & 0.8                                   & 0.8                                       \\
Primary                                 & 1          & 0.7                                   & \textcolor{red}{0}                        \\
Trunk                                   & 1          & 0.6                                   & \textcolor{red}{0}                        \\
Motorway                                & 1          & \textcolor{red}{0}                    & \textcolor{red}{0}                       
\end{longtable}

\noindent A weighted distance \(d_{w}\) for each road
segment is calculated as follows:\footnote{The \textbf{dodgr} r package
  (Padgham, 2019) is used to route cycling demand onto the road
  network. The package uses the OpenStreetMaps (OSM) road network and
  allows the user to assign weights to roads based on their type. The
  routing is done based on weighted shortest paths, with the distance
  along each road segment being divided by a factor to obtain the weighted
  distance for routing. It is more intuitive to multiply when weighting a
  network, but the dodgr package divides by numbers between 0 and 1, which
  achieves the same result (road types with a weight of 0 are not routed on).
  For the sake of reproducibility, we stick to
  the convention used in the package.}

\begin{equation}\label{eq:weight_distance}
    d_{w} = \frac{d_{unw}}{W}
\end{equation}

\noindent where \(d_{unw}\) is the unweighted distance and
\(W\) is the weight from Table \ref{table:weightprofiles}.

All weights are between 0 and 1, and the values in the
\textit{Weighted} profile are chosen to be inversely proportional to
the stress level experienced by cyclists on them. The
\textit{Unweighted} weighting profile is used to compare increases in
route length resulting from two different approaches:

\begin{enumerate}
\def\labelenumi{\arabic{enumi}.}
\tightlist
\item
  \textbf{Weighted}: Relatively high impedance on Primary and Trunk roads
  (to minimize cycling on them).
\item
  \textbf{Weighted\_2}: Avoiding Primary and Trunk Roads completely.
\end{enumerate}

\begin{figure}
\includegraphics[width=0.9\linewidth]{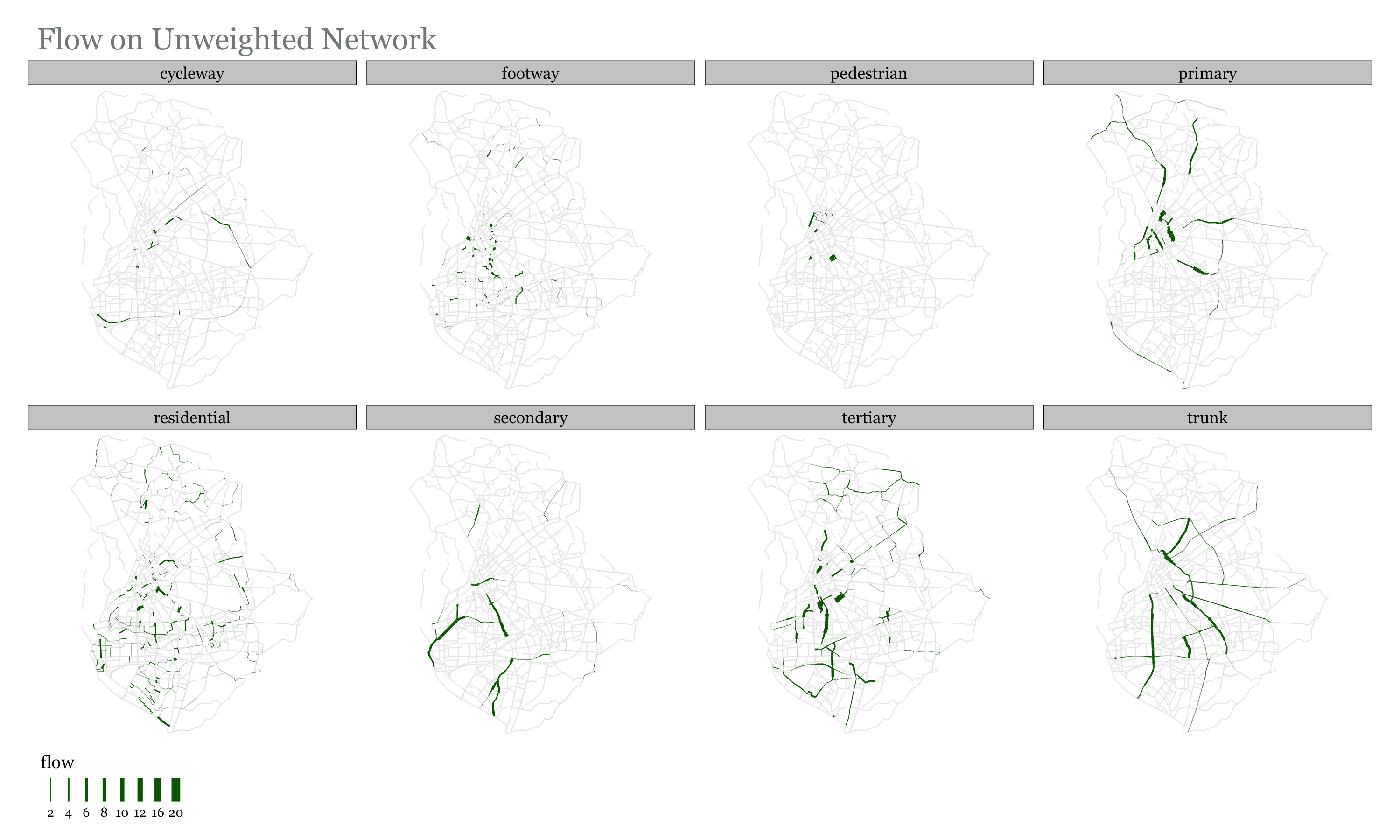} \caption{Flow results based on unweighted shortest paths (Manchester)}\label{fig:flowsfacetunweighted}
\end{figure}

\begin{figure}
\includegraphics[width=0.9\linewidth]{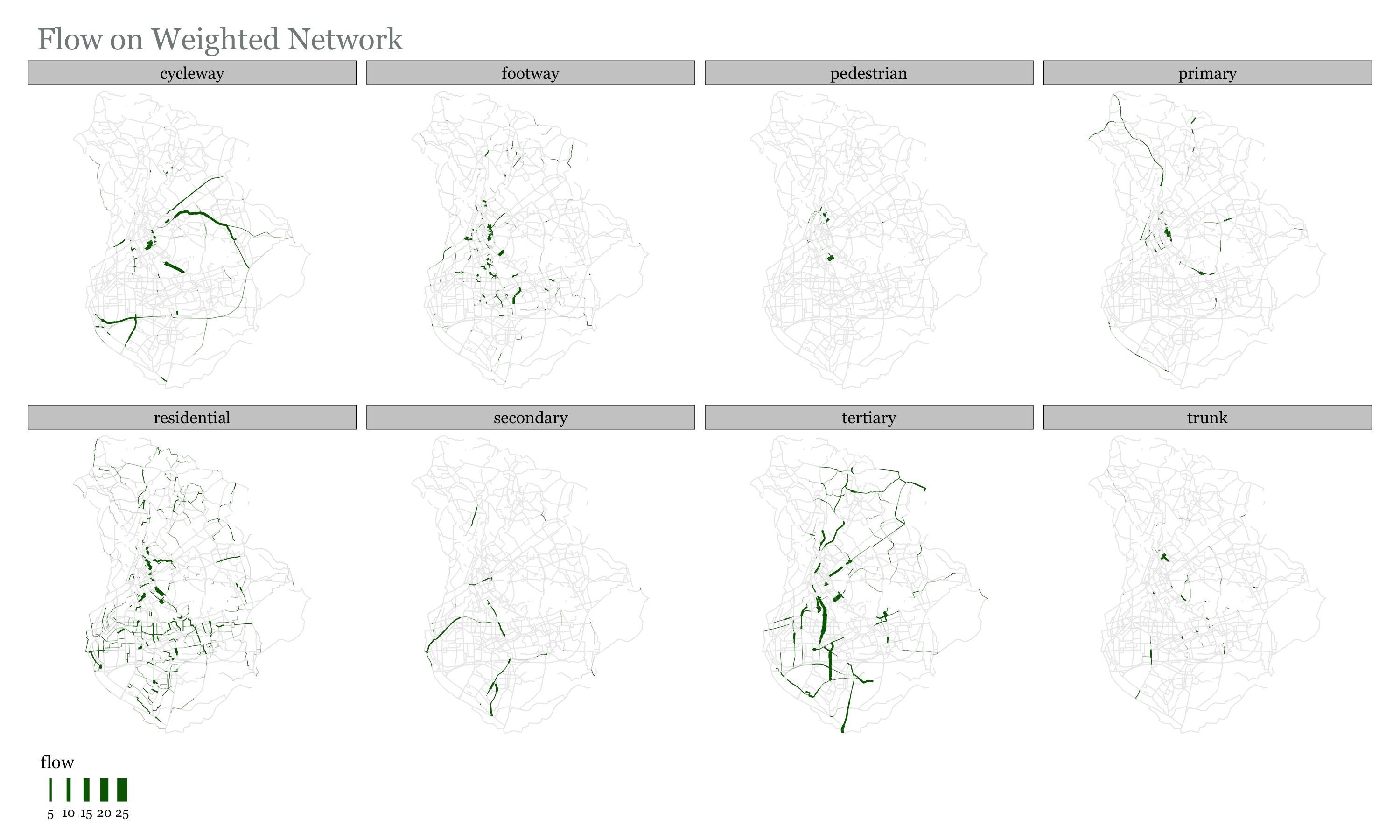} \caption{Flow results based on weighted shortest paths (Manchester)}\label{fig:flowsfacetweighted}
\end{figure}

Comparing the cycling demand routed on the weighted and unweighted road
network allows us to get a better understanding of the importance of
different road types. In the case of Manchester, trunk roads bisect the
city and are a major part of unweighted shortest paths (Figure
\ref{fig:flowsfacetunweighted}). On the other hand, cycleways are not
part of unweighted shortest paths, and so very little of the cycling
demand is routed through them. In the weighted network, cycleways are
much better utilized, and the majority of the cycling demand passes
through tertiary roads, as expected.

The results of routing potential cycling demand on the weighted and
unweighted networks are understandably quite different. From Figure
\ref{fig:flowsfacetunweighted} we can see that trunk and primary roads
are the most efficient means of traversing the road network of
Manchester. Both of these road types are classified as Primary A roads
according to the UK Department for Transport's road classification
(Table \ref{table:osmroadtypes}), and are therefore part of the Primary
Route Network (PRN) (DfT, 2012). The PRN has the widest,
most direct roads on the network, and carries most of the through
traffic. This includes freight, with all roads in the PRN being required
by law to provide unrestricted access to trucks up to 40 tonnes
(DfT, 2012).

We choose to avoid routing the potential cycling demand on Primary A
Roads for the following 2 reasons:

\begin{enumerate}
\def\labelenumi{\arabic{enumi}.}
\tightlist
\item
  \textbf{Logistical Difficulty}: Changes on these roads need to be agreed
  upon by all affected authorities (DfT, 2012), which
  may prove to be difficult.
\item
  \textbf{Low Traffic Neighborhoods (LTNs)}: The UK government is aiming to
  restrict access to motorized vehicles on residential roads to create
  LTNs (DfT, 2020a). This is part of a policy to
  prevent automobile rat-running and make streets more accessible to
  cyclists and pedestrians. Under such a policy, Primary A roads would
  become even more essential for motorized traffic and it would be
  more difficult to reallocate road space on these roads to cyclists.
\end{enumerate}

Figure \ref{fig:flowsfacetweighted} shows that routing on the weighted
network significantly reduces flow on the trunk and primary roads, but
does not eliminate it completely. This is intentional, as the impedance
on these roads is only slightly higher than the remaining road types (See
Table \ref{table:weightprofiles}). Potential cycling demand is only
routed on these roads if there are no routes through other roads that
offer comparable directness.

\begin{figure}

{\centering \includegraphics[width=0.5\linewidth]{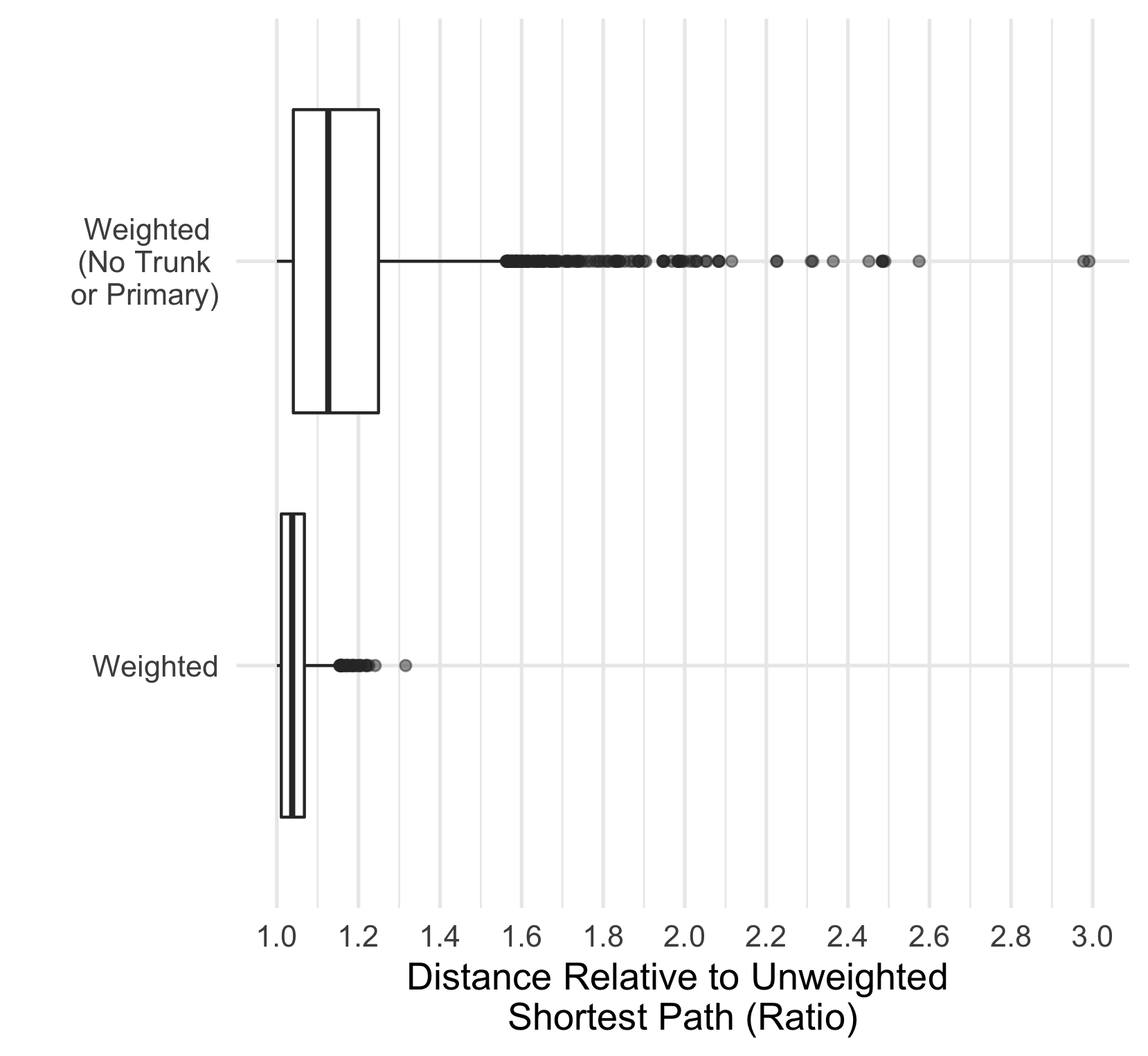} 

}

\caption{Effect of banning cyclists from trunk and primary roads for all OD pairs (Manchester)}\label{fig:boxplotcircuity}
\end{figure}

Banning cycling flow completely on trunk and primary roads may result in
excessively circuitous paths, as seen in Figure
\ref{fig:boxplotcircuity}. When routing using the \emph{weighted} weighting
profile in Table \ref{table:weightprofiles}, we see that shortest paths
increase by less than 5\% on average from unweighted shortest paths, with
the largest increases still below 30\%. When routing on primary and trunk
roads is banned (\emph{weighted\_2} profile in Table
\ref{table:weightprofiles}), the average increase relative to
unweighted shortest paths rises to 10\%, with certain locations
experiencing more significant negative effects on accessibility. Given
that cyclists will only deviate from shortest paths by a certain amount
to access better cycling infrastructure (as explained in Section
\ref{introduction}), allowing flow on some stretches of trunk and
primary roads is necessary to insure cycling uptake and equitable access
to cycling infrastructure. In its new vision for walking and cycling,
the Department for Transport acknowledges that minimal segregated
stretches of bicycle lanes on main roads will be necessary to avoid
circuitous cycling networks (DfT, 2020a).

Weighting the road network also allows us to better utilize existing
cycling infrastructure, as can be seen by the higher flow on cycleways
in Figure \ref{fig:flowsfacetweighted}. Again, the small differences in
impedance between cycleways and other road types mean that cycleways
that require significant deviation are not routed on.

It should be reiterated that the weighting profile used for routing has
been developed for the purposes of this study. It creates a hierarchy of
road preference that is grounded in cyclist preferences and government
plans to create LTNs. A wide range of weighting profiles could be used
to represent different types of cyclists and road environments, as
described in Gehrke et al. (2020) and Furth et al. (2016). A sensitivity
analysis could be done to determine an optimal weighting profile, but
given the variation in city road
networks, this would probably
require calibration to the specific city. More accurate routing could be
carried out given the availability of road-level data. In such cases we
would add additional impedance to specific roads, giving more useful
routing results than the current approach which considers all roads
of the same type to be equivalent.

One use-case of such granular data would be to identify roads that serve
schools. The Department of Transport notes that the number of school
children being driven to school has trebled over the past 40 years
(DfT, 2020a), and so having cycling infrastructure
serving schools is key to achieving the government target of getting
more children to cycle. This would not be difficult, as over 75\% of
children in the UK live within a 15 minute cycle from their school
(DfT, 2020b). Goodman et al. (2019) show that if
dutch levels of cycling were achieved in the UK, the \% of children
cycling to school could increase from 1.8\% to 41\%. In their typology of
cyclists, Dill and McNeil (2013) found that a majority of people who say they
would never cycle had never cycled to school, whereas confident cyclists
were those most likely to have cycled to school. Getting people to cycle
from a young age is therefore key to achieving societal change in
commuting habits.

\hypertarget{road-segment-prioritization}{%
\section{Road Segment Prioritization}\label{road-segment-prioritization}}

After routing the potential cycling demand onto the road network using
weighted shortest paths, we have estimates for the cumulative potential
cycling demand passing through all road segments. This cumulative demand
(referred to as \emph{flow}) is then used as a basis for identifying segments
that are in most need of investment in segregated cycling
infrastructure. In doing so, we must account for the motivations and
deterrents for cycling identified in Section \ref{introduction}, namely
direct and well-connected routes.

A range of algorithms could be used for prioritization. Because policy
priorities vary, we present two algorithms. Both utilize existing
infrastructure from the beginning and allow us to compare a solution
that focuses on utilitarianism to one that focuses on egalitarianism. In
both algorithms, links are selected iteratively and the iteration at
which each link is added to the solution is recorded. Investments in
cycling infrastructure can be limited by budget constraints, so it can
be useful to see where best to allocate a defined length of segregated
infrastructure. In order to incorporate egalitarian principles in our
approach, we use community detection to partition the study area and
evaluate the distribution of investment over the different subdivisions.

\hypertarget{community-detection}{%
\subsection{Community Detection}\label{community-detection}}

\begin{figure}

{\centering \includegraphics[width=0.4\linewidth]{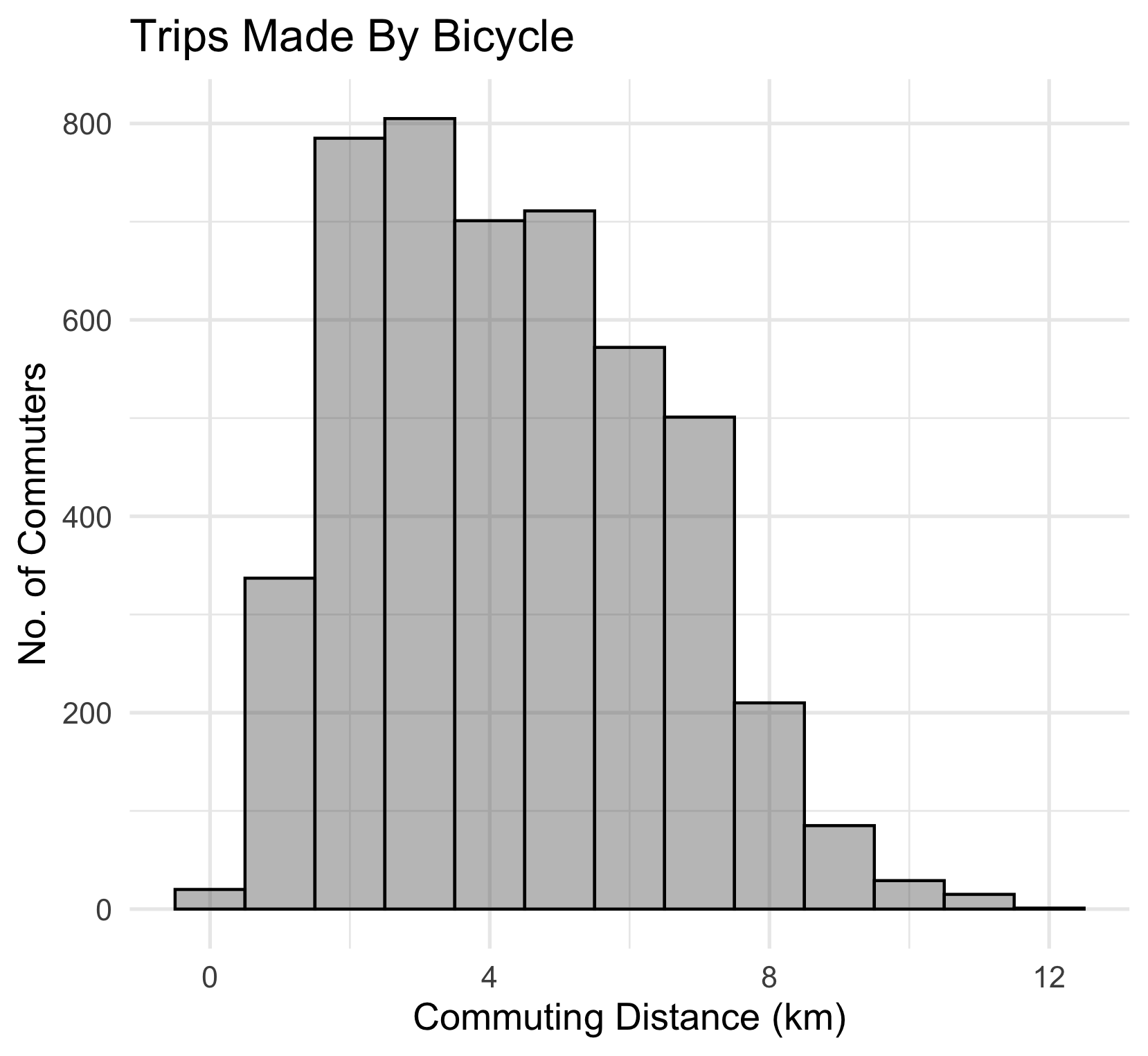} 

}

\caption{Cycling commuting distance - Manchester (2011 census data)}\label{fig:cyclinghistmanc}
\end{figure}

As explained in Section
\ref{ethical-underpinnings-and-proposed-approach}, a major challenge
facing `top-down' planning approaches is how to incorporate egalitarian
principles by fairly distributing investments in cycling infrastructure.
One way of quantifying this is to split up the city into smaller
geospatial areas and target equal investment in each of those areas.
This approach could also help ensure that on-the-ground surveys are made
by local stakeholders, an important component of the planning process
(Parkin, 2018). Community detection offers us a way to
delineate such a split; cyclists are limited in their commuting distance
(see Figure \ref{fig:cyclinghistmanc}), and so trip attractors are more
likely to have a local catchment area of cyclists.

In our case, the network is the city; the nodes are the
population-weighted MSOA centroids and the links connecting each MSOA
pair are weighted by the potential cycling demand between them. The
Louvain method (Blondel et al., 2008) is used to separate MSOAs into
communities. Potential cycling demand is used since we assume that this
is what the cycling demand will be once the cycling infrastructure is
added. To assign road links to communities, the following steps are
carried out:

\begin{verbatim}
1. Create links between MSOA centroids and weigh these links by potential cycling demand 
   between them.
2. Use the Louvain method to determine the optimal number of communities and assign each MSOA 
   centroid to a community.
3. Assign each road link to the same community as the closest MSOA centroid to it.
\end{verbatim}

\begin{figure}

{\centering \includegraphics[width=0.8\linewidth]{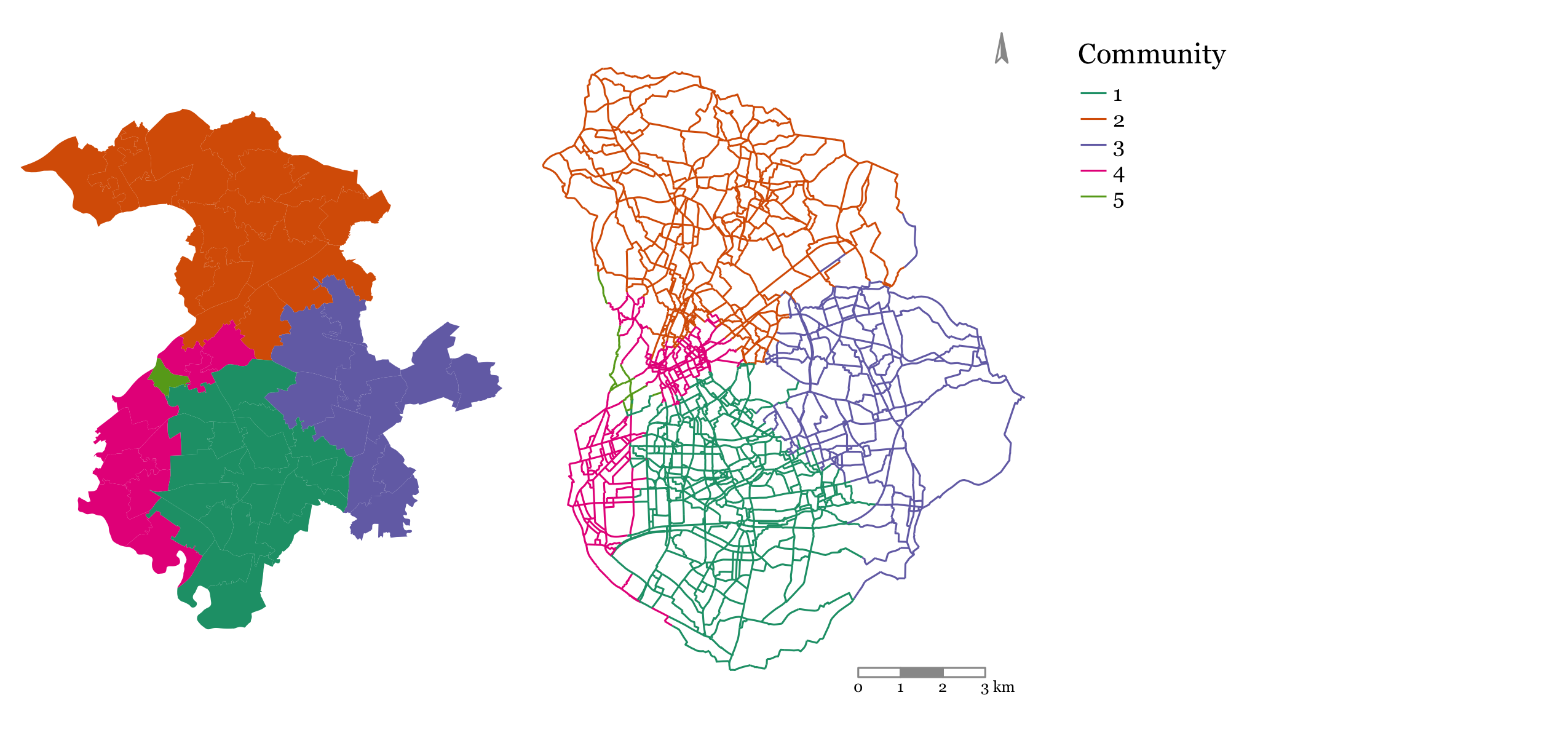} 

}

\caption{Communities based on potential cycling demand between MSOAs (Manchester)}\label{fig:communitiesmanchester}
\end{figure}

The results show that Manchester can be split into four large
communities and one small one (Figure \ref{fig:communitiesmanchester}).

~

\hypertarget{algorithm-1-utilitarian-expansion}{%
\subsection{Algorithm 1: Utilitarian Expansion}\label{algorithm-1-utilitarian-expansion}}

The algorithm logic is as follows:

\begin{verbatim}
1. Identify all links that have segregated cycling infrastructure and add them to the 
   initial solution
2. Identify all links that neighbor links in the current solution
3. Select neighboring link with highest flow and add it to the solution
4. Repeat steps 2 and 3 until all flow is satisfied or investment threshold is met
\end{verbatim}

This algorithm ensures that the resulting network is connected. It also
satisfies the directness criteria, since links on the weighted shortest
paths are those that have the highest flow passing through them (this is
a result of the routing in Section \ref{routing}.

\hypertarget{algorithm-2-egalitarian-expansion}{%
\subsection{Algorithm 2: Egalitarian Expansion}\label{algorithm-2-egalitarian-expansion}}

The first algorithm focuses on connectivity and directness, but not on
fairly distributing investment. The latter is not a requirement for
increasing cycling uptake, but it is fundamental for spatial equity, as
explained in Section \ref{ethical-underpinnings-and-proposed-approach}. This algorithm incorporates
the ideal of fair distribution by ensuring that investment is distributed between the defined communities.
This is done using the following logic:

\begin{verbatim}
1. Identify all links that have segregated cycling infrastructure and add them to the initial
   solution
2. Identify all links that neighbor links in the current solution
3. Select from each community one neighboring link with highest flow and add it to the
   solution
4. If there are no more neighboring links in a community, select the link with the highest
   flow in that community, regardless of connectivity, and add it to the solution
5. Repeat steps 2, 3 and 4 until all flow is satisfied or investment threshold is met
\end{verbatim}

Even though we may end up with a more disconnected network, we will have
separate connected networks in each community. Given that communities
are defined by having more internal flow than external flow, this is a
satisfactory solution.

The results of the community detection are used to evaluate the
algorithms. This is done by looking at the \emph{person-km satisfied} as
cycling infrastructure is added. Person-km is a measure of the total km
cycled on a road segment, so it is the product of the number of
potential commuters cycling on that road segment (\(flow\)) and the length
of the segment in km (\(l\)). For each road segment, the person-km is
equal to \(flow * l\). In the case of Manchester, Table
\ref{tab:personkmtable} shows that almost half of the person-km is in
community 1, while only 0.5\% of total person-km on the network is in
community 5.

\begin{table}[!h]

\caption{\label{tab:personkmtable}Total person-km in different communities (Manchester)}
\centering
\begin{tabular}[t]{l|l|r}
\hline
Community & Person-Km (Total) & Person-Km (\%)\\
\hline
1 & 284,458 & 44.4\\
\hline
2 & 163,877 & 25.6\\
\hline
3 & 79,218 & 12.4\\
\hline
4 & 109,635 & 17.1\\
\hline
5 & 3,317 & 0.5\\
\hline
\end{tabular}
\end{table}

\begin{figure}

{\centering \includegraphics[width=0.48\linewidth]{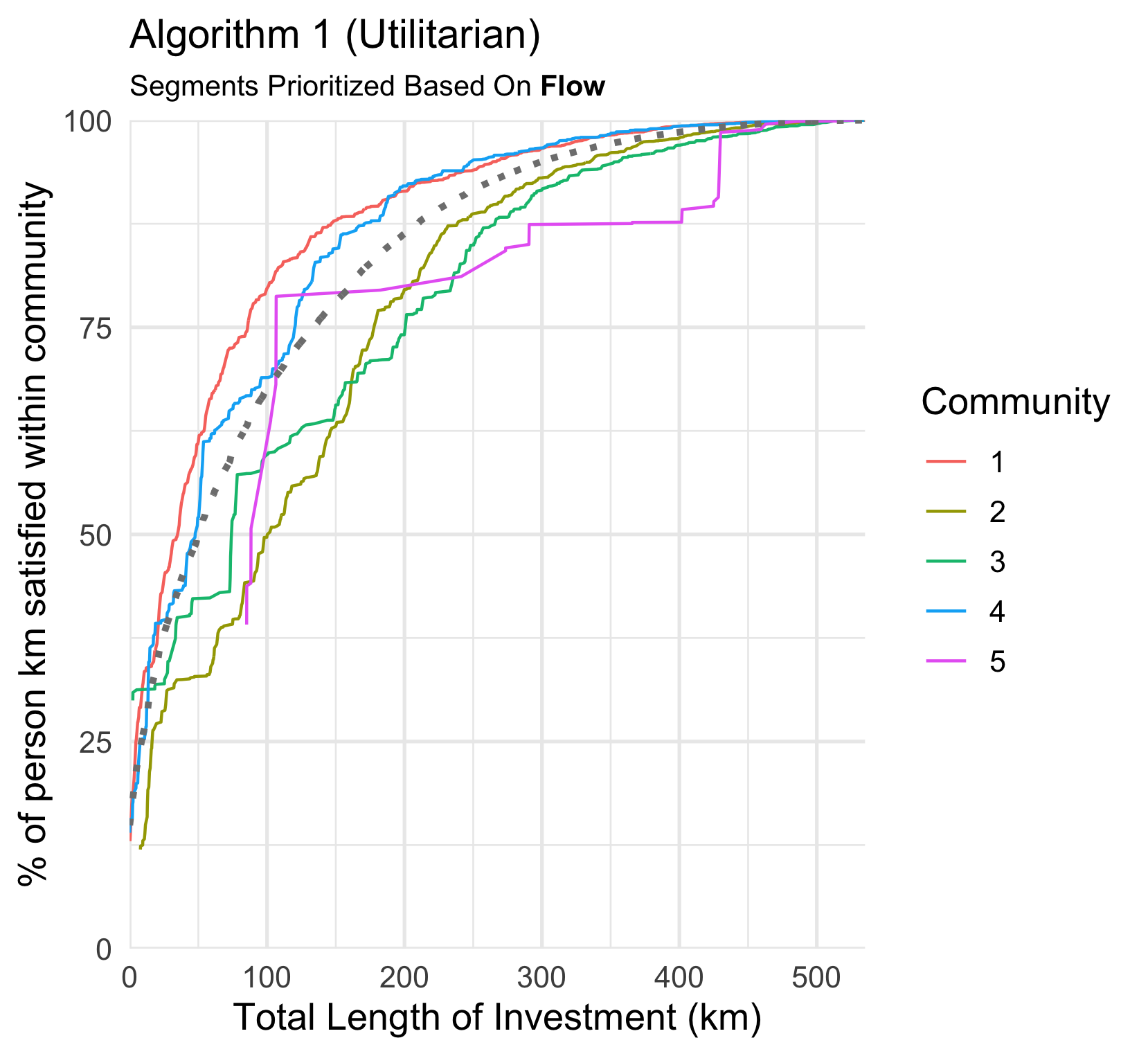} \includegraphics[width=0.48\linewidth]{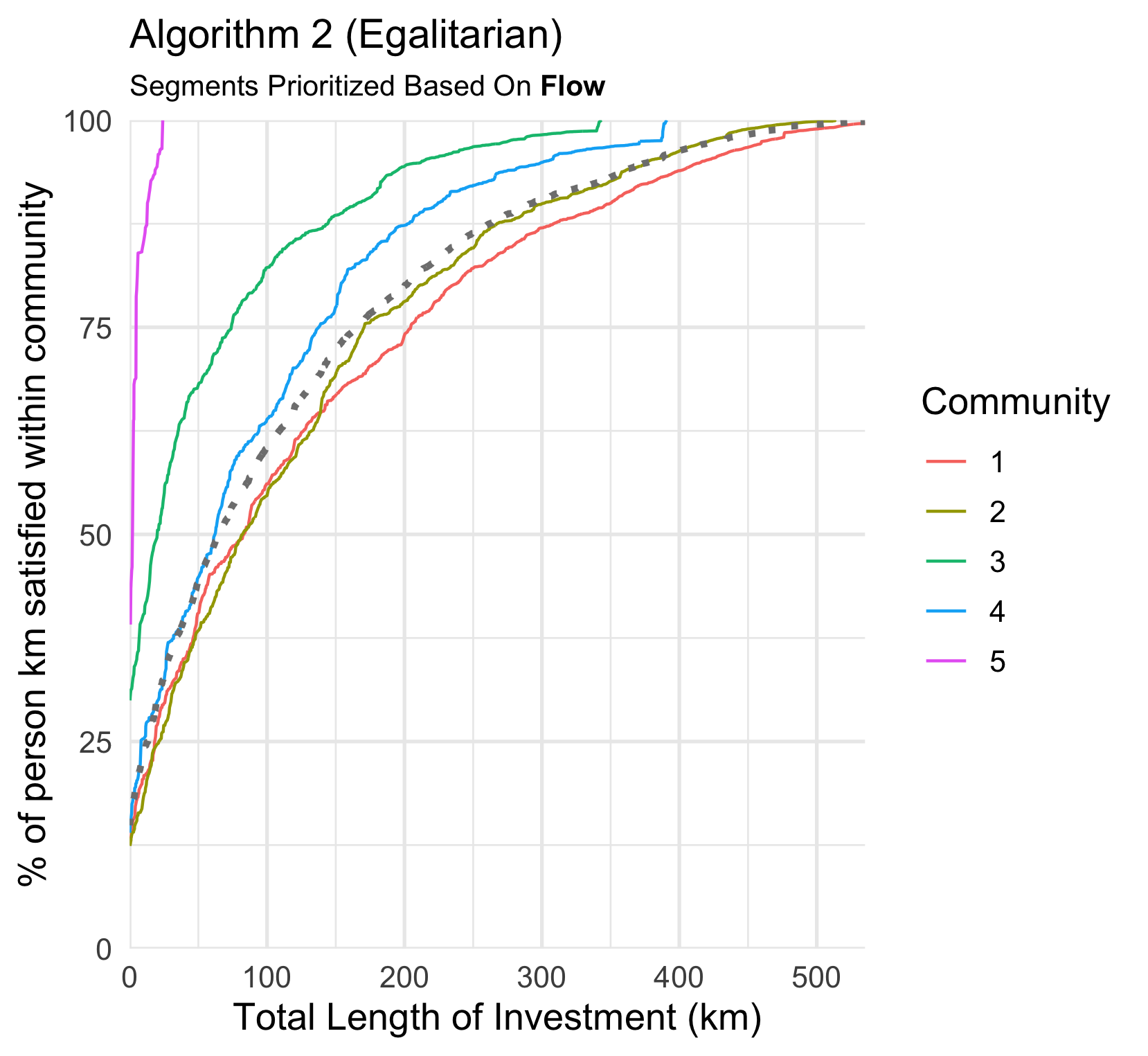} 

}

\caption{Comparing overall (dashed) and community level person-km satisfied (Manchester)}\label{fig:growthtotal}
\end{figure}

Looking at the person-km satisfied (Figure \ref{fig:growthtotal}), we
see that the incremental addition of cycling infrastructure is better
distributed between communities using Algorithm 2; equal distribution of
investment results in the gain in \% of person km satisfied in each
community being inversely correlated with the size of the community. In
addition, we find that the restrictions imposed by Algorithm 2 on the
network expansion do not seem to have a noticeable effect on the
city-wide \% of person-km satisfied. Comparing both algorithms, we can
see that Algorithm 1 provides only marginally quicker city-wide gains
than Algorithm 2.

\begin{figure}

{\centering \includegraphics[width=0.45\linewidth]{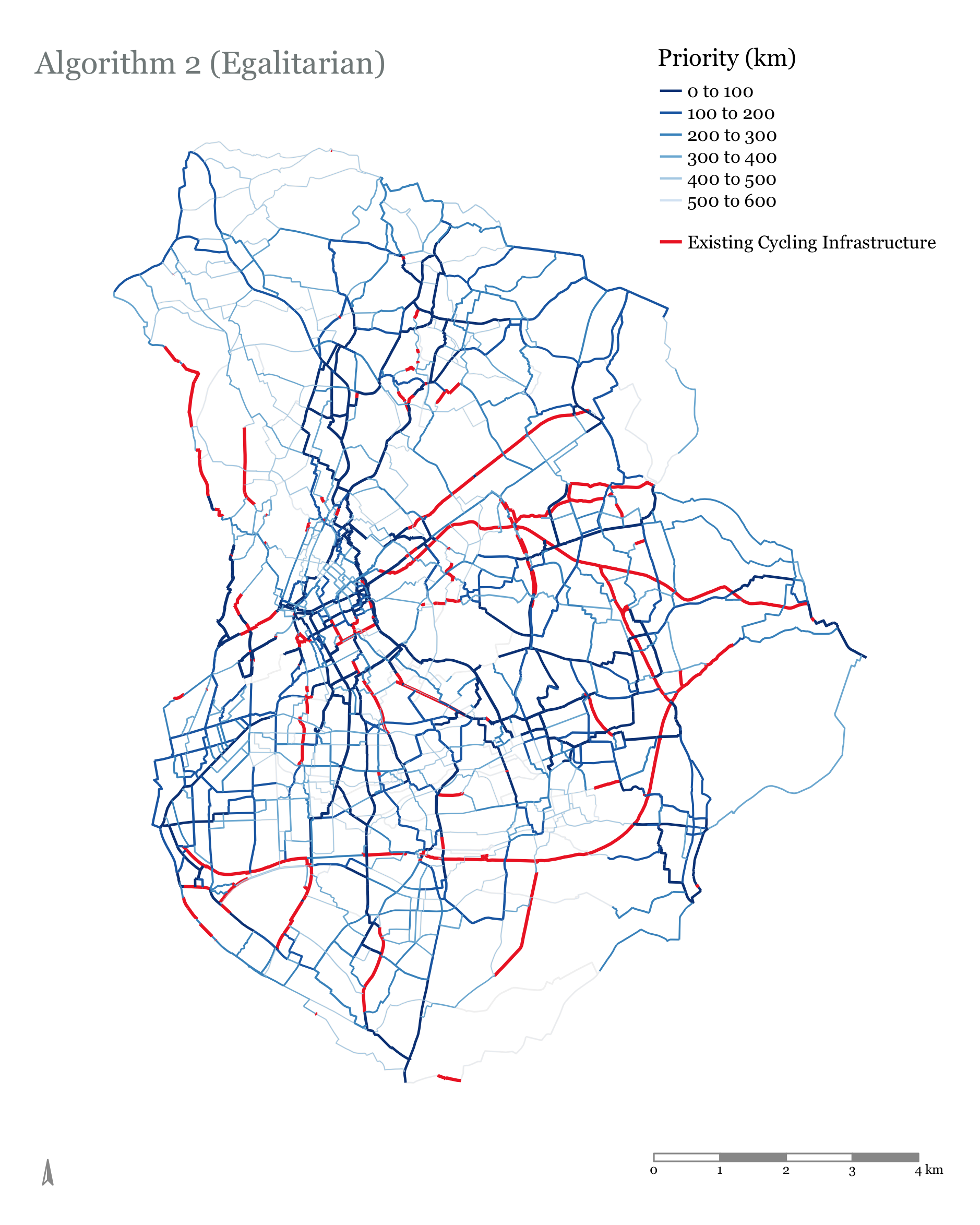} \includegraphics[width=0.45\linewidth]{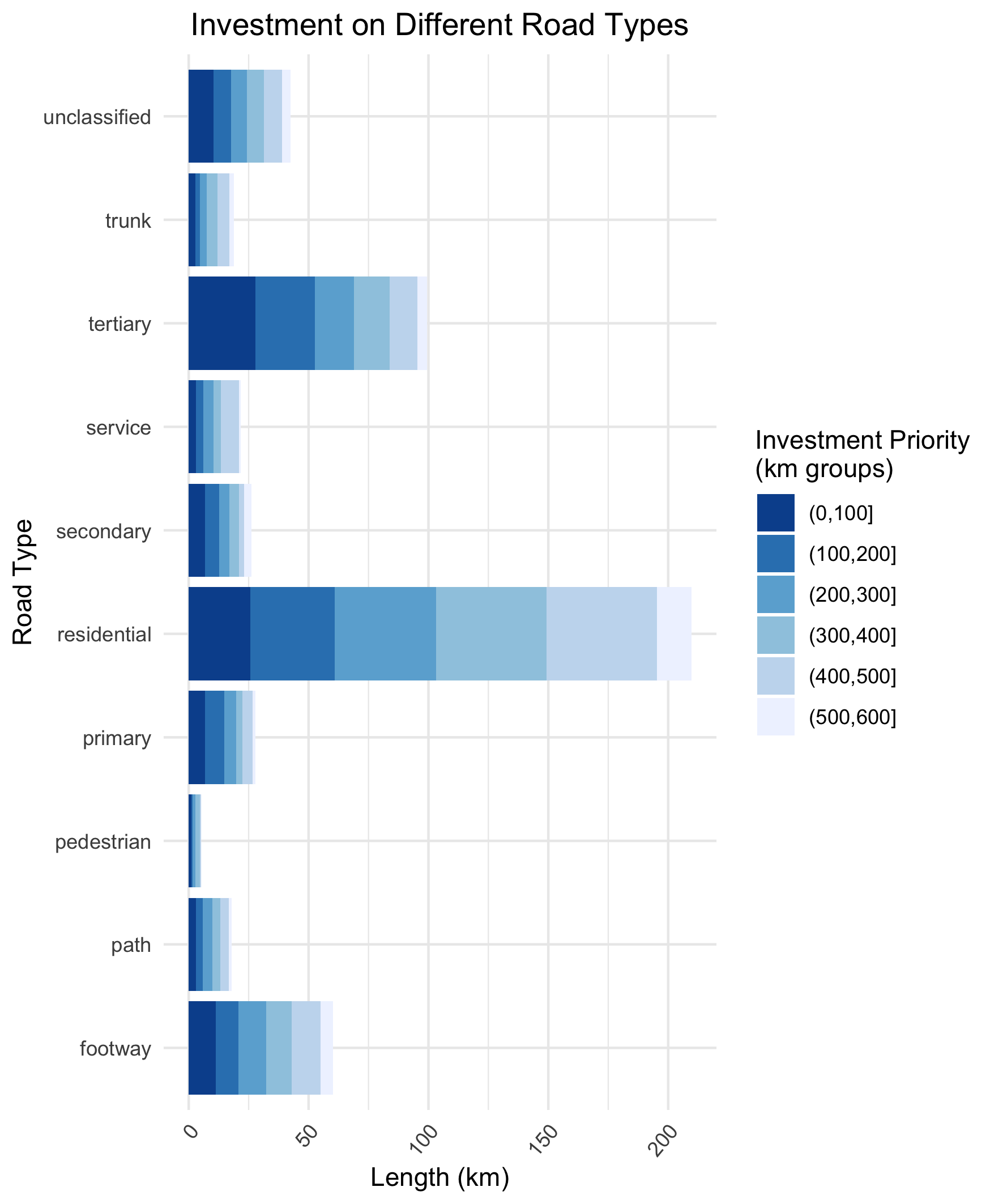} 

}

\caption{Road segment priority (left), disaggregated by road type (right) - egalitarian growth}\label{fig:growth3MapandBar}
\end{figure}

Figure \ref{fig:growth3MapandBar} gives us a geographic representation
of the results from Algorithm 2; it shows when each link was added to
the solution (first 100km, second 100km, etc). We can see that,
generally, road segments around cycling infrastructure are prioritized,
except for those neighboring cycling infrastructure on the very
periphery. The first 100km is also spatially distributed across the
city, with no apparent bias towards a particular area.

It is also important to understand how the different highway types
contribute to the proposed network. Figure \ref{fig:growth3MapandBar}
shows that most of the flow will be on residential and tertiary roads,
as expected from the weighting profile defined in Table
\ref{table:weightprofiles}.

\hypertarget{connectivity}{%
\subsection{Connectivity}\label{connectivity}}

\begin{figure}[H]

{\centering \includegraphics[width=0.85\linewidth]{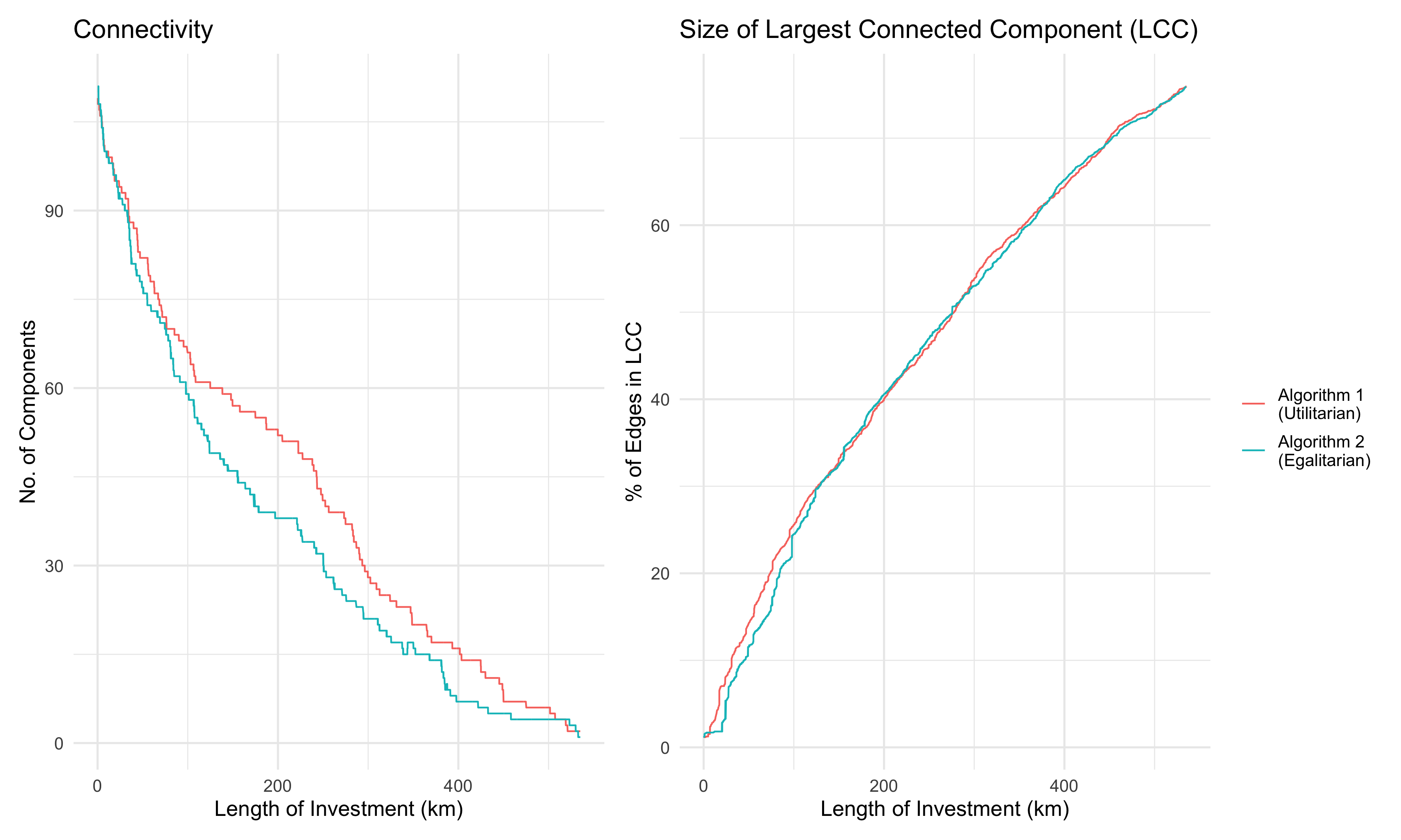} 

}

\caption{Network characteristics}\label{fig:componentsandGCC}
\end{figure}

Existing cycling infrastructure is made up of many disconnected
components. Both Algorithm 1 and 2 start with all existing segregated
cycling infrastructure and aim to create an efficient, connected
network. Figure \ref{fig:componentsandGCC} shows that both algorithms
gradually reduce the number of components as more infrastructure is
added, but Algorithm 2 is able to provide better connectivity with less
investment.

Consistent growth can also be seen for the size of the Largest Connected
Component in the proposed bicycle network (Figure
\ref{fig:componentsandGCC}). Here however, we find that there is little
difference between both Algorithms.

Overall, the improved distribution of investment shown by Algorithm 2 does not seem to come at the cost
of poorer connectivity or less city-wide gains when compared to Algorithm 1.
The results therefore advocate for the incorporation of egalitarian principles in bicycle network planning.

\hypertarget{discussion-and-conclusions}{%
\section{Discussion and conclusions}\label{discussion-and-conclusions}}

This paper demonstrates an approach for prioritizing
investments in city-wide cycleway networks. The approach aims to respect
both the needs of the users and the ambitions of stakeholders working at
local or regional levels. The results, including detailed route network
maps based on current travel behaviour derived from OD data, can provide
an evidence-base for designing direct, connected,
and low-stress networks.

Given that the ``most essential activity entailed in the design of
cycle-friendly infrastructure is developing a cycle network''
(Parkin, 2018), we believe that the approach has great
potential to inform investment in cities such as Manchester where there
is political will to invest in cycling long-term. A benefit of the
approach is that it has relatively modest data requirements: only the
road network (from OSM), its topography (from satellite imagery), and
OD data (from the national census in this case) are
needed, opening up the possibility of deploying the approach in other
cities.\footnote{The results are easily reproducible for all UK cities, and can also be reproduced for cities elsewhere given the availability of commuter data. Instructions for reproducing the results are detailed in the \textbf{README} file in the R folder of this \href{https://github.com/Hussein-Mahfouz/cycle-networks}{Github repository}}

The approach can identify not only where there is high
potential for cycling but also trade-offs between
stress levels (related to motor traffic) and directness.
The results can therefore be used as a basis for recommendations on
road space reallocation \emph{and} new infrastructure to unlock potential cycling demand.
The approach encourages consideration of a wider range of preferences
and needs than previous approaches that focus only on absolute potential.
Moreover, the inclusion of egalitarian principles in scenarios of change
encourages investment in cycling infrastructure to
increase the connectivity of existing cycling infrastructure
\emph{and} investment that addresses geographical and social inequalities.
This ability to address inequalities in network prioritization is particularly
important given research showing substantial \emph{inequalities}
around transport provision in general and cycling uptake and investment
in particular (Lucas et al., 2016; Vidal Tortosa et al., 2021).

The approach is not without limitations. The level of detail is only as
good as the granularity of the available data (in this case relatively
coarse zones). Iacono et al. (2010) note that such large travel zones
are not ideal for understanding route choice behaviour of cyclists and
pedestrians. They also give rise to an `ecological fallacy' whereby
average characteristics are assumed to apply to all residents of the
aggregated geographical area, suggesting a need for applying the methods
to more granular OD data (and for governments and other
data-collecting organisations to make OD data more
readily available).
The potential demand calculation is also based on cycling in the traditional sense, and does not consider the
effect of micro-mobility on reducing topology-related impedance to cycling. Given that we are proposing an approach which can accommodate any cycling uptake functions, this is an acceptable limitation.

The approach is also focused solely on the allocation of cycling
infrastructure, and does not consider the larger political and
regulatory environment that needs to exist to promote cycling; while
segregated, connected, and direct cycling infrastructure is key to
achieving high levels of cycling, research has shown that it cannot
exist in a vacuum. Wardman et al. (2007) developed a mode choice model
for the UK and their results showed that improved cycling infrastructure
on its own only had modest impacts on mode shift, and even the unlikely
scenario of all urban routes being serviced by segregated bike lanes was
forecast to increase cycling mode share by only 3\%. International
research shows that cities that invest in more comprehensive cycling
projects have a more significant increase in the number of cyclists as
well as the cycling mode share (Parkin, 2018; Pucher et al., 2010). These cities do not just focus on
infrastructure, but on general policies as well as restricting car use.
Evaluation of policies in Denmark and Germany and the Netherlands has
shown that their high cycling mode share is down to a broader set of
soft and hard policies. Hard policies include traffic calming, filtered permeability
interventions, cycling rights of way, bike parking, integration with the
public transport network, and making driving cars both expensive and
inconvenient, while soft policies include marketing and awareness campaigns (Gössling, 2013; Pucher and Buehler, 2008). While these policies are outside the scope of this research, it is important to recognize their key role in
bringing about an increase in levels of cycling.

Consideration of these limitations suggest future directions of research.
New datasets and continuously evolving computational (hardware and software)
capabilities should enable the data related issues to be overcome as open
datasets, and our ability to process them, improve.
Plugging-in alternative uptake models could help address the relatively narrow
definition of `potential' used in this paper, to consider broader social factors.
A promising area of future research in this direction could be
to integrate a wider range of modes, including e-bikes and e-scooters, into the
analysis. Currently scope for progress in this direction is
partially restricted by the lack of data on the
proliferation of these modes, raising the point that surveys could
replace the broad category of `Bicycle' with a range of small modes such as
`pedal cycles' (including tricycles and recumbent cycles), e-bikes
and other forms of micro-mobility.

Despite the rapid growth of these alternative modes, there is little doubt that
the humble bicycle is a key ingredient in the ongoing shift towards active transport.
Recognition of the importance of this shift for improved health and well-being
of citizens, and the wider challenge to decarbonise the global economy,
has grown with pressures on health and public transport systems during the pandemic.
The success of policies to accelerate cycling uptake depends on a range of
factors including, vitally, the design of the
cycling network for potential cyclists.
The approach presented in this paper provides a strong evidence base,
that considers both cycling potential and social equity,
for designing joined-up and cost-effective strategic cycle networks.

\clearpage

\hypertarget{references}{%
\section*{References}\label{references}}
\addcontentsline{toc}{section}{References}

\setlength{\parindent}{-0.5in}
\setlength{\leftskip}{0.5in}
\setlength{\parskip}{8pt}

\hypertarget{refs}{}
\begin{CSLReferences}{1}{0}
\leavevmode\hypertarget{ref-akbarzadeh2018designing}{}%
Akbarzadeh, M., Mohri, S.S., Yazdian, E., 2018. Designing bike networks using the concept of network clusters. Applied network science 3, 12.

\leavevmode\hypertarget{ref-aldred2019impacts}{}%
Aldred, R., Croft, J., Goodman, A., 2019. Impacts of an active travel intervention with a cycling focus in a suburban context: One-year findings from an evaluation of london's in-progress mini-hollands programme. Transportation research part A: policy and practice 123, 147--169.

\leavevmode\hypertarget{ref-bao2017planning}{}%
Bao, J., He, T., Ruan, S., Li, Y., Zheng, Y., 2017. Planning bike lanes based on sharing-bikes' trajectories, in: Proceedings of the 23rd ACM SIGKDD International Conference on Knowledge Discovery and Data Mining. pp. 1377--1386.

\leavevmode\hypertarget{ref-black1995spatial}{}%
Black, W.R., 1995. Spatial interaction modeling using artificial neural networks. Journal of Transport Geography 3, 159--166.

\leavevmode\hypertarget{ref-blondel2008fast}{}%
Blondel, V.D., Guillaume, J.-L., Lambiotte, R., Lefebvre, E., 2008. Fast unfolding of communities in large networks. Journal of statistical mechanics: theory and experiment 2008, P10008.

\leavevmode\hypertarget{ref-brand_climate_2020}{}%
Brand, C., Dons, E., Anaya-Boig, E., Avila-Palencia, I., Clark, A., de Nazelle, A., Gascon, M., Gaupp-Berghausen, M., Gerike, R., Gotschi, T., 2020. The climate change mitigation effects of active travel. Preprint: researchsquare.com.

\leavevmode\hypertarget{ref-broach2011bicycle}{}%
Broach, J., Gliebe, J., Dill, J., 2011. Bicycle route choice model developed using revealed preference GPS data, in: 90th Annual Meeting of the Transportation Research Board, Washington, DC.

\leavevmode\hypertarget{ref-caulfield2012determining}{}%
Caulfield, B., Brick, E., McCarthy, O.T., 2012. Determining bicycle infrastructure preferences--a case study of dublin. Transportation research part D: transport and environment 17, 413--417.

\leavevmode\hypertarget{ref-celis2017association}{}%
Celis-Morales, C.A., Lyall, D.M., Welsh, P., Anderson, J., Steell, L., Guo, Y., Maldonado, R., Mackay, D.F., Pell, J.P., Sattar, N., others, 2017. Association between active commuting and incident cardiovascular disease, cancer, and mortality: Prospective cohort study. bmj 357, j1456.

\leavevmode\hypertarget{ref-crane2017longitudinal}{}%
Crane, M., Rissel, C., Standen, C., Ellison, A., Ellison, R., Wen, L.M., Greaves, S., 2017. Longitudinal evaluation of travel and health outcomes in relation to new bicycle infrastructure, sydney, australia. Journal of Transport \& Health 6, 386--395.

\leavevmode\hypertarget{ref-departmentgearchange2020}{}%
DfT, 2020a. Gear change: A bold vision for cycling and walking.

\leavevmode\hypertarget{ref-departmentcycleinfradesign2020}{}%
DfT, 2020b. Cycling infrastructure design.

\leavevmode\hypertarget{ref-department2012guidance}{}%
DfT, 2012. Guidance on road classification and the primary route network.

\leavevmode\hypertarget{ref-dill2013four}{}%
Dill, J., McNeil, N., 2013. Four types of cyclists? Examination of typology for better understanding of bicycling behavior and potential. Transportation Research Record 2387, 129--138.

\leavevmode\hypertarget{ref-duthie2014optimization}{}%
Duthie, J., Unnikrishnan, A., 2014. Optimization framework for bicycle network design. Journal of Transportation engineering 140, 04014028.

\leavevmode\hypertarget{ref-furth2016network}{}%
Furth, P.G., Mekuria, M.C., Nixon, H., 2016. Network connectivity for low-stress bicycling. Transportation research record 2587, 41--49.

\leavevmode\hypertarget{ref-gehrke2020cycling}{}%
Gehrke, S.R., Akhavan, A., Furth, P.G., Wang, Q., Reardon, T.G., 2020. A cycling-focused accessibility tool to support regional bike network connectivity. Transportation research part D: transport and environment 85, 102388.

\leavevmode\hypertarget{ref-goodman2019scenarios}{}%
Goodman, A., Rojas, I.F., Woodcock, J., Aldred, R., Berkoff, N., Morgan, M., Abbas, A., Lovelace, R., 2019. Scenarios of cycling to school in england, and associated health and carbon impacts: Application of the {`propensity to cycle tool.'} Journal of Transport \& Health 12, 263--278.

\leavevmode\hypertarget{ref-goodman2014new}{}%
Goodman, A., Sahlqvist, S., Ogilvie, D., Consortium, iConnect, 2014. New walking and cycling routes and increased physical activity: One-and 2-year findings from the UK iConnect study. American journal of public health 104, e38--e46.

\leavevmode\hypertarget{ref-gossling2013urban}{}%
Gössling, S., 2013. Urban transport transitions: Copenhagen, city of cyclists. Journal of Transport Geography 33, 196--206.

\leavevmode\hypertarget{ref-iacono2010measuring}{}%
Iacono, M., Krizek, K.J., El-Geneidy, A., 2010. Measuring non-motorized accessibility: Issues, alternatives, and execution. Journal of Transport Geography 18, 133--140.

\leavevmode\hypertarget{ref-jafino2020transport}{}%
Jafino, B.A., Kwakkel, J., Verbraeck, A., 2020. Transport network criticality metrics: A comparative analysis and a guideline for selection. Transport Reviews 40, 241--264.

\leavevmode\hypertarget{ref-jarrett2012effect}{}%
Jarrett, J., Woodcock, J., Griffiths, U.K., Chalabi, Z., Edwards, P., Roberts, I., Haines, A., 2012. Effect of increasing active travel in urban england and wales on costs to the national health service. The Lancet 379, 2198--2205.

\leavevmode\hypertarget{ref-kohl2012pandemic}{}%
Kohl 3rd, H.W., Craig, C.L., Lambert, E.V., Inoue, S., Alkandari, J.R., Leetongin, G., Kahlmeier, S., Group, L.P.A.S.W., others, 2012. The pandemic of physical inactivity: Global action for public health. The lancet 380, 294--305.

\leavevmode\hypertarget{ref-lovelace2017propensity}{}%
Lovelace, R., Goodman, A., Aldred, R., Berkoff, N., Abbas, A., Woodcock, J., 2017. The propensity to cycle tool: An open source online system for sustainable transport planning. Journal of transport and land use 10, 505--528.

\leavevmode\hypertarget{ref-lucas2016method}{}%
Lucas, K., Van Wee, B., Maat, K., 2016. A method to evaluate equitable accessibility: Combining ethical theories and accessibility-based approaches. Transportation 43, 473--490.

\leavevmode\hypertarget{ref-marques2015infrastructure}{}%
Marqués, R., Hernández-Herrador, V., Calvo-Salazar, M., García-Cebrián, J., 2015. How infrastructure can promote cycling in cities: Lessons from seville. Research in Transportation Economics 53, 31--44.

\leavevmode\hypertarget{ref-martinez2013new}{}%
Martínez, L.M., Viegas, J.M., 2013. A new approach to modelling distance-decay functions for accessibility assessment in transport studies. Journal of Transport Geography 26, 87--96.

\leavevmode\hypertarget{ref-mauttone2017bicycle}{}%
Mauttone, A., Mercadante, G., Rabaza, M., Toledo, F., 2017. Bicycle network design: Model and solution algorithm. Transportation research procedia 27, 969--976.

\leavevmode\hypertarget{ref-mesbah2012bilevel}{}%
Mesbah, M., Thompson, R., Moridpour, S., 2012. Bilevel optimization approach to design of network of bike lanes. Transportation research record 2284, 21--28.

\leavevmode\hypertarget{ref-nahmias2017integrating}{}%
Nahmias-Biran, B., Martens, K., Shiftan, Y., 2017. Integrating equity in transportation project assessment: A philosophical exploration and its practical implications. Transport reviews 37, 192--210.

\leavevmode\hypertarget{ref-natera2019data}{}%
Natera, L., Battiston, F., Iñiguez, G., Szell, M., 2019. Data-driven strategies for optimal bicycle network growth. arXiv preprint arXiv:1907.07080.

\leavevmode\hypertarget{ref-nello2020environmental}{}%
Nello-Deakin, S., 2020. Environmental determinants of cycling: Not seeing the forest for the trees? Journal of transport geography 85, 102704.

\leavevmode\hypertarget{ref-olmos2020data}{}%
Olmos, L.E., Tadeo, M.S., Vlachogiannis, D., Alhasoun, F., Alegre, X.E., Ochoa, C., Targa, F., González, M.C., 2020. A data science framework for planning the growth of bicycle infrastructures. Transportation Research Part C: Emerging Technologies 115, 102640.

\leavevmode\hypertarget{ref-ofn2018population}{}%
ONS, 2018. Population estimates for the UK, england and wales, scotland and northern ireland: Mid-2017. Hampshire: Office for National Statistics.

\leavevmode\hypertarget{ref-ONS2011flowdata}{}%
ONS, 2011. 2011 census: Special workplace statistics (united kingdom).

\leavevmode\hypertarget{ref-padgham2019dodgr}{}%
Padgham, M., 2019. Dodgr: An r package for network flow aggregation. Transport Findings. Network Design Lab.

\leavevmode\hypertarget{ref-parkin2018designing}{}%
Parkin, J., 2018. Designing for cycle traffic. ICE Publishing. \url{https://doi.org/10.1680/dfct.63495}

\leavevmode\hypertarget{ref-patterson2020associations}{}%
Patterson, R., Panter, J., Vamos, E.P., Cummins, S., Millett, C., Laverty, A.A., 2020. Associations between commute mode and cardiovascular disease, cancer, and all-cause mortality, and cancer incidence, using linked census data over 25 years in england and wales: A cohort study. The Lancet Planetary Health 4, e186--e194.

\leavevmode\hypertarget{ref-pereira2017distributive}{}%
Pereira, R.H., Schwanen, T., Banister, D., 2017. Distributive justice and equity in transportation. Transport reviews 37, 170--191.

\leavevmode\hypertarget{ref-pucher2008making}{}%
Pucher, J., Buehler, R., 2008. Making cycling irresistible: Lessons from the netherlands, denmark and germany. Transport reviews 28, 495--528.

\leavevmode\hypertarget{ref-pucher2010infrastructure}{}%
Pucher, J., Dill, J., Handy, S., 2010. Infrastructure, programs, and policies to increase bicycling: An international review. Preventive medicine 50, S106--S125.

\leavevmode\hypertarget{ref-schoner2014missing}{}%
Schoner, J.E., Levinson, D.M., 2014. The missing link: Bicycle infrastructure networks and ridership in 74 US cities. Transportation 41, 1187--1204.

\leavevmode\hypertarget{ref-sorton1994bicycle}{}%
Sorton, A., Walsh, T., 1994. Bicycle stress level as a tool to evaluate urban and suburban bicycle compatibility. Transportation Research Record 17--17.

\leavevmode\hypertarget{ref-stinson2003commuter}{}%
Stinson, M.A., Bhat, C.R., 2003. Commuter bicyclist route choice: Analysis using a stated preference survey. Transportation Research Record 1828, 107--115.

\leavevmode\hypertarget{ref-agreement2015paris}{}%
UN, 2015. Paris agreement, in: Report of the Conference of the Parties to the United Nations Framework Convention on Climate Change (21st Session, 2015: Paris). Retrived December. HeinOnline, p. 2017.

\leavevmode\hypertarget{ref-vidaltortosa_infrastructure_2020}{}%
Vidal Tortosa, E., Lovelace, R., Heinen, E., Mann, R.P., 2021. Infrastructure is not enough: Interactions between the environment, socioeconomic disadvantage and cycling participation in {England}. Journal of Transport and Land Use.

\leavevmode\hypertarget{ref-wardman2007factors}{}%
Wardman, M., Tight, M., Page, M., 2007. Factors influencing the propensity to cycle to work. Transportation Research Part A: Policy and Practice 41, 339--350.

\leavevmode\hypertarget{ref-wilson1971family}{}%
Wilson, A.G., 1971. A family of spatial interaction models, and associated developments. Environment and Planning A 3, 1--32.

\leavevmode\hypertarget{ref-winters2011motivators}{}%
Winters, M., Davidson, G., Kao, D., Teschke, K., 2011. Motivators and deterrents of bicycling: Comparing influences on decisions to ride. Transportation 38, 153--168.

\leavevmode\hypertarget{ref-winters2010far}{}%
Winters, M., Teschke, K., Grant, M., Setton, E.M., Brauer, M., 2010. How far out of the way will we travel? Built environment influences on route selection for bicycle and car travel. Transportation Research Record 2190, 1--10.

\end{CSLReferences}

\hypertarget{appendix-appendix}{%
\appendix}

\clearpage

\hypertarget{appendix-reproducing-the-results-for-other-cities}{%
\section{Appendix: Reproducing the results for other cities}\label{appendix-reproducing-the-results-for-other-cities}}

To reproduce the results presented for other cities in the UK there are a few computing pre-requisites:

\begin{itemize}
\tightlist
\item
  A reasonably powerful computer, e.g.~with 8 GB RAM
\item
  The statistical programming language R installed (we recommend running it in the RStudio editor)
\item
  The source code of the paper repository, which can be download from this \href{https://github.com/Hussein-Mahfouz/cycle-networks}{Github repository}{]}
\end{itemize}

To ensure you have the necessary packages installed, open the \texttt{cycle-networks.Rproj} file in RStudio and run the following commands to ensure you have the necessary packages installed:

\begin{verbatim}
install.packages("renv")
renv::restore()
\end{verbatim}

The \texttt{build.R} script contains all of the necessary steps to reproduce the results for any city in the UK.
Open this file with the following command and run the code line-by-line.

\begin{verbatim}
file.edit("R/build.R")
\end{verbatim}

You can select a city to build by executing the following command:

\begin{verbatim}
chosen_city <- "Leeds"
\end{verbatim}

Replacing \texttt{Leeds} with any other city will make the script work for that city.
Note: this only works for Local Authority names in England and Wales listed below.

In this script, you choose which city you wish to run the analysis from
this list of available towns and cities:

\begin{verbatim}
##   [1] "Barnsley"             "Basildon"             "Basingstoke"
##   [4] "Bath"                 "Bedford"              "Birkenhead"
##   [7] "Birmingham"           "Blackburn"            "Blackpool"
##  [10] "Bolton"               "Bournemouth"          "Bracknell"
##  [13] "Bradford"             "Brighton and Hove"    "Bristol"
##  [16] "Burnley"              "Burton upon Trent"    "Bury"
##  [19] "Cambridge"            "Cardiff"              "Carlisle"
##  [22] "Chatham"              "Chelmsford"           "Cheltenham"
##  [25] "Chester"              "Chesterfield"         "Colchester"
##  [28] "Coventry"             "Crawley"              "Darlington"
##  [31] "Derby"                "Doncaster"            "Dudley"
##  [34] "Eastbourne"           "Exeter"               "Gateshead"
##  [37] "Gillingham"           "Gloucester"           "Grimsby"
##  [40] "Guildford"            "Halifax"              "Harlow"
##  [43] "Harrogate"            "Hartlepool"           "Hastings"
##  [46] "Hemel Hempstead"      "High Wycombe"         "Huddersfield"
##  [49] "Ipswich"              "Kingston upon Hull"   "Leeds"
##  [52] "Leicester"            "Lincoln"              "Liverpool"
##  [55] "London"               "Luton"                "Maidstone"
##  [58] "Manchester"           "Mansfield"            "Middlesbrough"
##  [61] "Milton Keynes"        "Newcastle upon Tyne"  "Newcastle-under-Lyme"
##  [64] "Newport"              "Northampton"          "Norwich"
##  [67] "Nottingham"           "Nuneaton"             "Oldham"
##  [70] "Oxford"               "Peterborough"         "Plymouth"
##  [73] "Poole"                "Portsmouth"           "Preston"
##  [76] "Reading"              "Redditch"             "Rochdale"
##  [79] "Rotherham"            "Salford"              "Scunthorpe"
##  [82] "Sheffield"            "Shrewsbury"           "Slough"
##  [85] "Solihull"             "South Shields"        "Southampton"
##  [88] "Southend-on-Sea"      "Southport"            "St Albans"
##  [91] "St Helens"            "Stevenage"            "Stockport"
##  [94] "Stockton-on-Tees"     "Stoke-on-Trent"       "Sunderland"
##  [97] "Sutton Coldfield"     "Swansea"              "Swindon"
## [100] "Telford"              "Wakefield"            "Walsall"
## [103] "Warrington"           "Watford"              "West Bromwich"
## [106] "Weston-Super-Mare"    "Wigan"                "Woking"
## [109] "Wolverhampton"        "Worcester"            "Worthing"
## [112] "York"
\end{verbatim}

\newpage

\hypertarget{birmingham}{%
\subsection{Birmingham}\label{birmingham}}

\subsubsection{Potential Demand}

\begin{figure}
\includegraphics[width=0.3\linewidth]{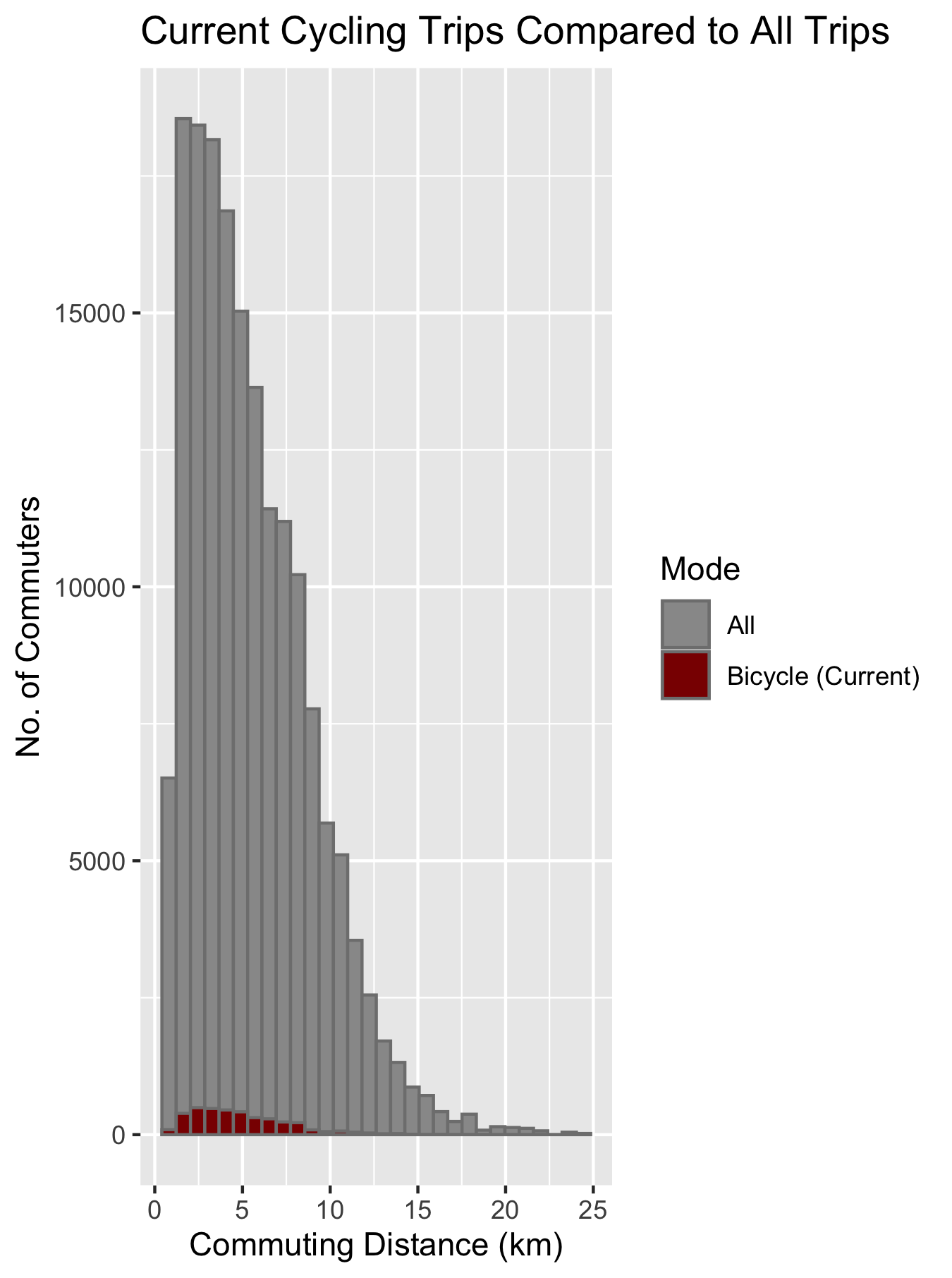} \includegraphics[width=0.3\linewidth]{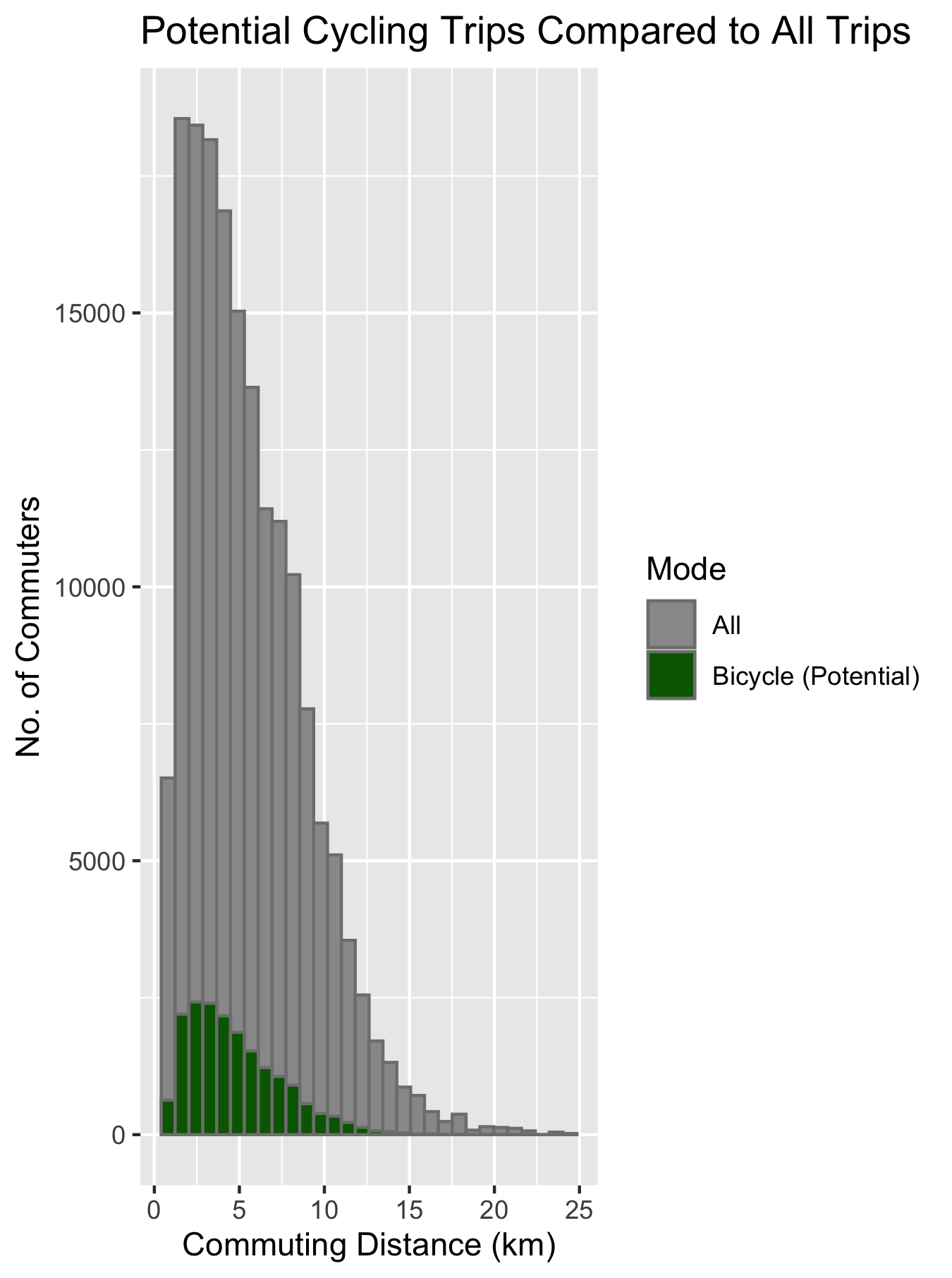} \includegraphics[width=0.3\linewidth]{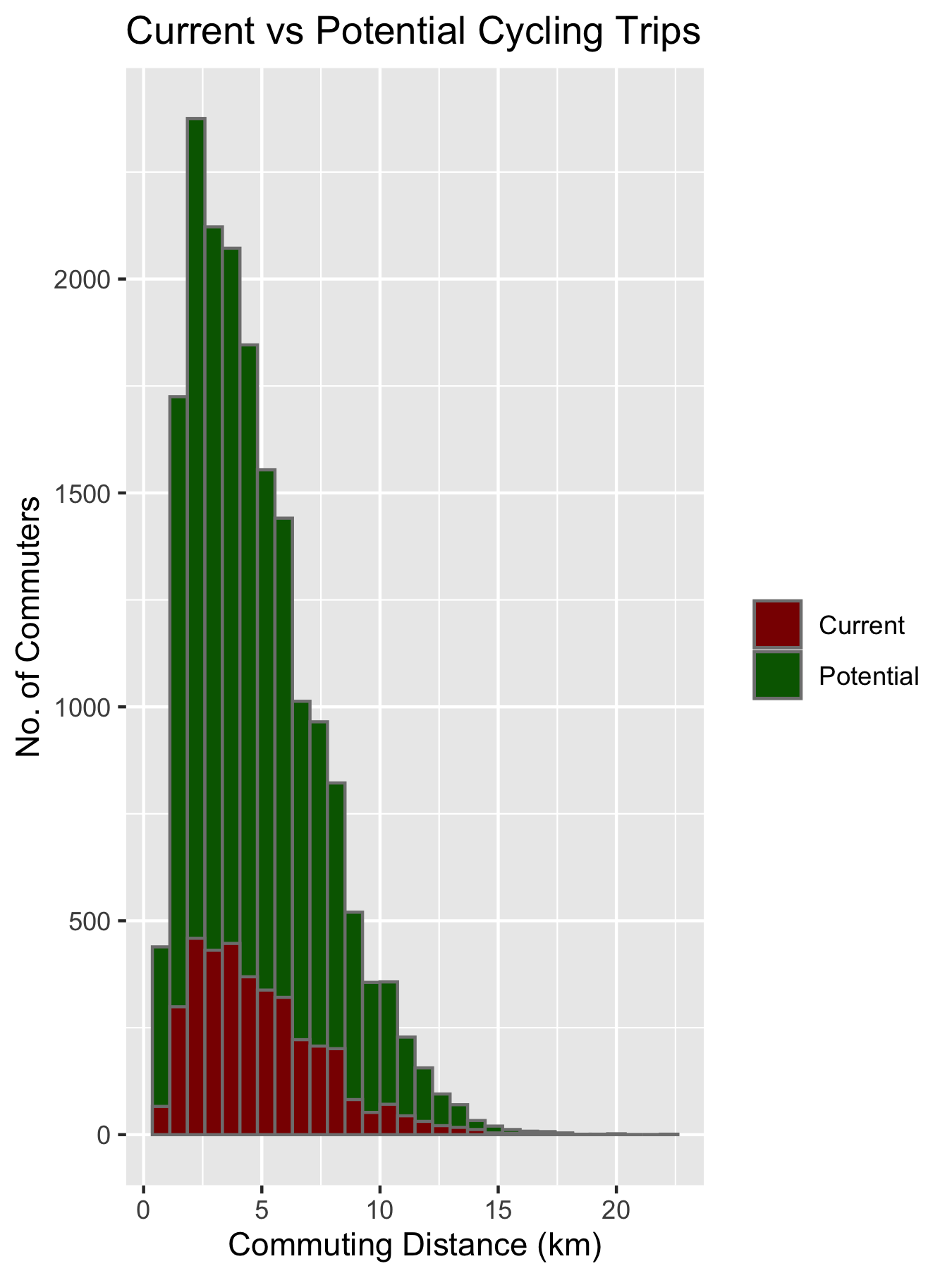} \caption{Distribution of Potential Cycling Demand}\label{fig:potdemhistogramsBirmingham}
\end{figure}

\begin{figure}[H]

{\centering \includegraphics[width=0.8\linewidth]{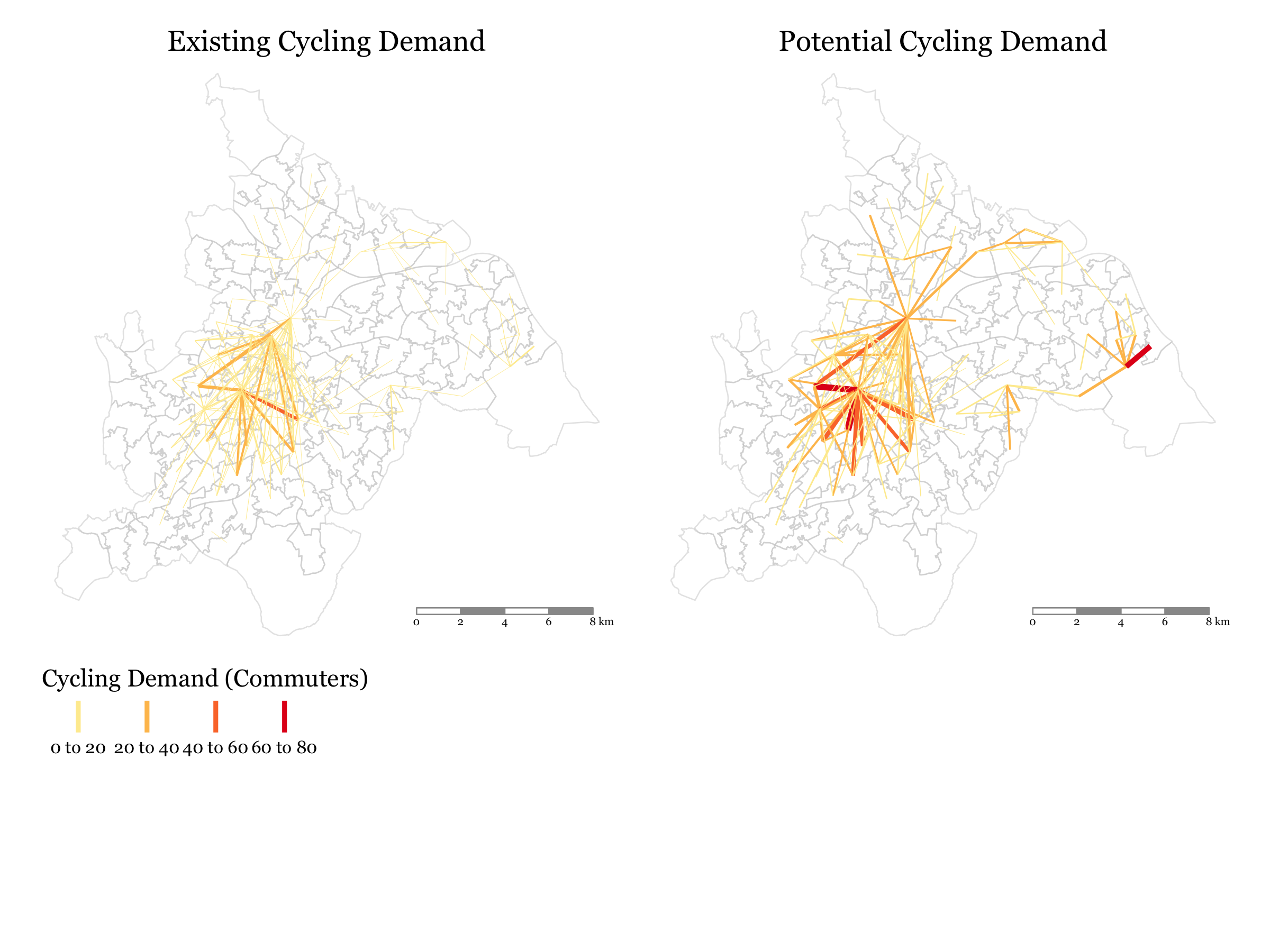} 

}

\caption{Current and Potential Cycling Demand}\label{fig:desirefacetcyclingBirmingham}
\end{figure}

\clearpage

\subsubsection{Routing Cycling Flows}

\begin{figure}

{\centering \includegraphics[width=0.75\linewidth]{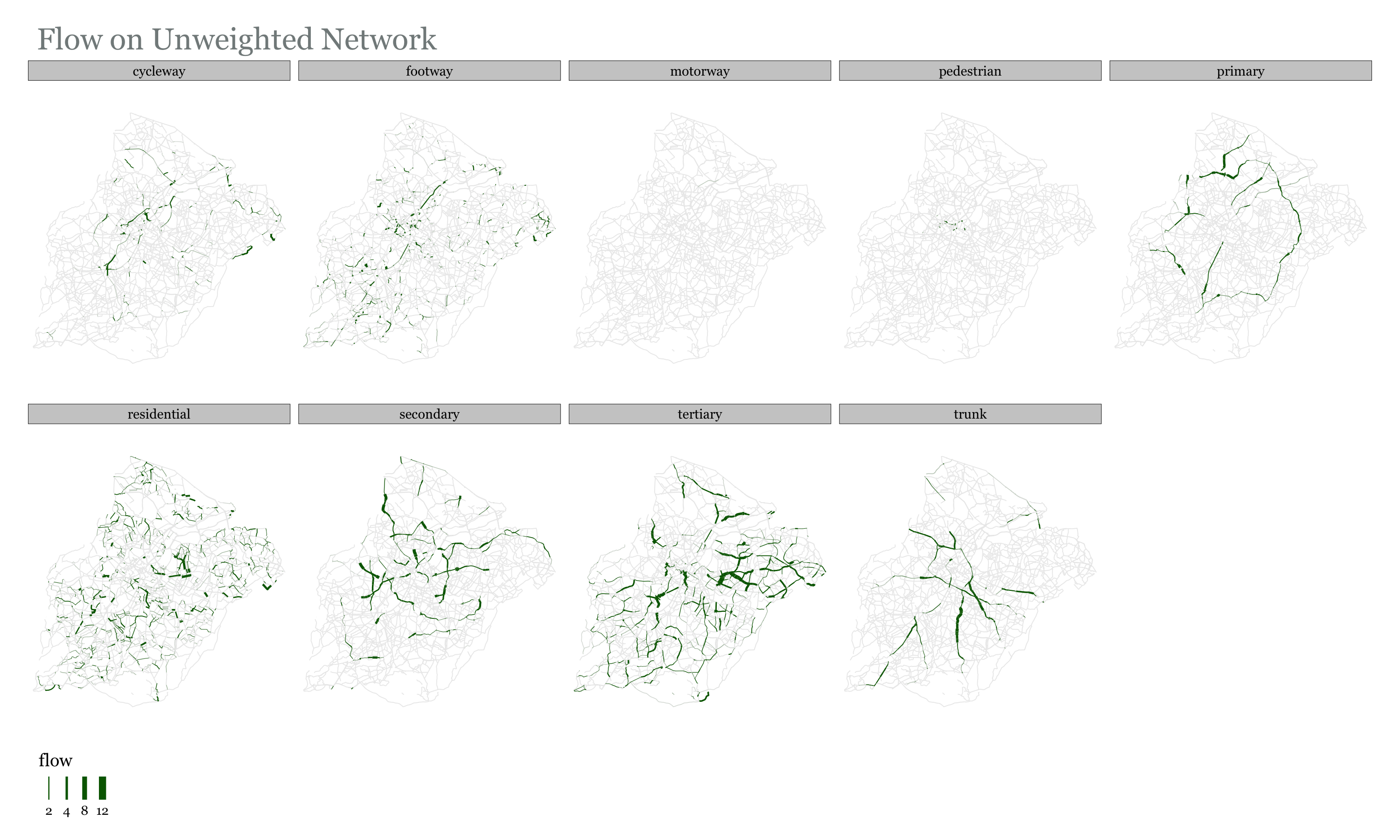} 

}

\caption{Flow Results Based on Unweighted Shortest Paths (Birmingham)}\label{fig:flowsfacetunweightedBirmingham}
\end{figure}

\begin{figure}

{\centering \includegraphics[width=0.75\linewidth]{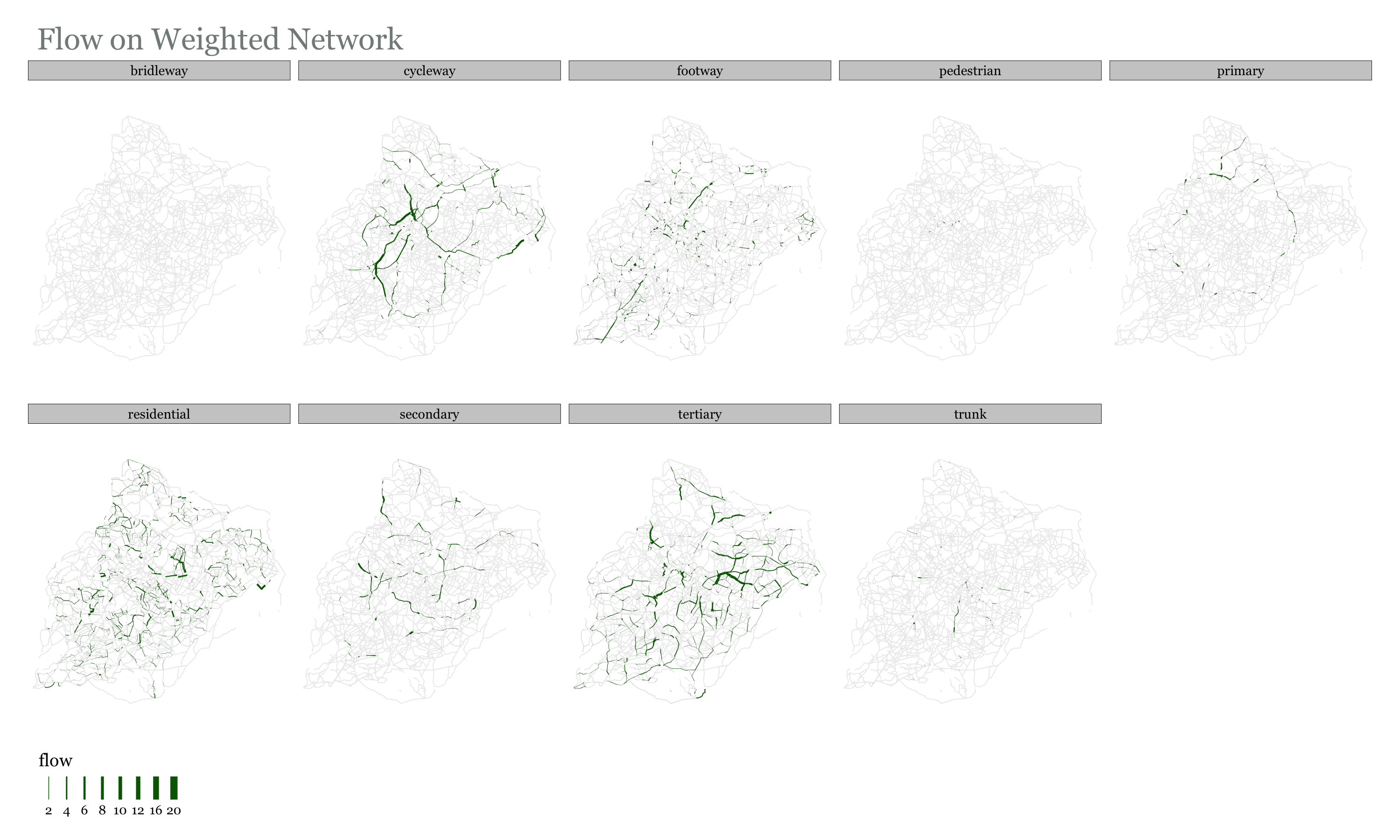} 

}

\caption{Flow Results Based on Weighted Shortest Paths (Birmingham)}\label{fig:flowsfacetweightedBirmingham}
\end{figure}

\clearpage

\subsubsection{Community Detection}

\begin{figure}

{\centering \includegraphics[width=0.9\linewidth]{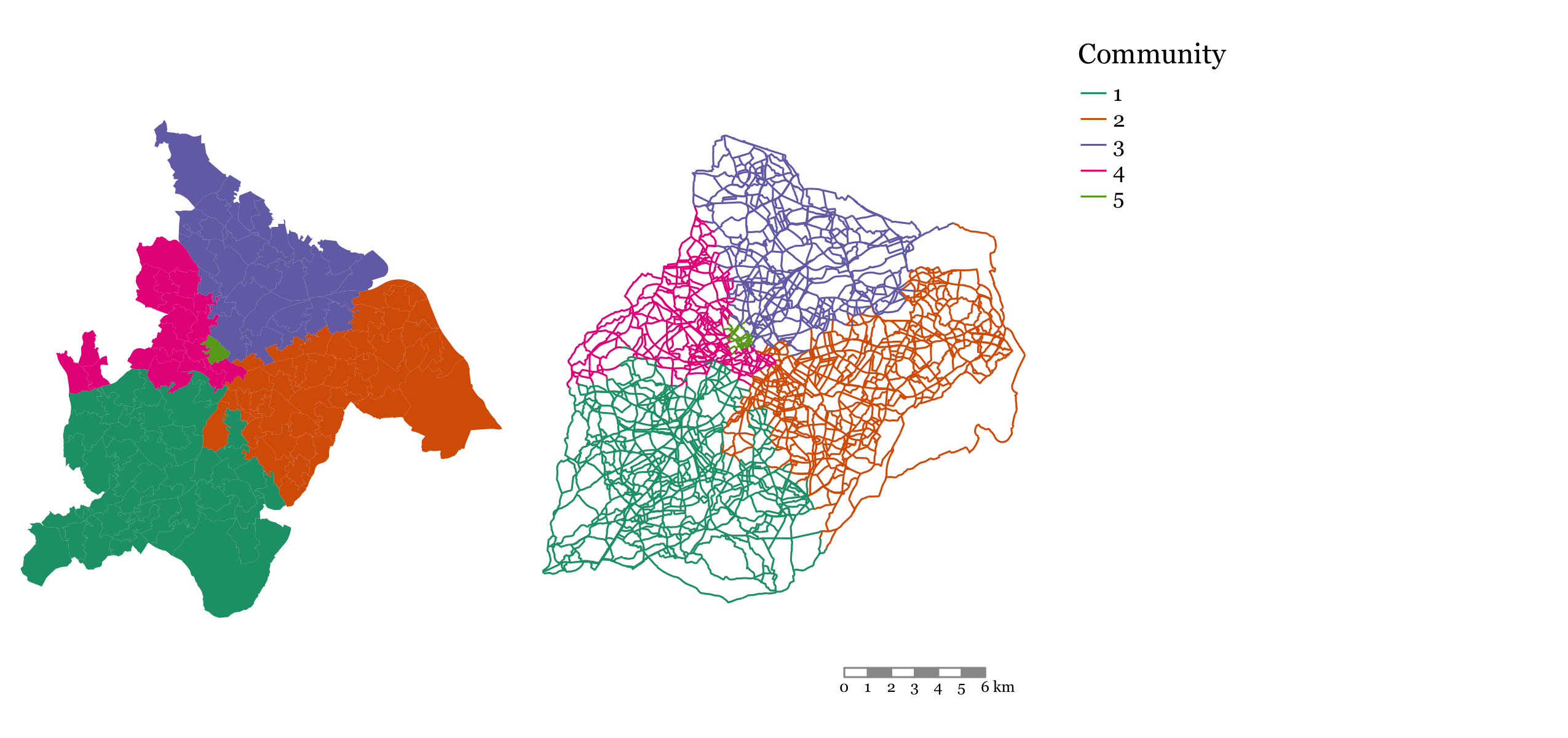} 

}

\caption{Communities Based on Potential Cycling Demand (Birmingham)}\label{fig:communitiesBirmingham}
\end{figure}

\subsubsection{Network Expansion Algorithms}

\begin{figure}[H]

{\centering \includegraphics[width=0.45\linewidth]{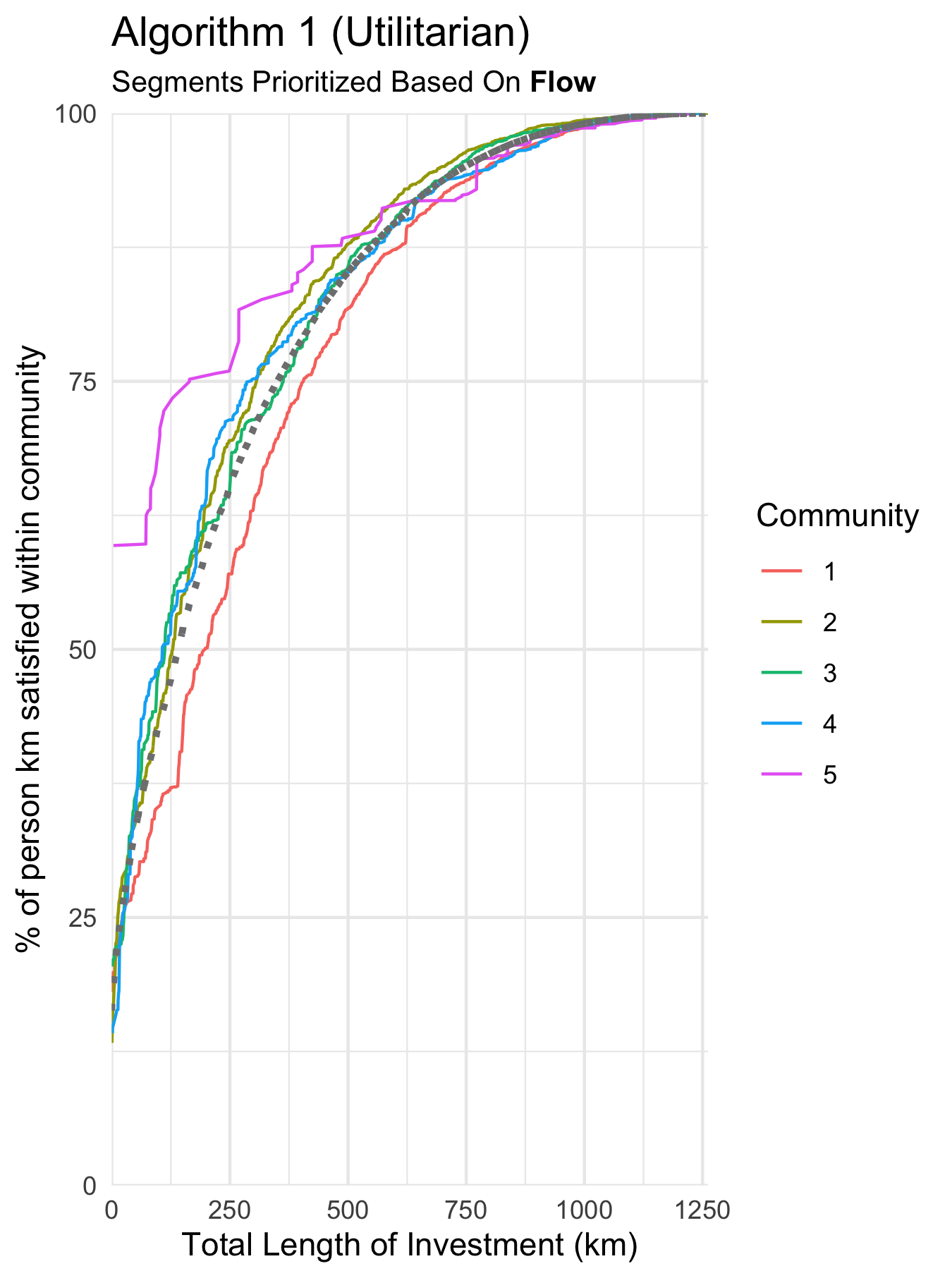} \includegraphics[width=0.45\linewidth]{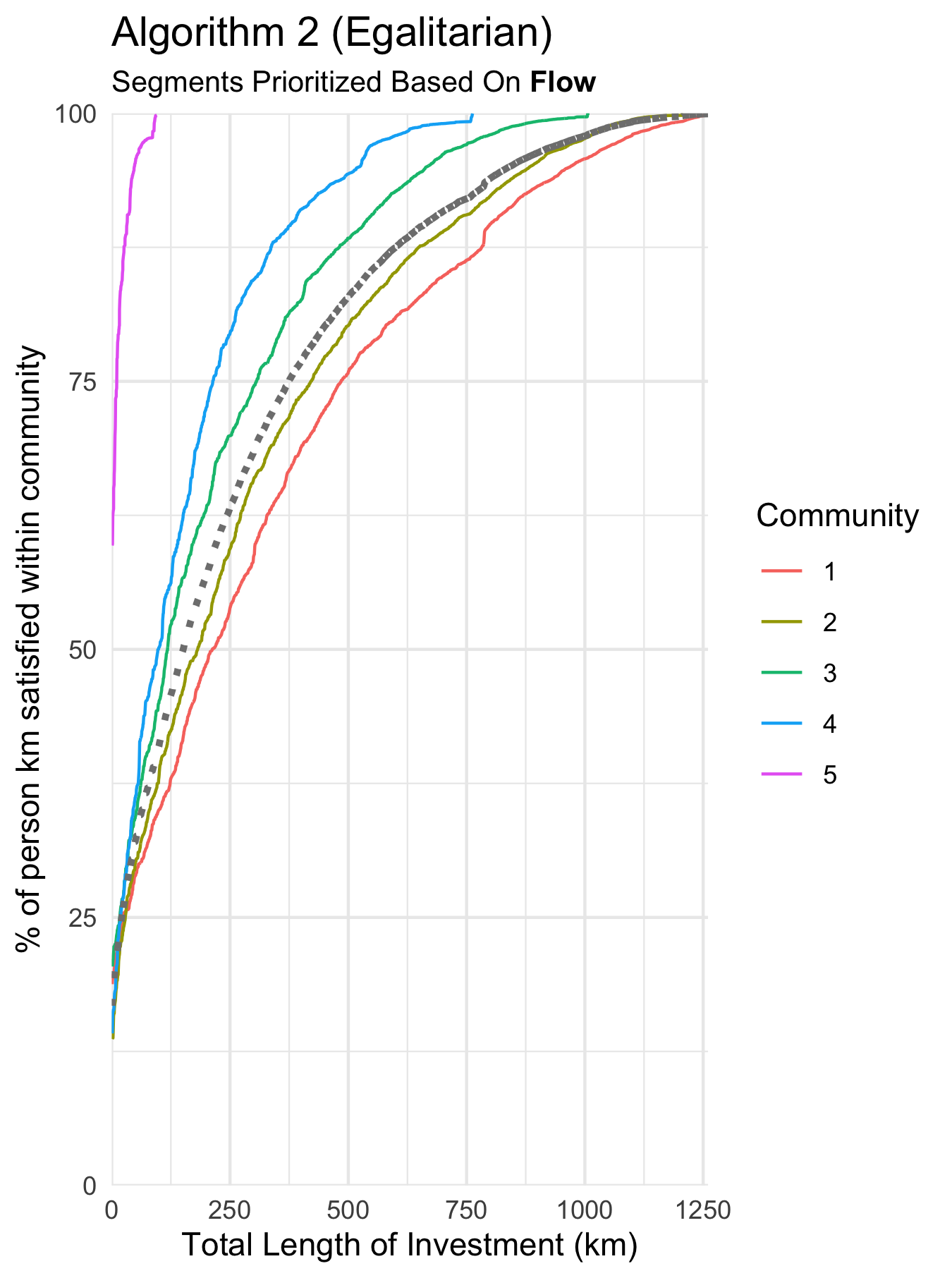} 

}

\caption{Comparing Overall (Dashed) and Community Level Person-Km Satisfied (Birmingham)}\label{fig:growthtotalBirmingham}
\end{figure}

\begin{figure}[H]

{\centering \includegraphics[width=0.45\linewidth]{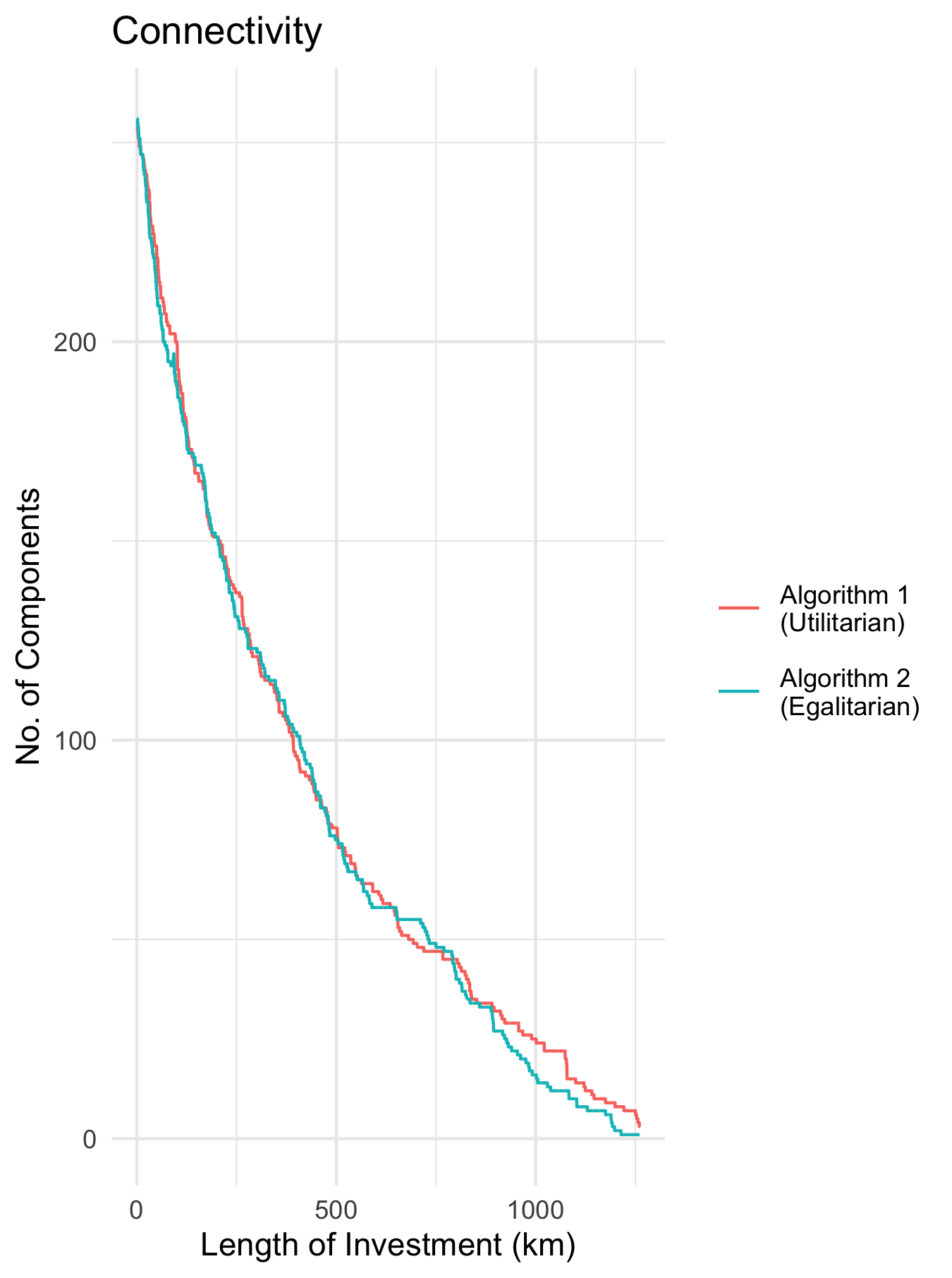} \includegraphics[width=0.45\linewidth]{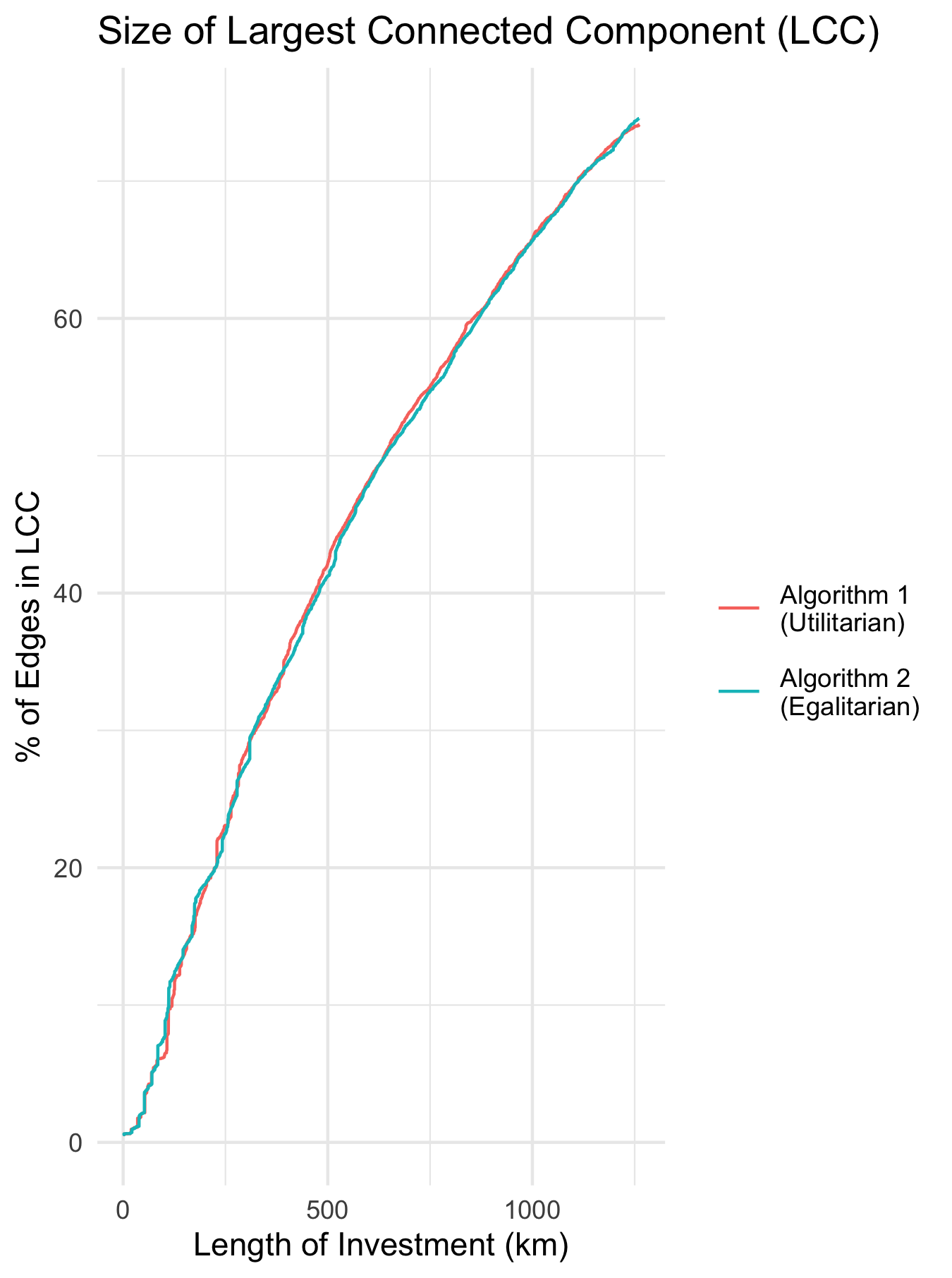} 

}

\caption{Network Characteristics}\label{fig:componentsandGCCBirmingham}
\end{figure}

\begin{figure}

{\centering \includegraphics[width=0.45\linewidth]{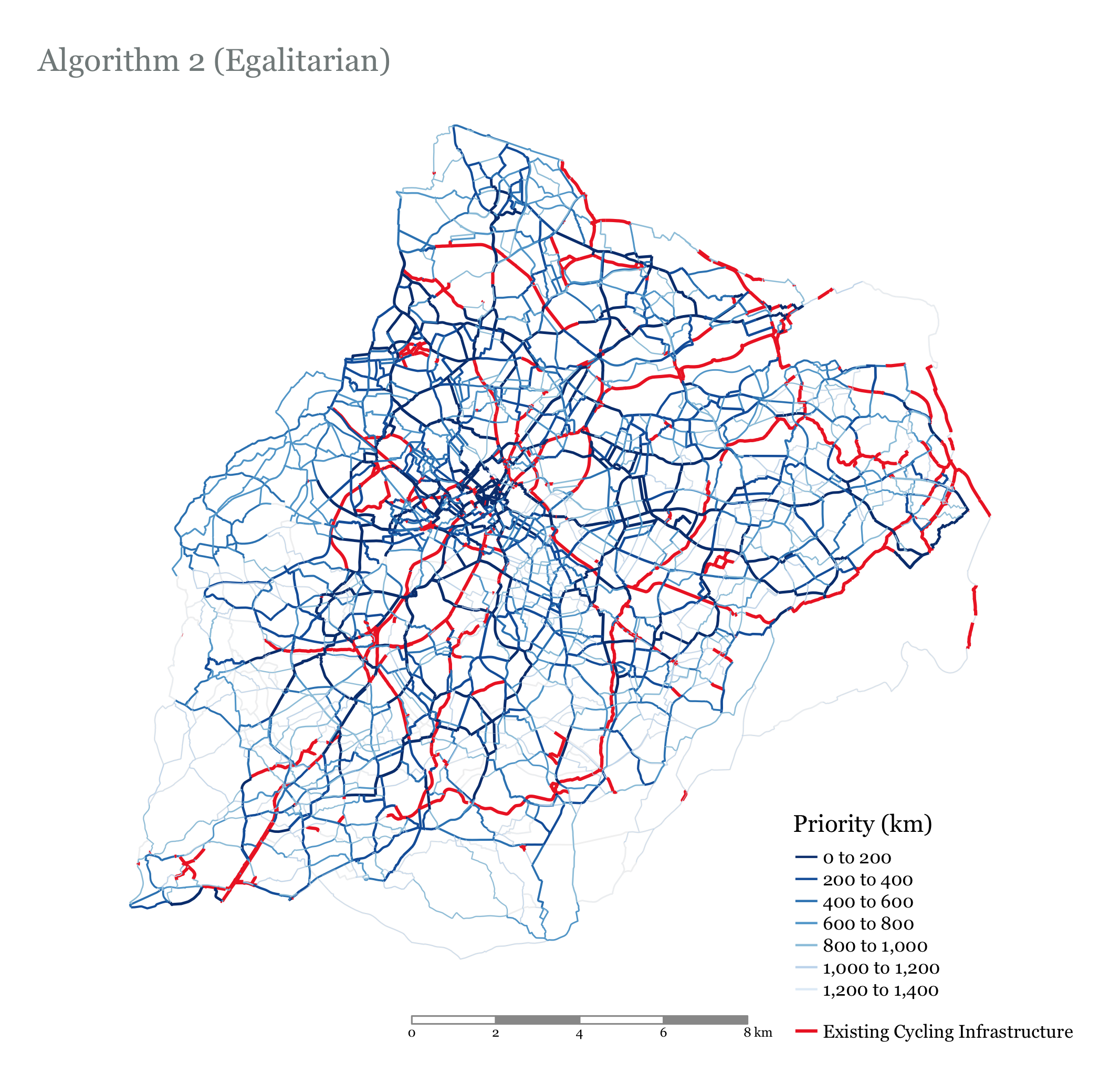} 

}

\caption{Road Segment Priority (left), disaggregated by road type (right) - Egalitarian Growth}\label{fig:growth3MapandBarBirmingham}
\end{figure}

\clearpage

\hypertarget{leeds}{%
\subsection{Leeds}\label{leeds}}

\subsubsection{Potential Demand}

\begin{figure}
\includegraphics[width=0.3\linewidth]{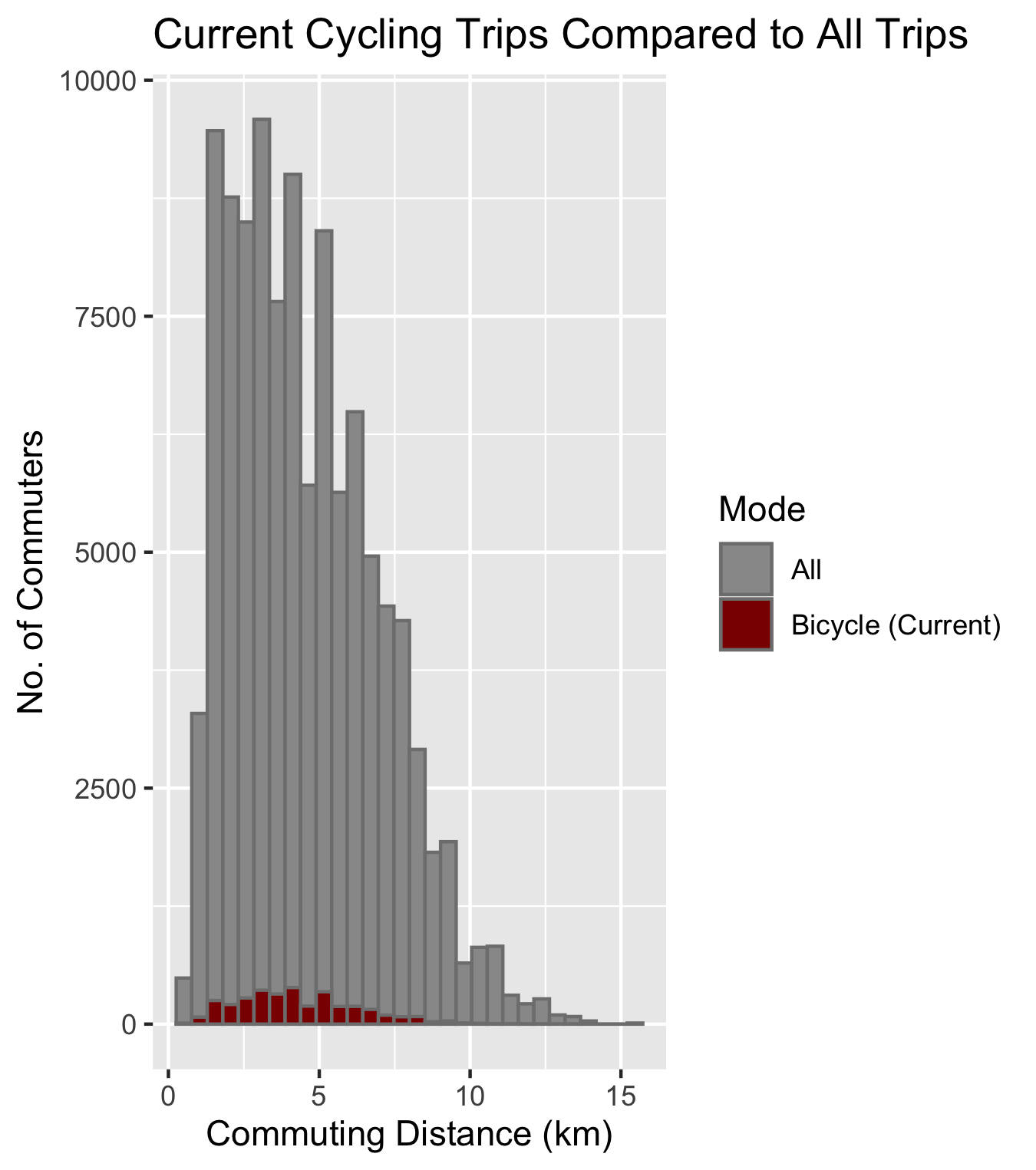} \includegraphics[width=0.3\linewidth]{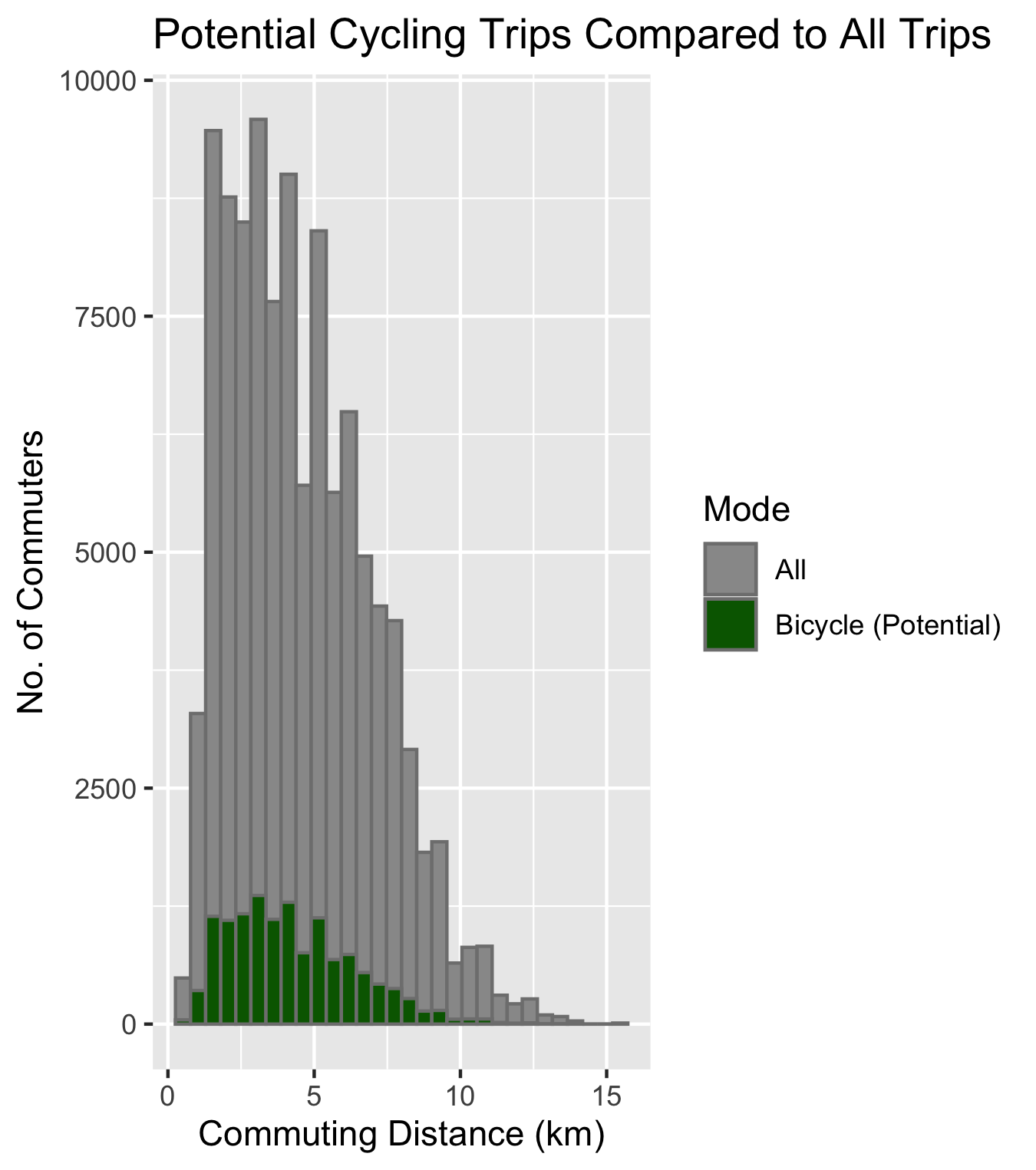} \includegraphics[width=0.3\linewidth]{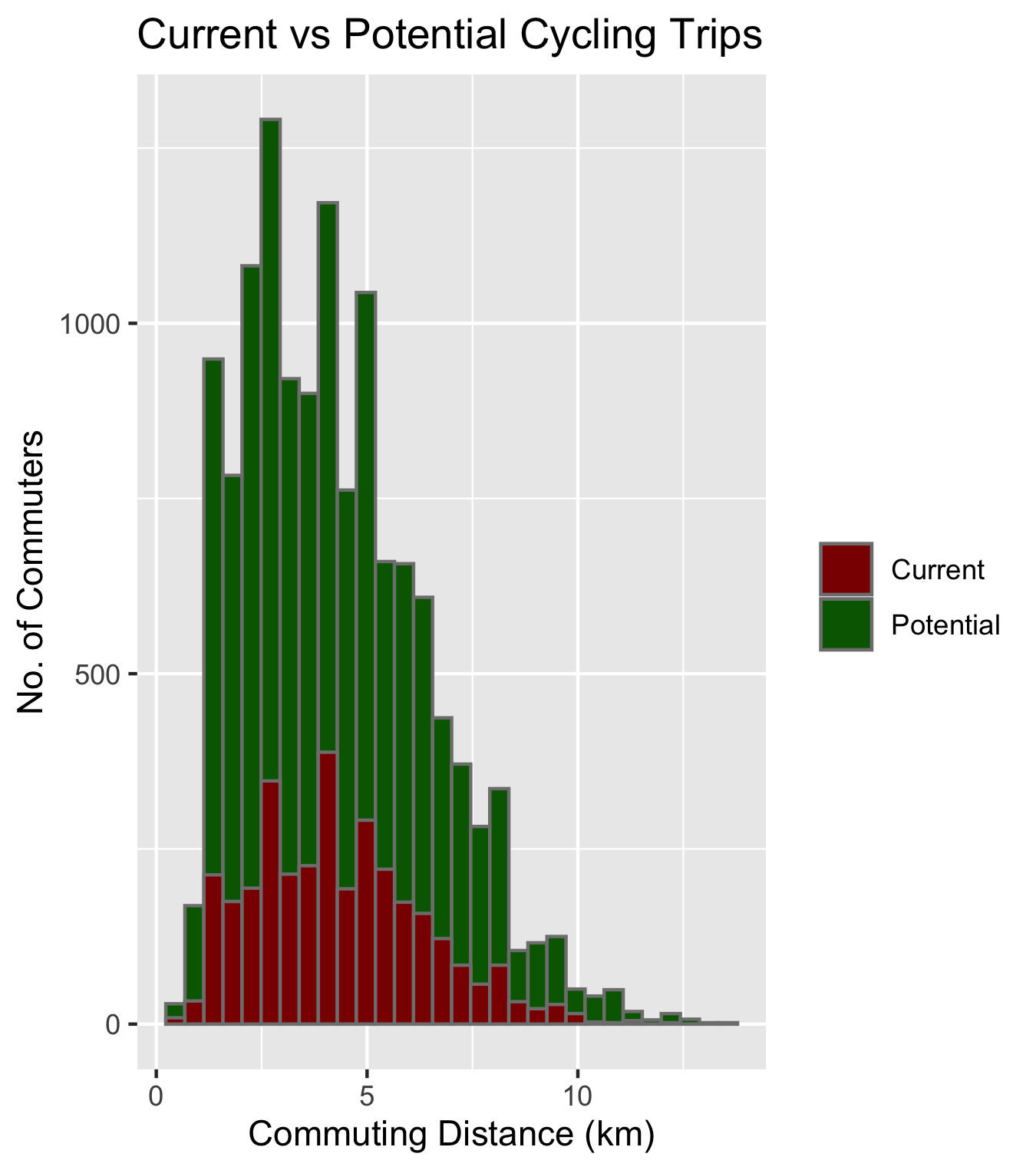} \caption{Distribution of Potential Cycling Demand}\label{fig:potdemhistogramsLeeds}
\end{figure}

\begin{figure}[H]

{\centering \includegraphics[width=0.8\linewidth]{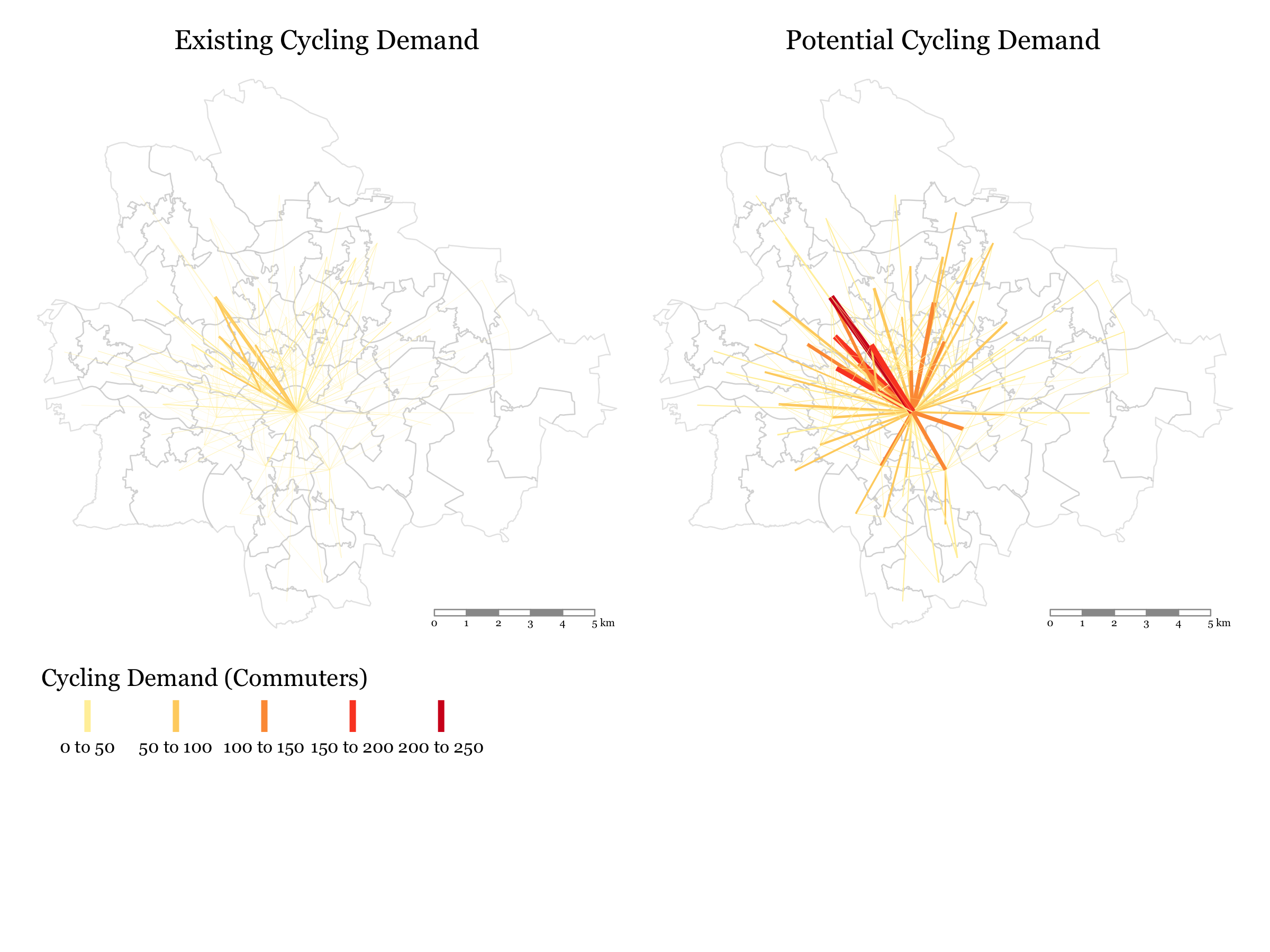} 

}

\caption{Current and Potential Cycling Demand}\label{fig:desirefacetcyclingLeeds}
\end{figure}

\clearpage

\subsubsection{Routing Cycling Flows}

\begin{figure}

{\centering \includegraphics[width=0.75\linewidth]{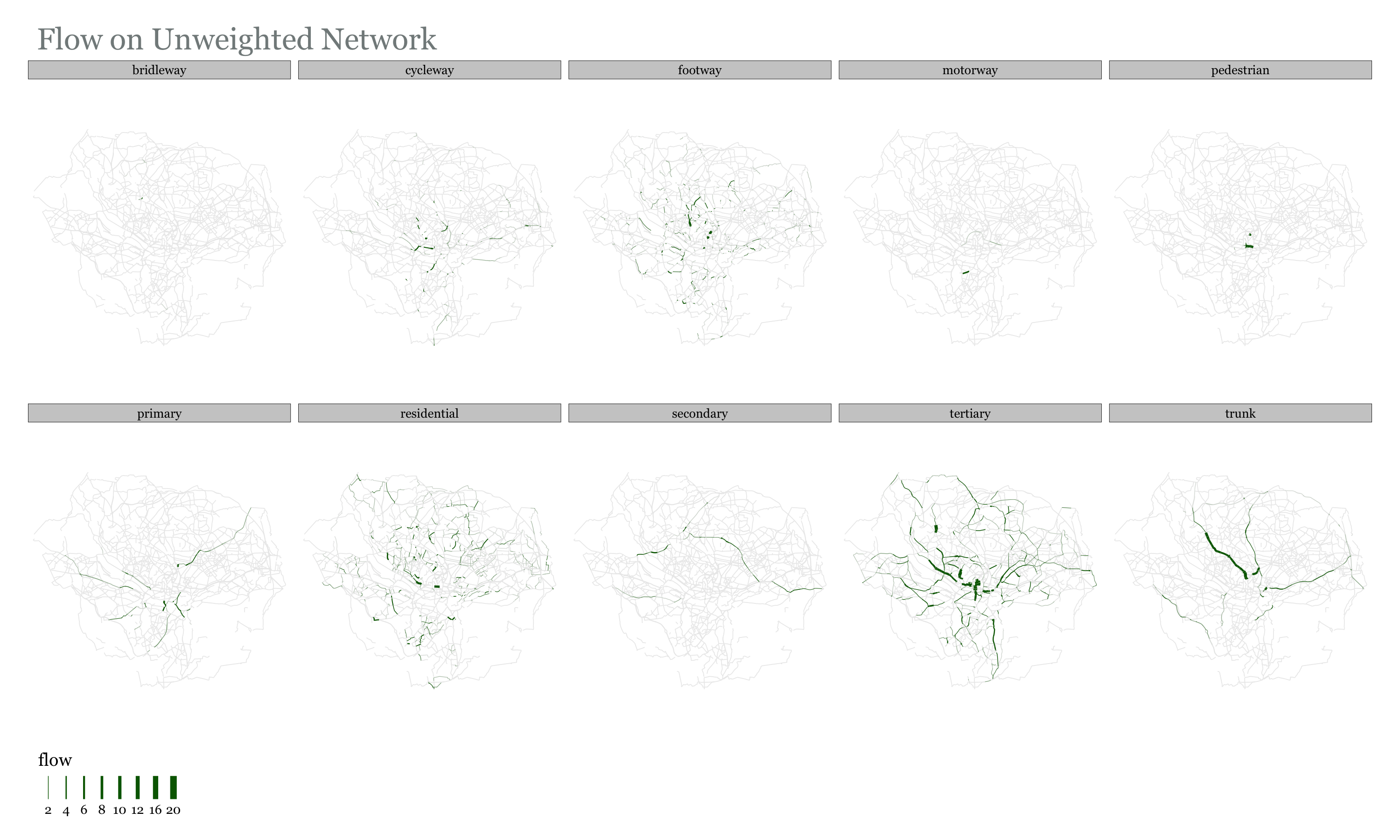} 

}

\caption{Flow Results Based on Unweighted Shortest Paths (Leeds)}\label{fig:flowsfacetunweightedLeeds}
\end{figure}

\begin{figure}

{\centering \includegraphics[width=0.75\linewidth]{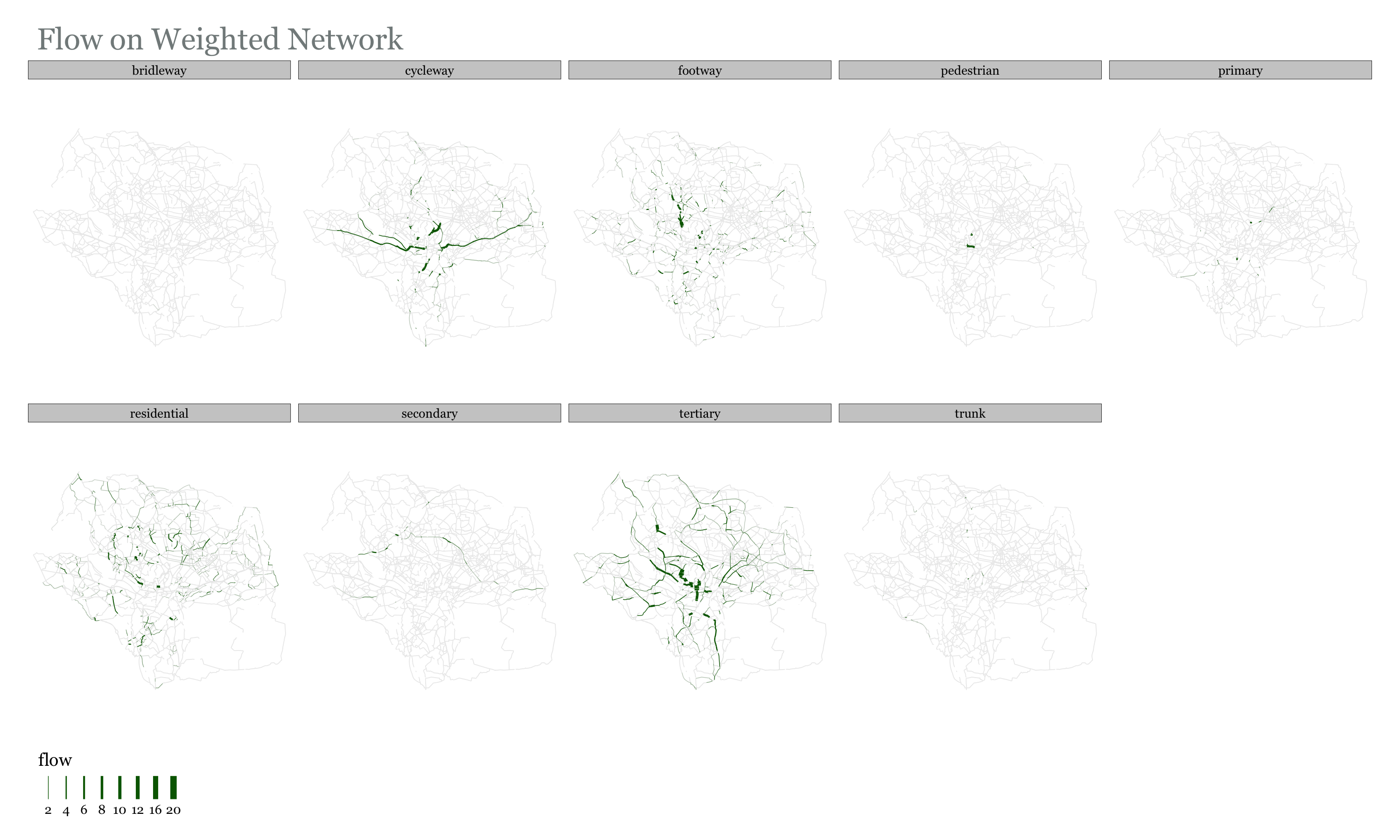} 

}

\caption{Flow Results Based on Weighted Shortest Paths (Leeds)}\label{fig:flowsfacetweightedLeeds}
\end{figure}

\clearpage

\subsubsection{Community Detection}

\begin{figure}

{\centering \includegraphics[width=0.9\linewidth]{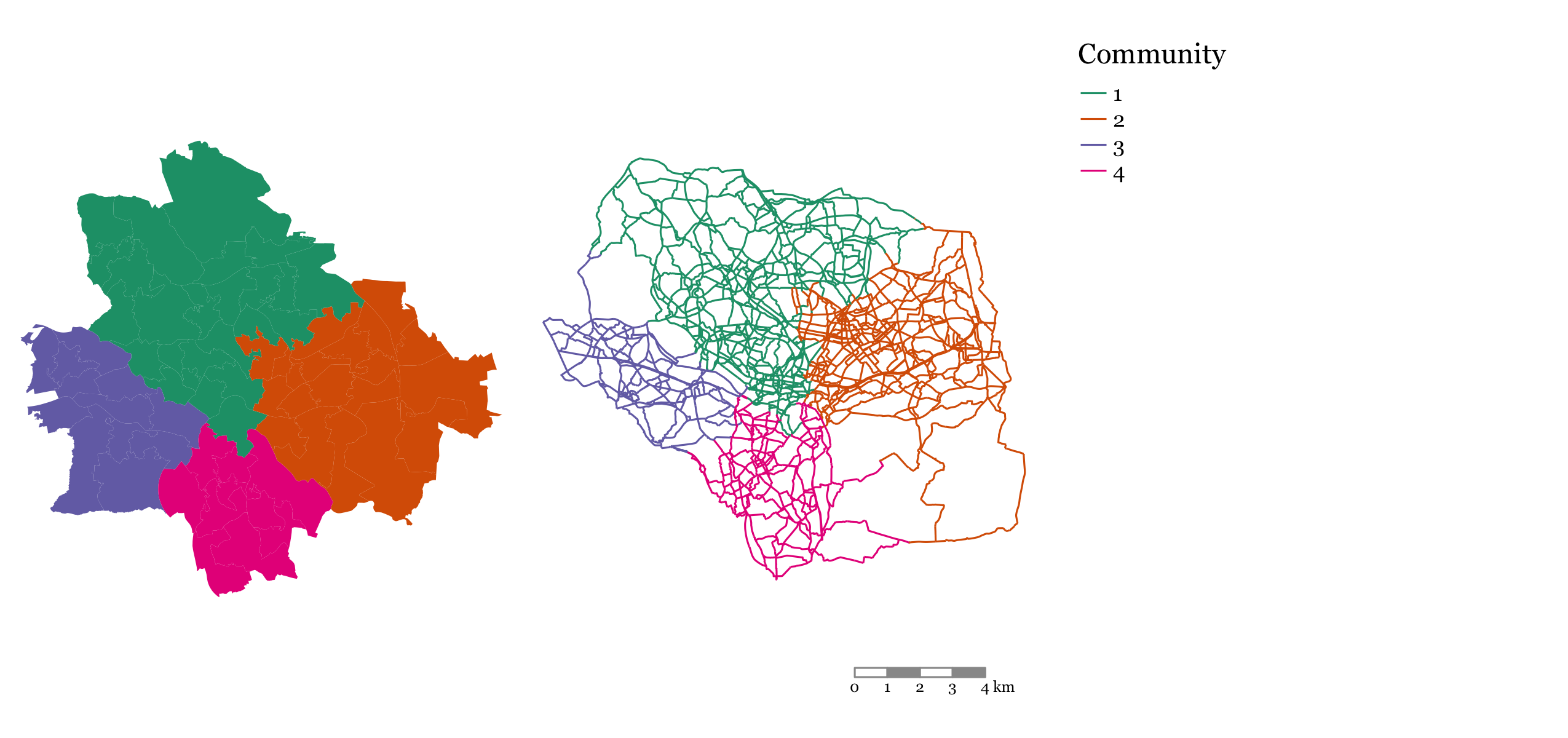} 

}

\caption{Communities Based on Potential Cycling Demand (Leeds)}\label{fig:communitiesLeeds}
\end{figure}

\subsubsection{Network Expansion Algorithms}

\begin{figure}[H]

{\centering \includegraphics[width=0.45\linewidth]{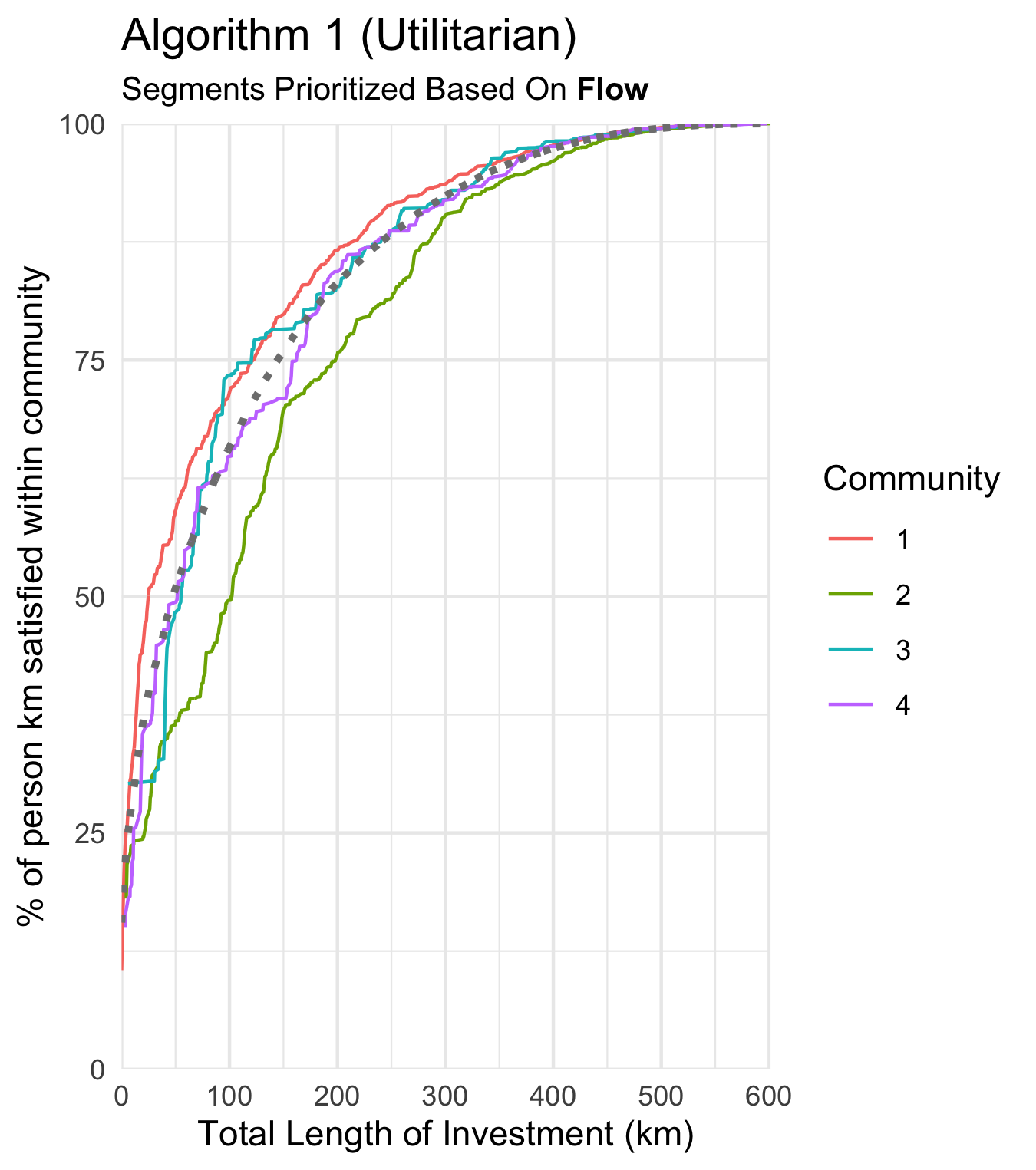} \includegraphics[width=0.45\linewidth]{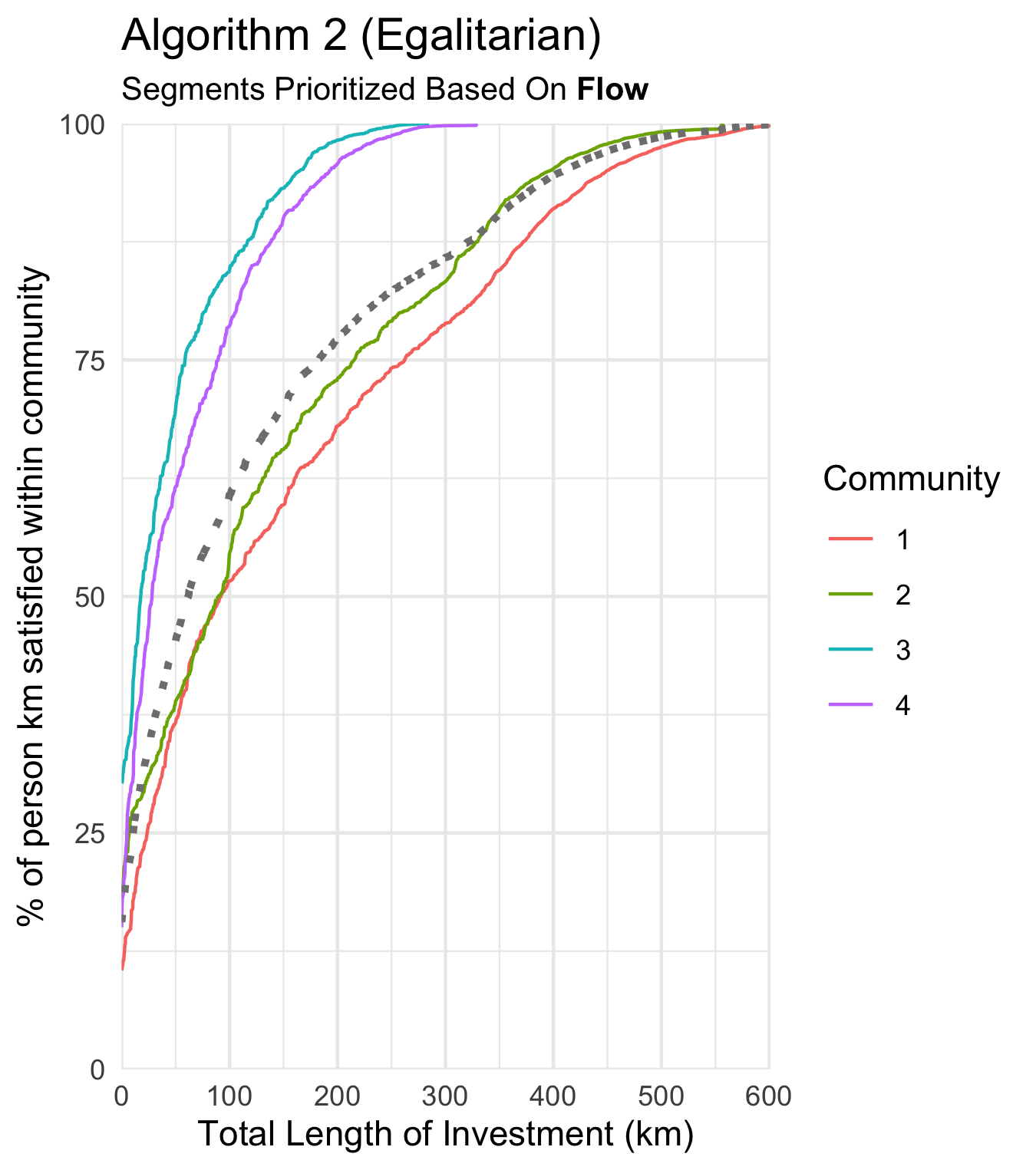} 

}

\caption{Comparing Overall (Dashed) and Community Level Person-Km Satisfied (Leeds)}\label{fig:growthtotalLeeds}
\end{figure}

\begin{figure}[H]

{\centering \includegraphics[width=0.45\linewidth]{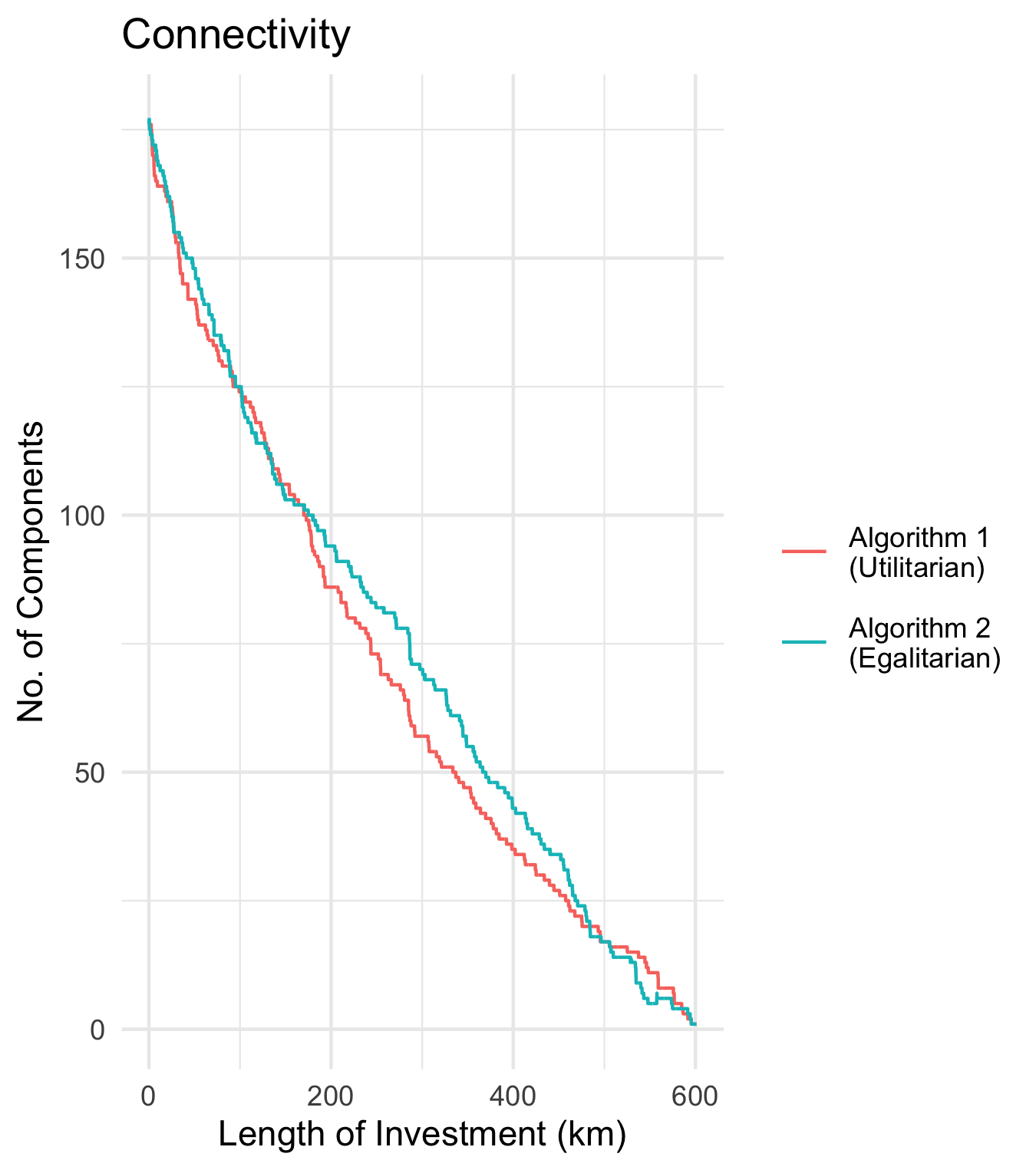} \includegraphics[width=0.45\linewidth]{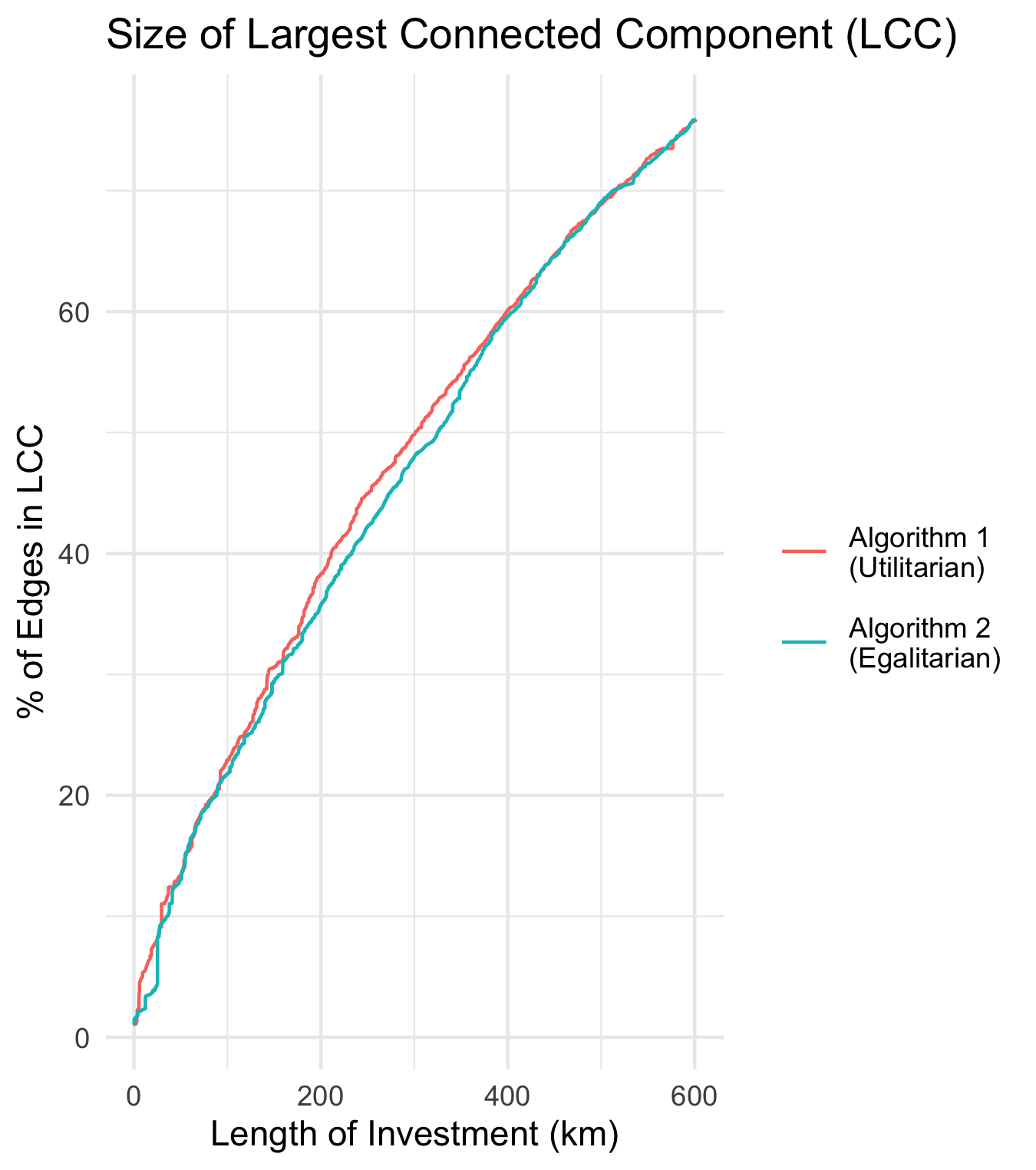} 

}

\caption{Network Characteristics}\label{fig:componentsandGCCLeeds}
\end{figure}

\begin{figure}

{\centering \includegraphics[width=0.45\linewidth]{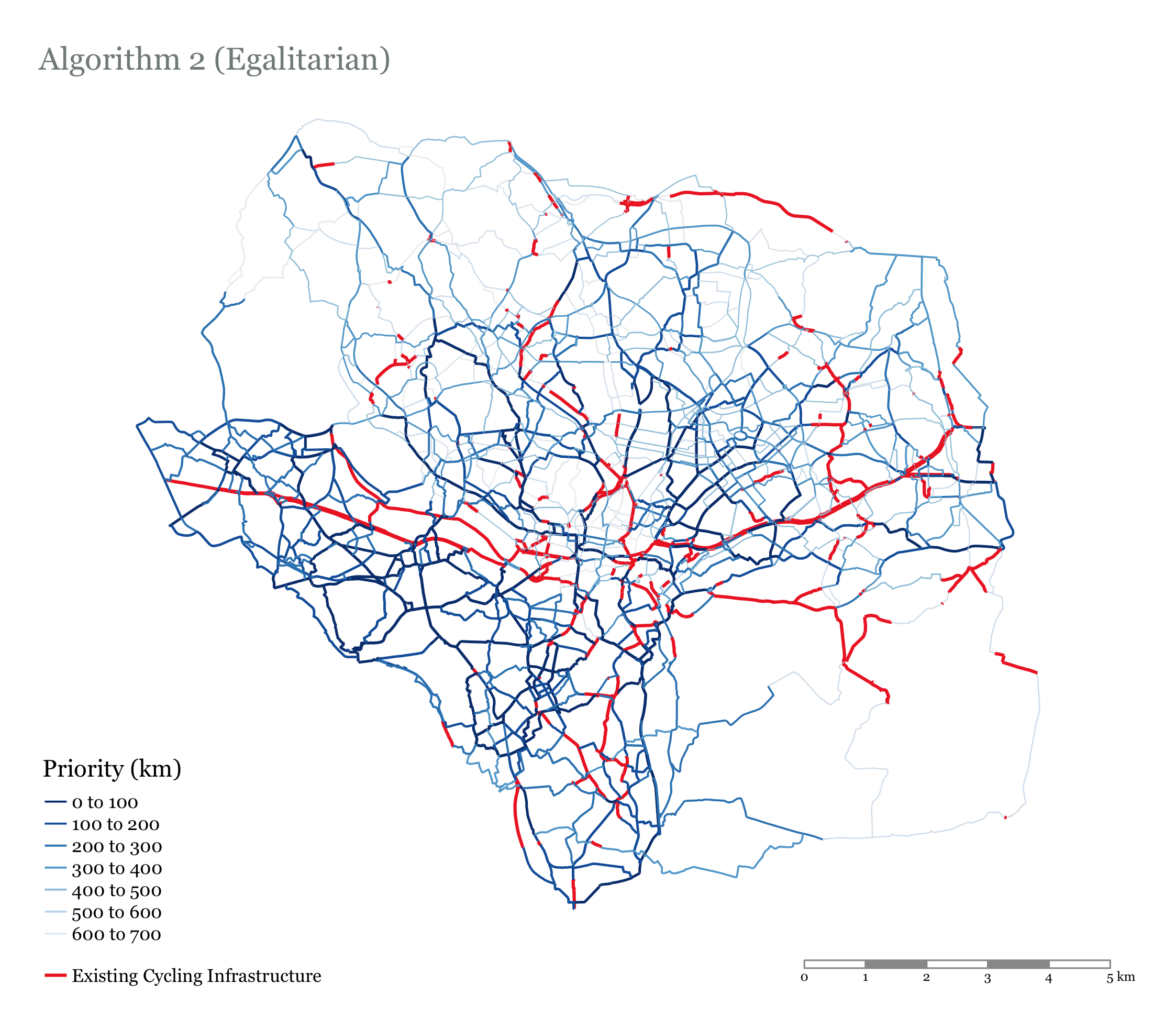} \includegraphics[width=0.45\linewidth]{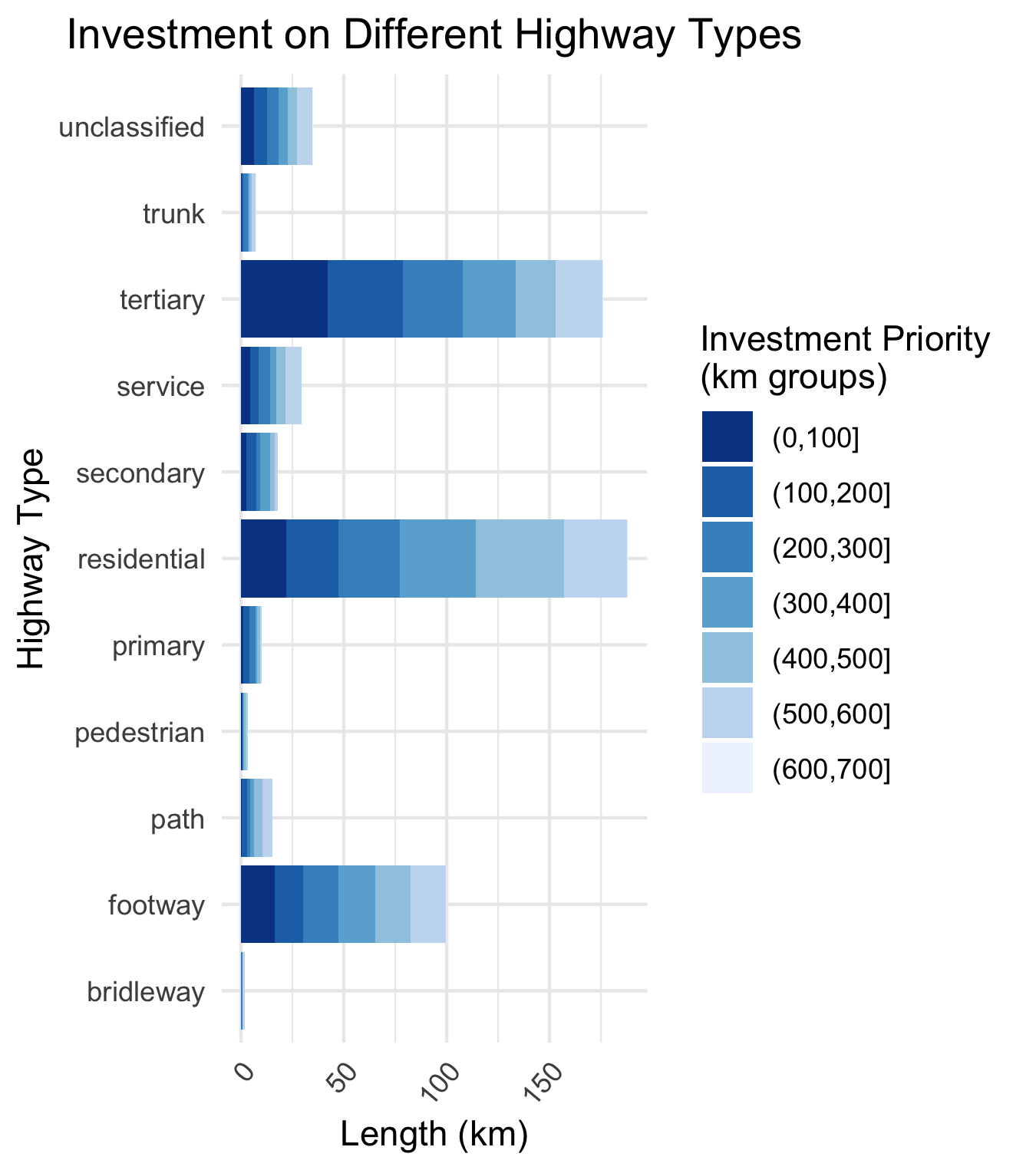} 

}

\caption{Road Segment Priority (left), disaggregated by road type (right) - Egalitarian Growth}\label{fig:growth3MapandBarLeeds}
\end{figure}

\clearpage

\hypertarget{nottingham}{%
\subsection{Nottingham}\label{nottingham}}

\subsubsection{Potential Demand}

\begin{figure}
\includegraphics[width=0.3\linewidth]{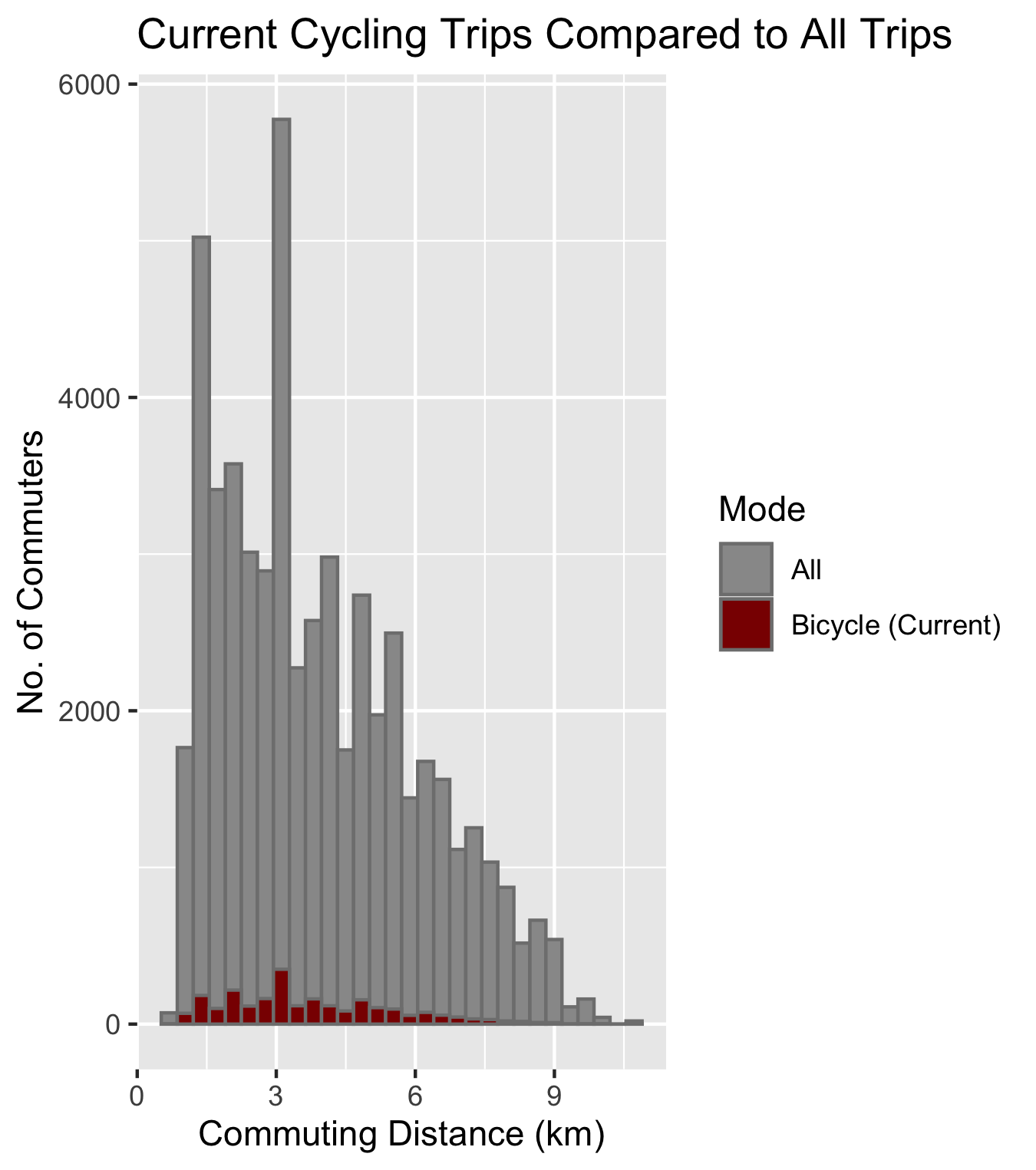} \includegraphics[width=0.3\linewidth]{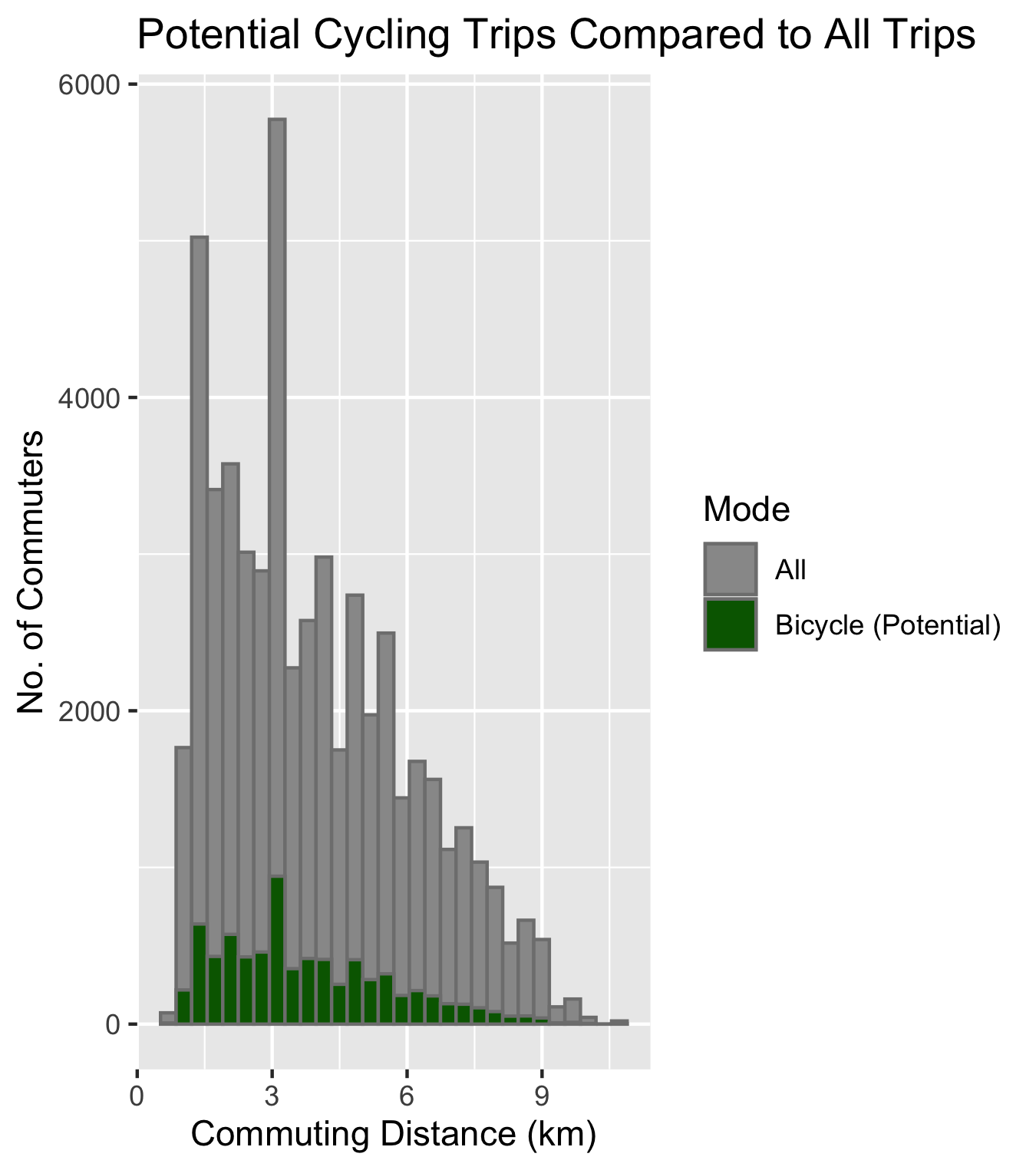} \includegraphics[width=0.3\linewidth]{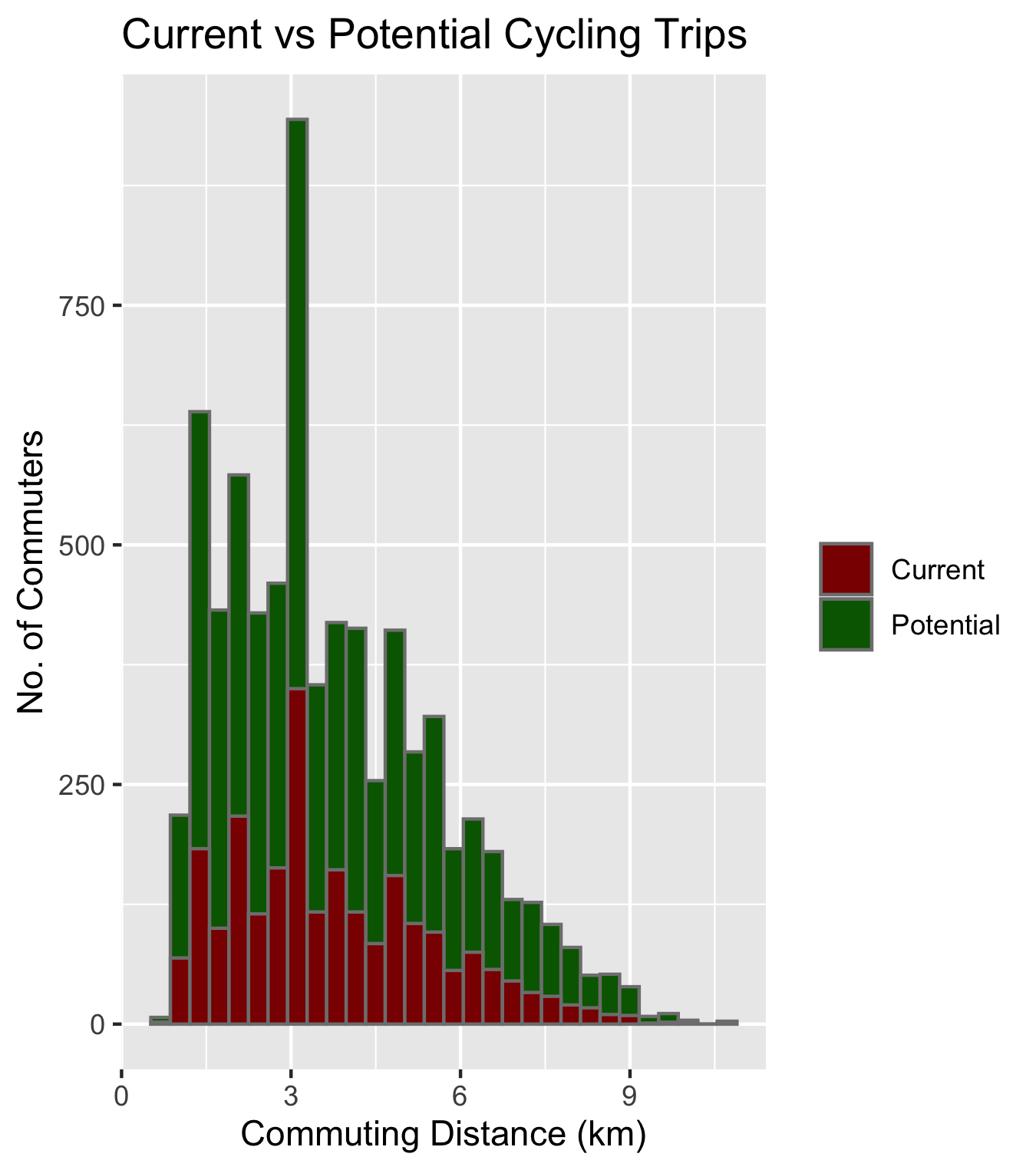} \caption{Distribution of Potential Cycling Demand}\label{fig:potdemhistogramsNottingham}
\end{figure}

\begin{figure}[H]

{\centering \includegraphics[width=0.8\linewidth]{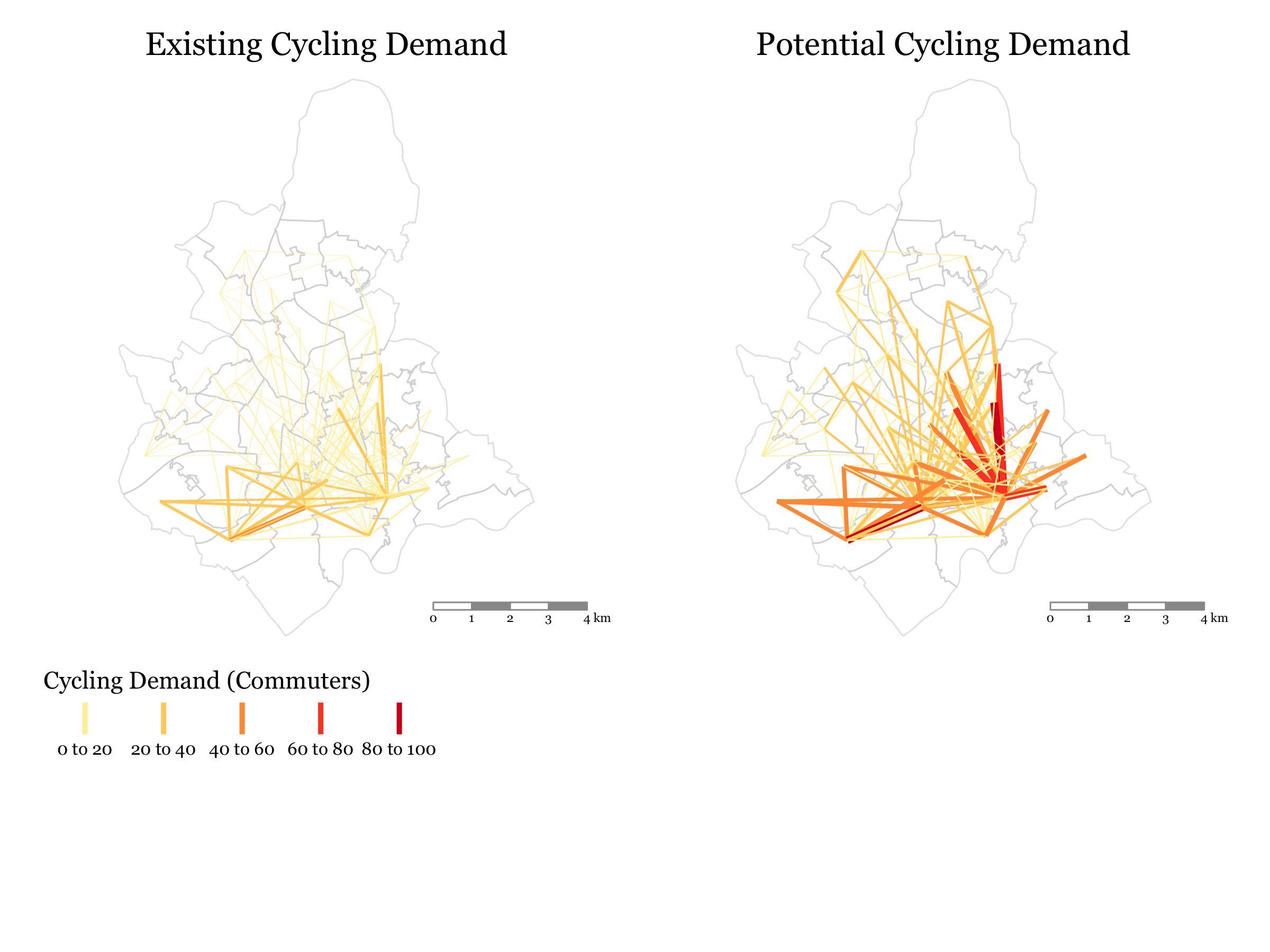} 

}

\caption{Current and Potential Cycling Demand}\label{fig:desirefacetcyclingNottingham}
\end{figure}

\clearpage

\subsubsection{Routing Cycling Flows}

\begin{figure}

{\centering \includegraphics[width=0.75\linewidth]{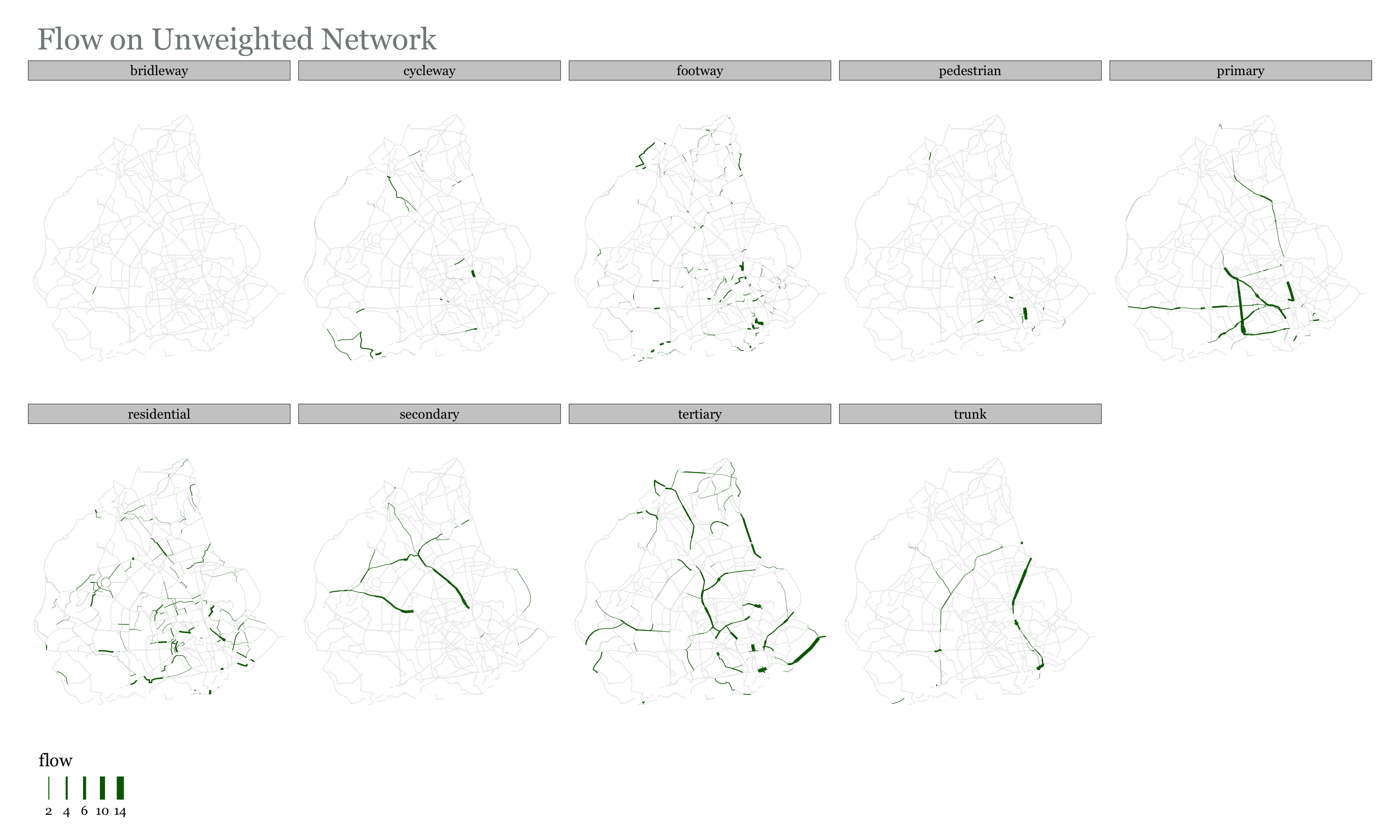} 

}

\caption{Flow Results Based on Unweighted Shortest Paths (Nottingham)}\label{fig:flowsfacetunweightedNottingham}
\end{figure}

\begin{figure}

{\centering \includegraphics[width=0.75\linewidth]{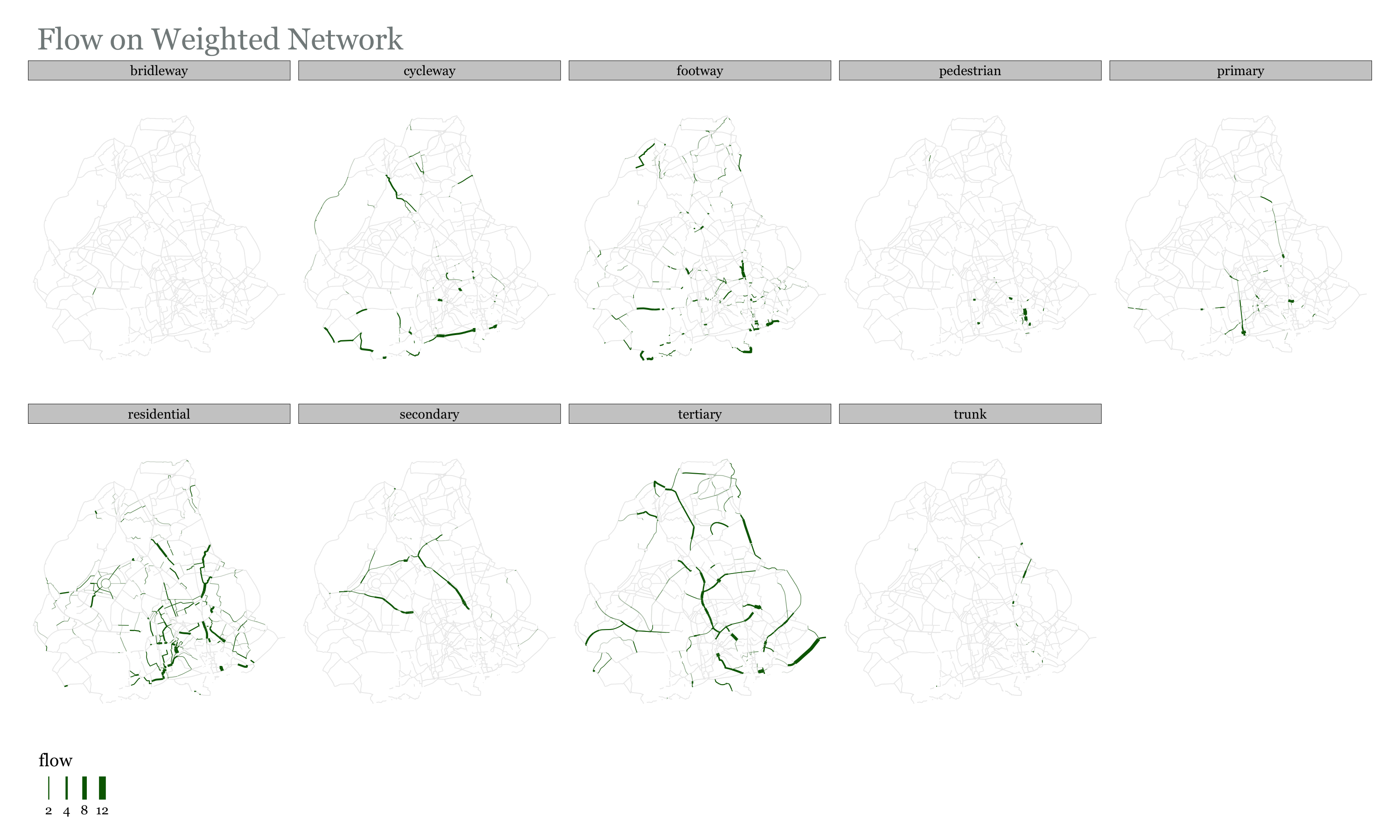} 

}

\caption{Flow Results Based on Weighted Shortest Paths (Nottingham)}\label{fig:flowsfacetweightedNottingham}
\end{figure}

\clearpage

\subsubsection{Community Detection}

\begin{figure}

{\centering \includegraphics[width=0.9\linewidth]{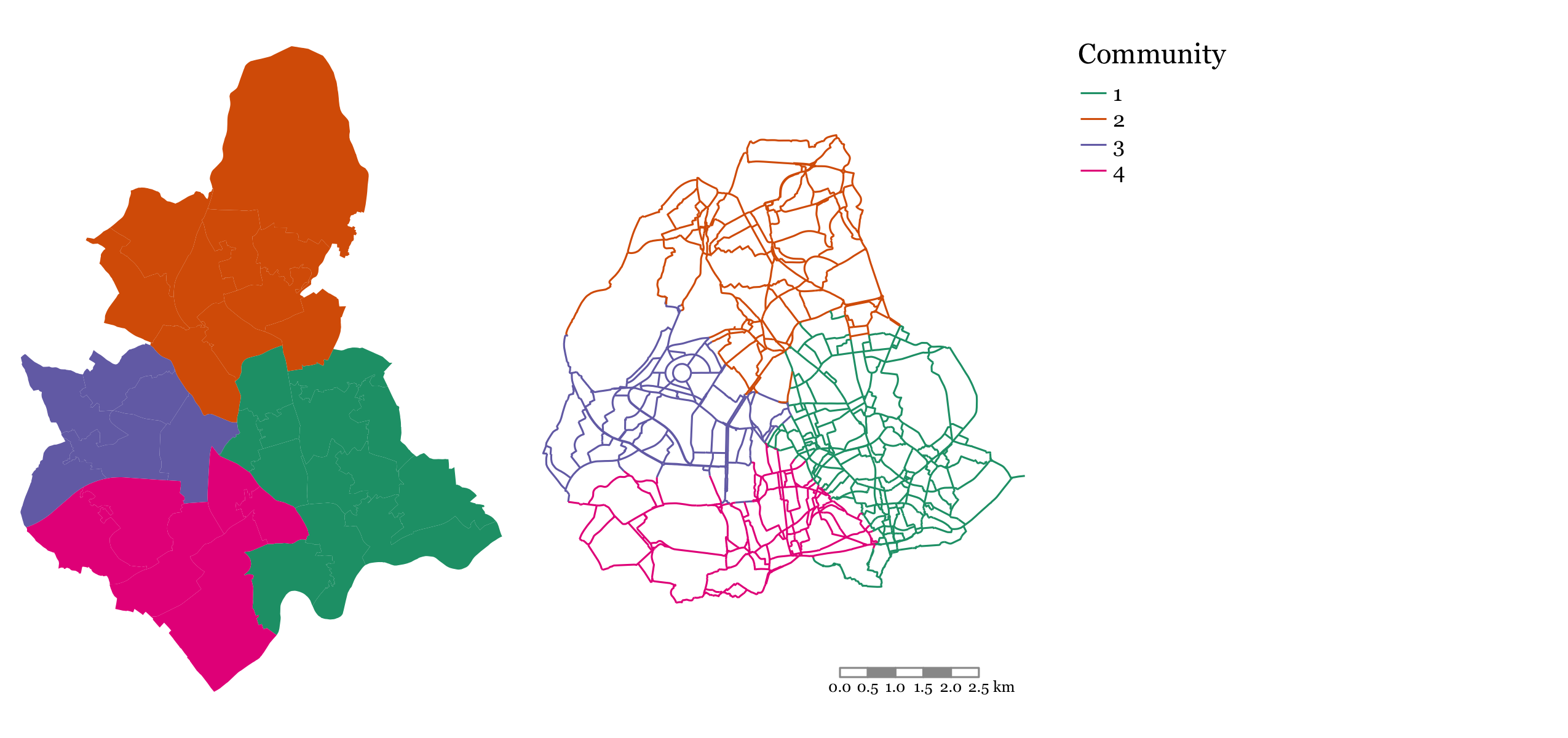} 

}

\caption{Communities Based on Potential Cycling Demand (Nottingham)}\label{fig:communitiesNottingham}
\end{figure}

\subsubsection{Network Expansion Algorithms}

\begin{figure}[H]
\includegraphics[width=0.45\linewidth]{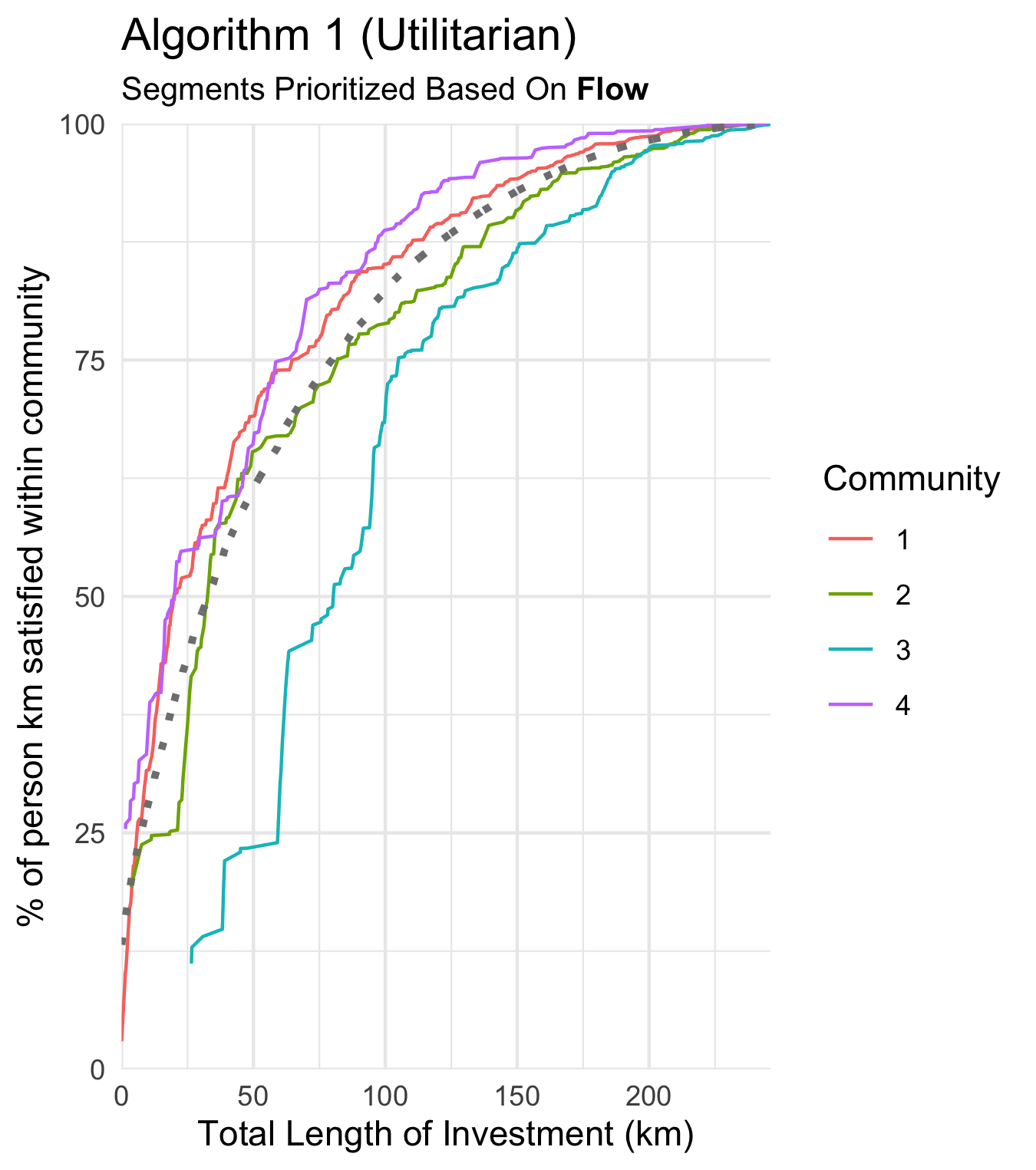} \includegraphics[width=0.45\linewidth]{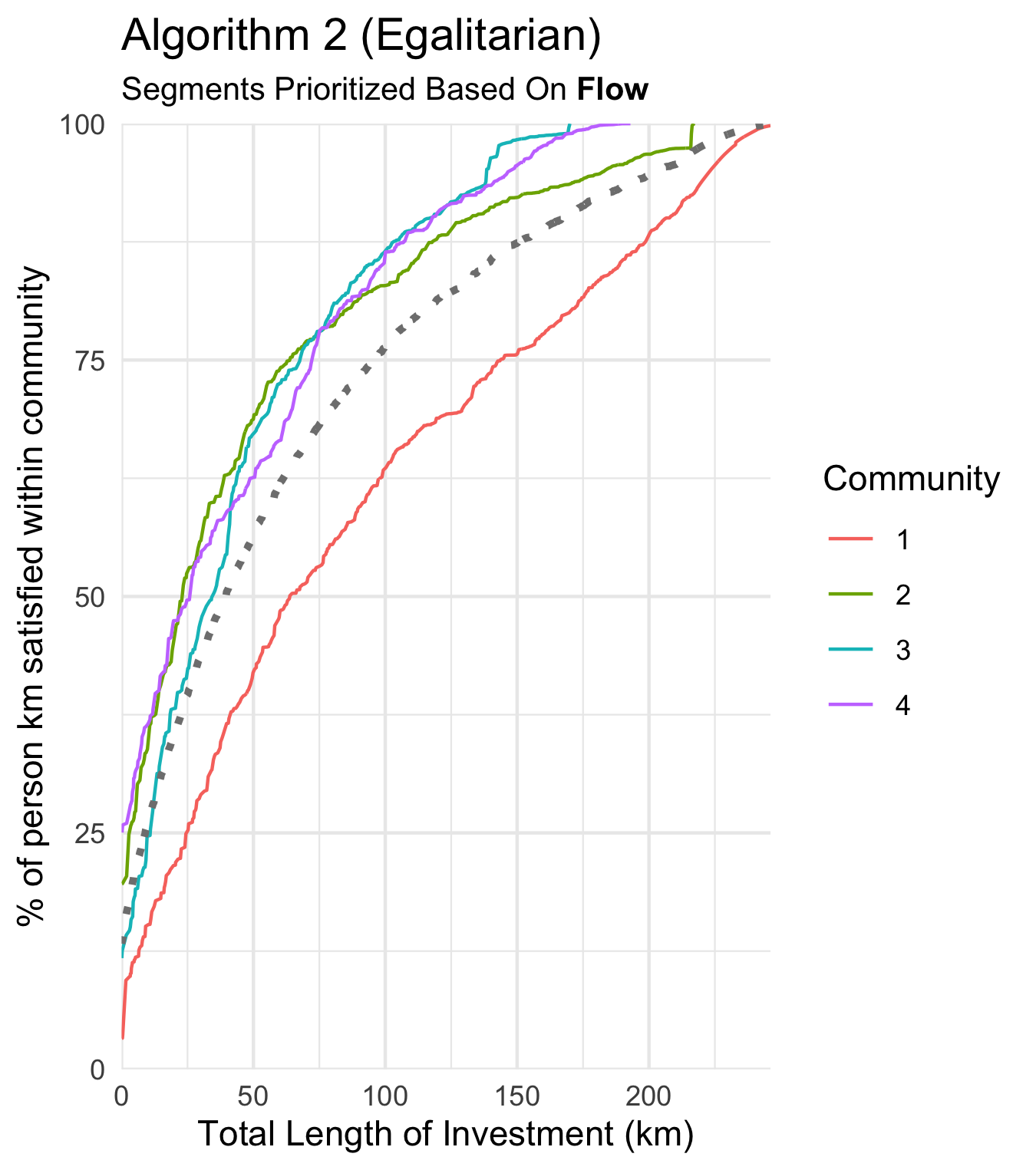} \caption{Comparing Overall (Dashed) and Community Level Person-Km Satisfied (Nottingham)}\label{fig:growthtotalNottingham}
\end{figure}

\begin{figure}[H]

{\centering \includegraphics[width=0.45\linewidth]{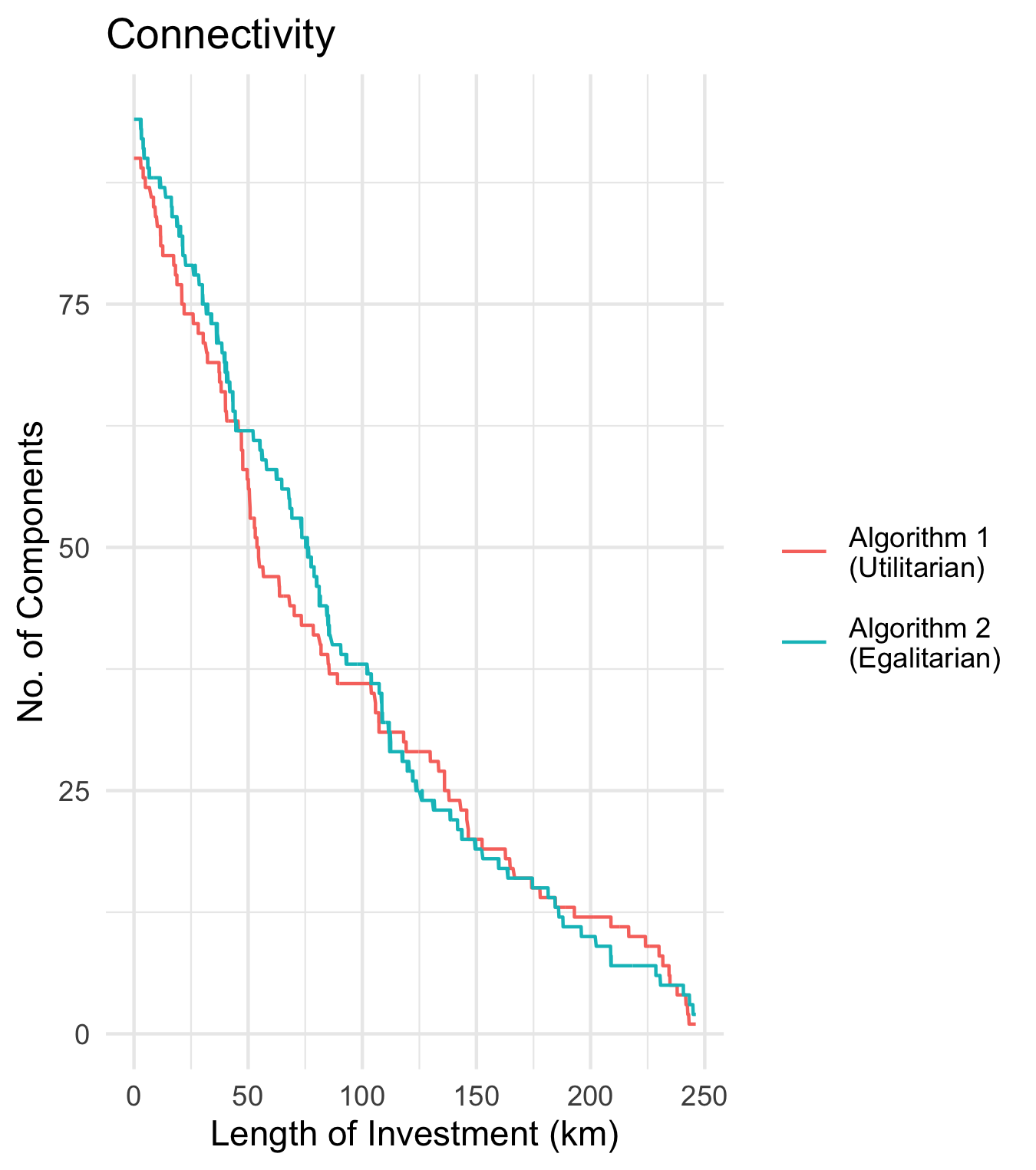} \includegraphics[width=0.45\linewidth]{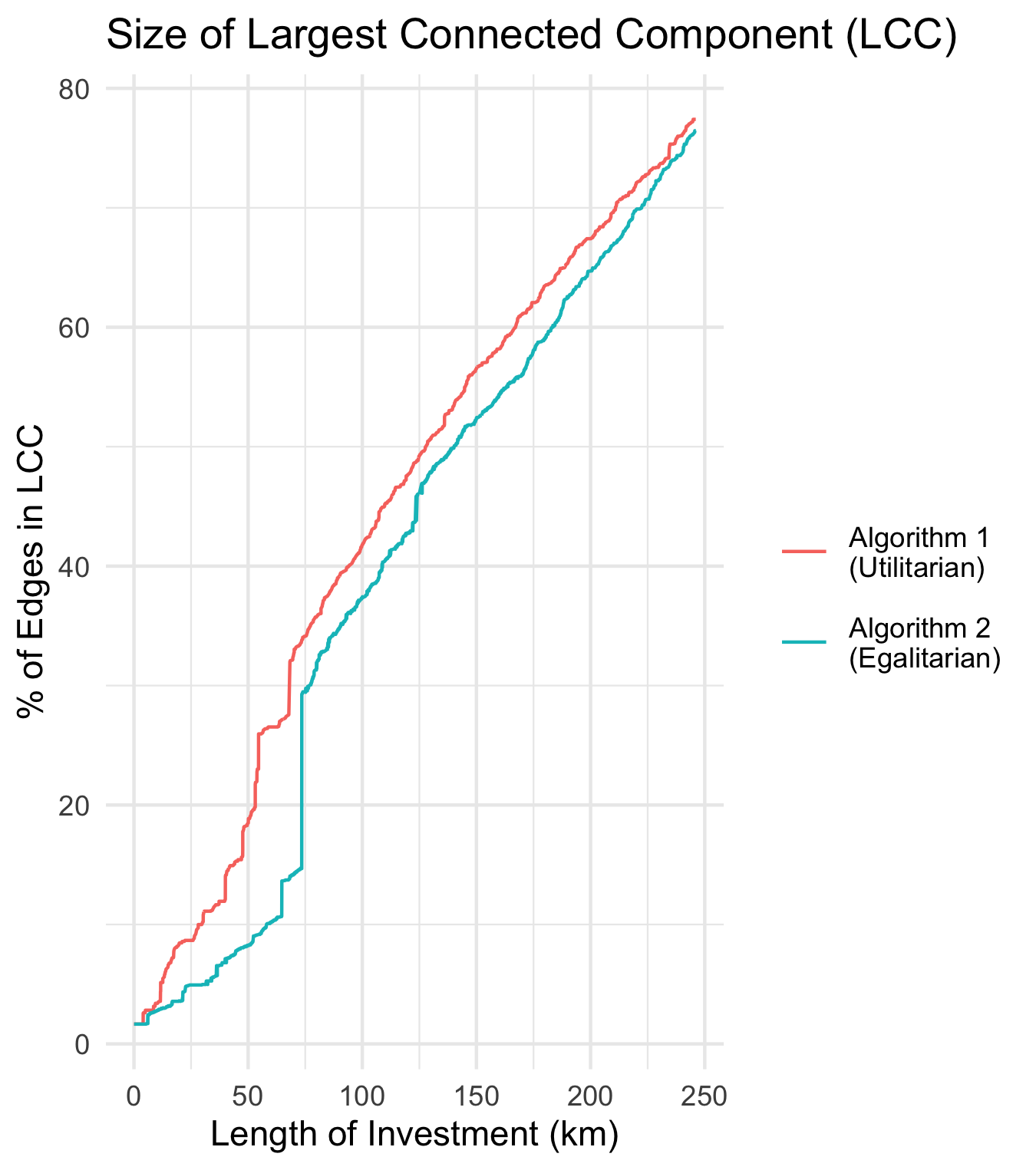} 

}

\caption{Network Characteristics}\label{fig:componentsandGCCNottingham}
\end{figure}

\begin{figure}

{\centering \includegraphics[width=0.45\linewidth]{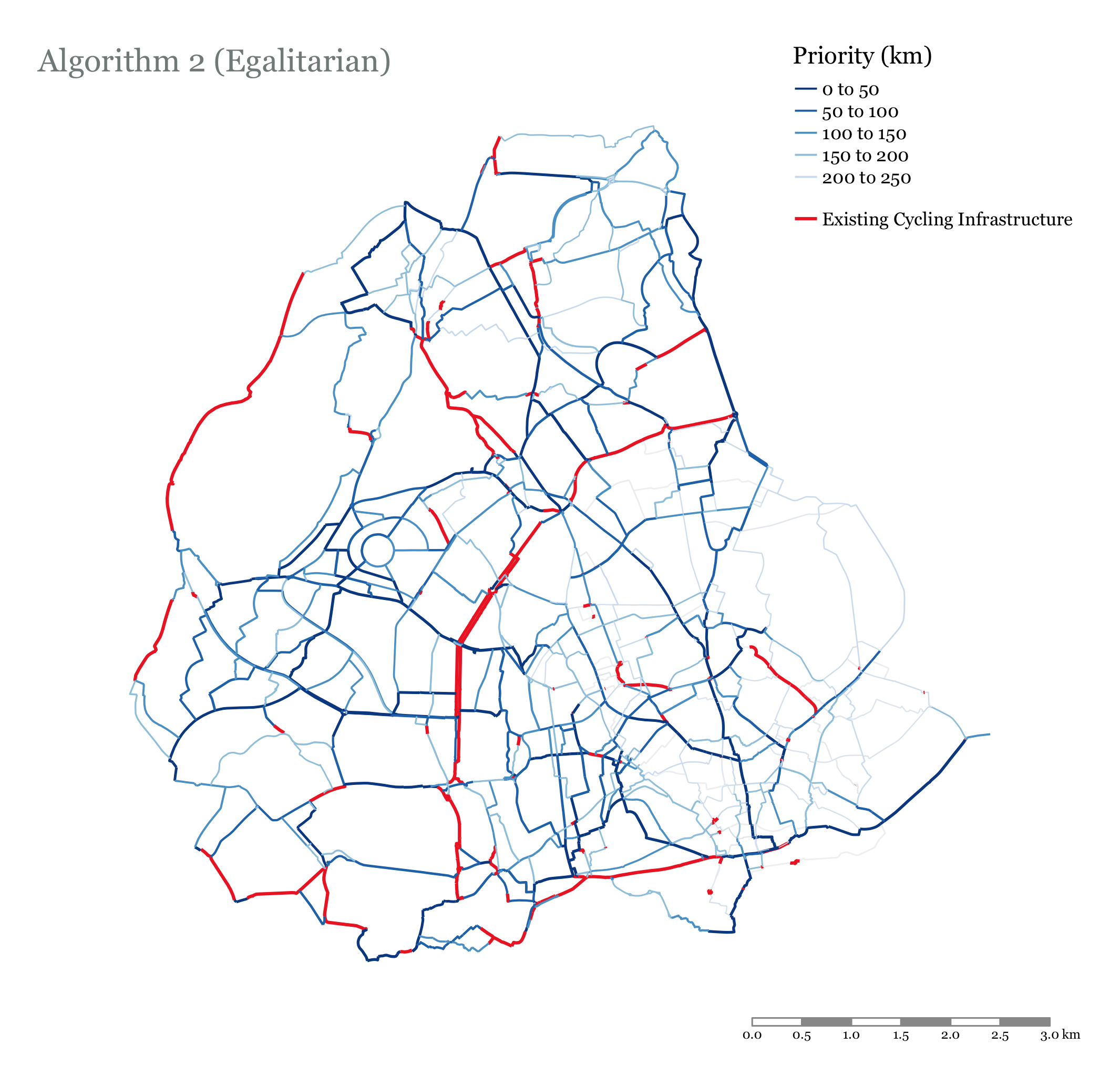} \includegraphics[width=0.45\linewidth]{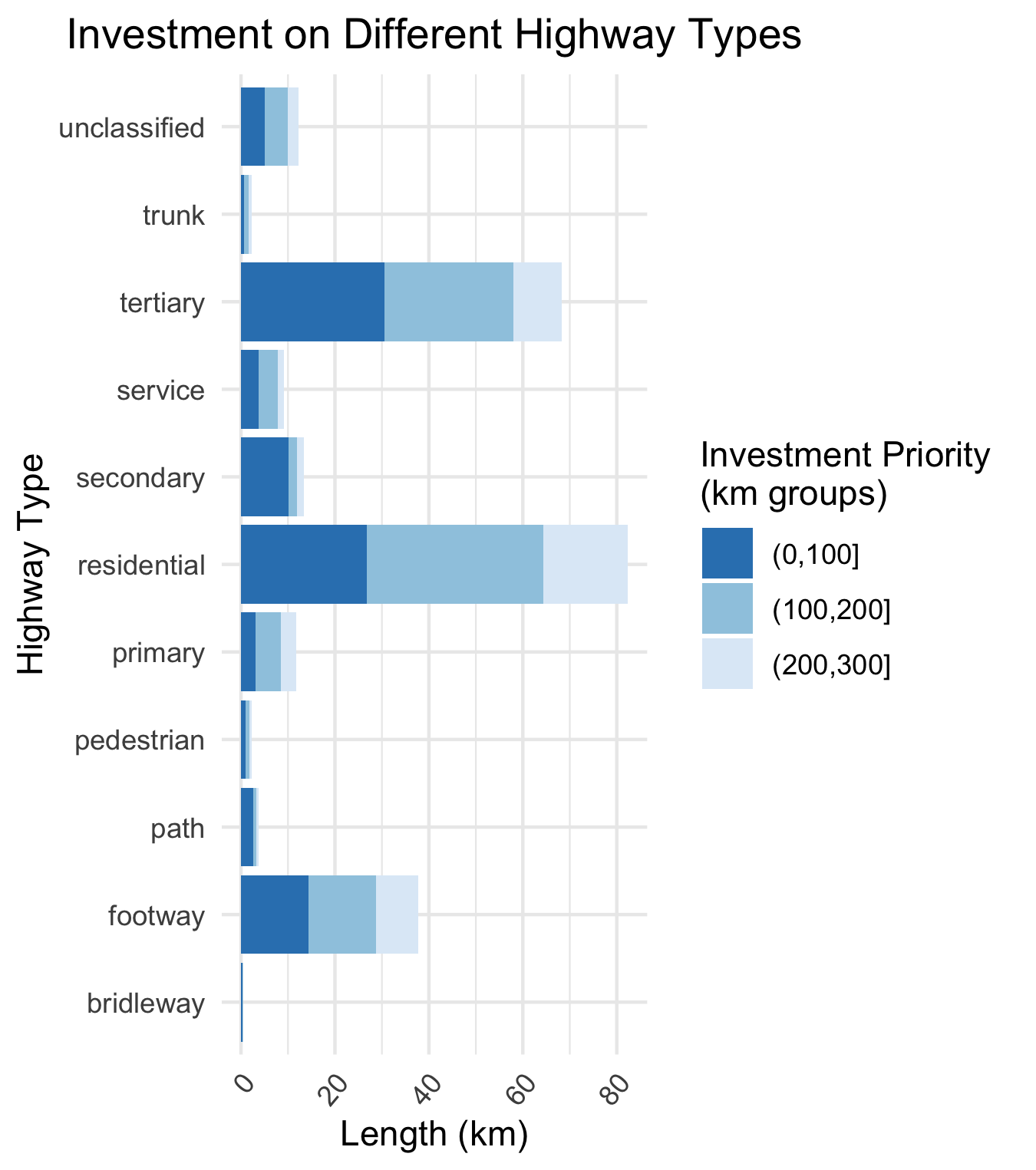} 

}

\caption{Road Segment Priority (left), disaggregated by road type (right) - Egalitarian Growth}\label{fig:growth3MapandBarNottingham}
\end{figure}

\clearpage

\end{document}